\documentclass[iicol,sn-mathphys-num]{sn-jnl}
\usepackage{amsmath} 
\usepackage{lineno,hyperref}
\modulolinenumbers[5]

\usepackage[linesnumbered,ruled]{algorithm2e}
\usepackage{amssymb}
\usepackage{array}
\usepackage{bm}
\usepackage{booktabs}
\usepackage{cases}
\usepackage{caption}
\usepackage{enumitem}
\usepackage{longtable}
\usepackage{lscape}
\usepackage{makecell}
\usepackage{mathtools}
\usepackage{multirow}
\usepackage{standalone}
\usepackage{subcaption}	
\usepackage{tabularx}
\usepackage{titlesec}
\usepackage{xfp}

\makeatletter
\let\cline\@cline 
\makeatother

\newtheorem{remark}{Remark}

\DeclareMathOperator*{\argmax}{arg\,max}

\usepackage[final]{changes} 
\usepackage{environ}
\definechangesauthor[name={Benjamin}, color=purple]{BM}
\definechangesauthor[name={Chen}, color=yellow]{CM}
\definechangesauthor[name={Florian}, color=green]{FK}
\definechangesauthor[name={Michael}, color=blue]{ML}
\definechangesauthor[name={Toprak}, color=violet]{TT}

\usepackage{tikz}
\usepackage{pgfplots}
\usepackage{pgfplotstable}
\usepackage{standalone} 
\usepgfplotslibrary{external}
\usetikzlibrary{pgfplots.groupplots}
\usetikzlibrary{positioning} 
\def\tikzfontsizetiny{\tiny}

\pgfplotsset{
  mystyle/.style ={%
    grid = major,
    every tick label/.append style={font=\scriptsize},
    every axis label/.append style={font=\scriptsize},
    legend style={font=\scriptsize},
    label style={font=\scriptsize},
    title style={font=\scriptsize},
    /pgf/number format/set thousands separator = {}, 
  },
}

\pgfplotsset{select coords between index/.style 2 args={
		x filter/.code={
			\ifnum\coordindex<#1\fi
			\ifnum\coordindex>#2\fi
		}
}}

\definecolor{colorAlgoim}{rgb}{0.00,0.45,0.74}
\definecolor{colorBoSSS}{rgb}{0.85,0.33,0.10}
\definecolor{colorFcmlab}{rgb}{0.93,0.69,0.13}
\definecolor{colorGinkgo}{rgb}{0.49,0.18,0.56}
\definecolor{colorGridap}{rgb}{0.47,0.67,0.19}
\definecolor{colorMlhp}{rgb}{0.49,0.18,0.56}
\definecolor{colorNgsxfem}{rgb}{0.30,0.75,0.93}
\definecolor{colorNutils}{rgb}{0.64,0.08,0.18}
\definecolor{colorQuahog}{rgb}{0.00,0.45,0.74}
\definecolor{colorQuahogPE}{rgb}{0.85,0.33,0.10}
\definecolor{colorQueso}{rgb}{0.49,0.18,0.56}

\pgfplotsset{
	styleAlgoim/.style={color=colorAlgoim, mark=x, solid},
	styleBoSSS/.style={color=colorBoSSS, mark=|},
	styleBoSSS1/.style={color=colorBoSSS, mark=|, dotted, mark options={solid}},
	styleBoSSS2/.style={color=colorBoSSS, mark=|, dashed, mark options={solid}},
	styleBoSSS3/.style={color=colorBoSSS, mark=|, dashdotted, mark options={solid}},
	styleFCMLab/.style={color=colorFcmlab, mark=otimes},
	styleFCMLab1/.style={color=colorFcmlab, mark=otimes, dotted, mark options={solid}},
	styleFCMLab2/.style={color=colorFcmlab, mark=otimes, dashed, mark options={solid}},
	styleFCMLab3/.style={color=colorFcmlab, mark=otimes, dashdotted, mark options={solid}},
	styleGinkgo/.style={color=colorGinkgo, mark=square},
	styleGridap/.style={color=colorGridap, mark=star},
	styleGridap1/.style={color=colorGridap, mark=star, dotted, mark options={solid}},
	styleMlhp/.style={color=colorMlhp, mark=o, solid},
	styleMlhp1/.style={color=colorMlhp, mark=o, dotted, mark options={solid}},
	styleMlhp2/.style={color=colorMlhp, mark=o, dashed, mark options={solid}},
	styleNgsxfem/.style={color=colorNgsxfem, solid, mark=triangle},
	styleNgsxfem1/.style={color=colorNgsxfem, mark=triangle, dotted, mark options={solid}},
	styleNgsxfem2/.style={color=colorNgsxfem, mark=triangle, dashed, mark options={solid}},
	styleNgsxfem3/.style={color=colorNgsxfem, mark=triangle, dashdotted, mark options={solid}},
	styleNutils/.style={color=colorNutils, mark=diamond, solid},
	styleNutils1/.style={color=colorNutils, mark=diamond, dotted,  mark options={solid}},
	styleNutils2/.style={color=colorNutils, mark=diamond, dashed,  mark options={solid}},
	styleNutils3/.style={color=colorNutils, mark=diamond, dashdotted,  mark options={solid}},
	styleQuahog/.style={color=colorQuahog, solid, mark=pentagon},
	styleQuahogPE/.style={color=colorQuahogPE, solid, mark=Mercedes star},
	styleQueso/.style={color=colorQueso, mark=square, solid},
	styleQueso1/.style={color=colorQueso, mark=square, dotted, mark options={solid}},
}

\pgfplotscreateplotcyclelist{mycyclelist}{
	colorAlgoim,every mark/.append style={fill=white,scale=1.2},mark=x\\%
	colorBoSSS,every mark/.append style={fill=white,scale=1.1},mark=|\\%
	colorFcmlab,every mark/.append style={fill=white,scale=1.5},mark=otimes\\%
	colorGinkgo,every mark/.append style={fill=white,scale=1.3},mark=square\\%
	colorGridap,every mark/.append style={fill=white,scale=1.4},mark=star\\%
	colorNgsxfem,every mark/.append style={solid,fill=white,scale=1.2},mark=o\\%
	colorNutils,every mark/.append style={scale=0.6},mark=triangle\\%
	colorQuahog,every mark/.append style={scale=0.55},mark=diamond\\%
}

\def\trianglecolor{black}
\newcommand{\upperSlopeTriangle}[4] 	
{
	\addplot[forget plot, domain=#3:#4,color=\trianglecolor,samples=2]{  #2 / (x^#1) } node (A1) [pos=1] {}; 
	\addplot[forget plot, domain=#3:#4,color=\trianglecolor,samples=2]{   #2  / (#3^#1)} node (A2) [pos=1] {} node [anchor=south,pos=0.5,black] {\tikzfontsizetiny $1$};
	\draw[color=\trianglecolor] (A1.center) -- (A2.center) node [anchor=west,pos=0.5,black] {\tikzfontsizetiny #1};
}
\newcommand{\upperSlopeTriangleFlip}[4] 	
{
	\addplot[forget plot, domain=#3:#4,color=\trianglecolor,samples=2]{  #2 * (x^#1) } node (A1) [pos=1] {};
	\addplot[forget plot, domain=#3:#4,color=\trianglecolor,samples=2]{   #2  * (#3^#1)} node (A2) [pos=1] {} node [anchor=south,pos=0.5,black] {\tikzfontsizetiny $1$};
	\draw[color=\trianglecolor] (A1.center) -- (A2.center) node [anchor=west,pos=0.5,black] {\tikzfontsizetiny #1};
}
\newcommand{\lowerSlopeTriangle}[4] 	
{
	\addplot[forget plot, domain=#3:#4,color=\trianglecolor,samples=2]{  #2 / (x^#1) } node (A1) [pos=0] {}; 
	\addplot[forget plot, domain=#3:#4,color=\trianglecolor,samples=2]{   #2  / (#4^#1)} node (A2) [pos=0] {} node [anchor=north,pos=0.5,black] {\tikzfontsizetiny $1$};
	\draw[color=\trianglecolor] (A1.center) -- (A2.center) node [anchor=east,pos=0.5,black] {\tikzfontsizetiny #1};
}
\newcommand{\lowerSlopeTriangleFlip}[4] 	
{
	\addplot[forget plot, domain=#3:#4,color=\trianglecolor,samples=2]{  #2 * (x^#1) } node (A1) [pos=0] {}; 
	\addplot[forget plot, domain=#3:#4,color=\trianglecolor,samples=2]{   #2  * (#4^#1)} node (A2) [pos=0] {} node [anchor=south,pos=0.5,black] {\tikzfontsizetiny $1$};
	\draw[color=\trianglecolor] (A1.center) -- (A2.center) node [anchor=east,pos=0.5,black] {\tikzfontsizetiny #1};
}

\usepackage{tikzfigwrapper}

\newcommand{\integral}{I}
\newcommand{\integrand}{f}
\newcommand{\integrandQuadDomain}{\hat{f}}
\newcommand{\integrandDegree}{q}
\newcommand{\polynomial}{q}

\newcommand{\polynomialSpace}{\mathbb{P}}
\newcommand{\polynomialInterpolation}{\mathcal{P}}
\newcommand{\xQuadDomain}{\hat{\bm{x}}}

\newcommand{\xElemDomain}{\bm{x}}

\newcommand{\diffOf}[1]{\mathop{}\!\mathrm{d}#1}
\newcommand{\wQuadDomain}{\hat{w}}

\newcommand{\curveDegree}{p}
\newcommand{\reparamDegree}{q_{rep}}
\newcommand{\quadDegreeLower}{k}
\newcommand{\quadDegree}{k_{q}}
\newcommand{\numQuadPoints}{n_{QP}}
\newcommand{\numQuadPointsSetting}{n_{QP,set}}
\newcommand{\numSubLevel}{\ell}

\newcommand{\elemDomain}{K}
\newcommand{\quadDomain}{\hat{K}}
\newcommand{\elemParametrization}{F}
\newcommand{\EuclideanSpace}{\mathbb{R}}
\newcommand{\dimEuclideanSpace}{d}
\newcommand{\dimElem}{{d^e}}

\newcommand{\numElem}{n_{ele}}
\newcommand{\intErrorQuadDomain}{\hat{E}}

\newcommand{\nQuadPoints}{\numQuadPoints}
\newcommand{\TK}{\mathbf{T}_{\elemDomain}} 
\newcommand{\TKa}{\mathbf{T}^a_{\elemDomain}} 

\newcommand{\JK}{\nabla\TK}   
\newcommand{\JKa}{\nabla\TKa}

\newcommand{\detOf}[1]{\det\left(#1\right)}

\newcommand{\polyIntDegreeDomain}[2]{\polynomialInterpolation^{#1}_{#2}}
\newcommand{\quadFunc}{\polyIntDegreeDomain{\quadDegree}{\quadDomain}\integrandQuadDomain}

\newcommand{\semiNormWspk}[2]{\lvert#1\rvert_{#2}}

\newcommand{\elemSize}{h}
\newcommand{\numIntegral}{\integral_{n}}

\newcommand{\elemDomainApprox}{\tilde{K}}
\newcommand{\degreeApprox}{q}
\newcommand{\TKApprox}{\tilde{\mathbf{T}}_{\elemDomain}}
\newcommand{\elemDomainApproxOne}{\tilde{K}_{1}}
\newcommand{\TKApproxOne}{\tilde{\mathbf{T}}_{\elemDomainApproxOne}}
\newcommand{\pApproxOne}{p}
\newcommand{\elemDomainApproxOneB}{\tilde{K}_{\bar{1}}}
\newcommand{\TKApproxOneB}{\tilde{\mathbf{T}}_{\elemDomainApproxOneB}}
\newcommand{\pApproxOneB}{p_{2}}
\newcommand{\elemDomainApproxDeg}{\tilde{K}_{0}}
\newcommand{\TKApproxDeg}{\tilde{\mathbf{T}}_{\elemDomainApproxDeg}}
\newcommand{\pApproxDeg}{p^v}

\newcommand{\JKApprox}{\nabla\TKApprox} 
\newcommand{\JKApproxOne}{\nabla\TKApproxOne} 
\newcommand{\JKApproxOneB}{\nabla\TKApproxOneB} 
\newcommand{\JKApproxDeg}{\nabla\TKApproxDeg} 
\newcommand{\dMax}{d_{\text{max}}}
\newcommand{\gExact}{\integrand\left(\TK{\left(\xQuadDomain\right)}\right)}
\newcommand{\gApprox}{\integrand\left(\TKApprox\left(\xQuadDomain\right)\right)}
\newcommand{\gApproxNumInt}{\polyIntDegreeDomain{\quadDegree}{\elemDomain}\gApprox}
\newcommand{\detJExact}{\det\left(\JK(\xQuadDomain)\right)}
\newcommand{\detJApprox}{\det\left(\JKApprox(\xQuadDomain)\right)}
\newcommand{\integralApprox}{\tilde{\integral}}
\newcommand{\numIntegralApprox}{\integralApprox_n}

\newcommand{\minOrder}{\kappa}

\newcommand{\Q}[1]{\mathbb{Q}_{#1}}      
     
\makeatletter
\let\oldtheequation\theequation
\@ifundefined{tagform@}{%
  \renewcommand\theequation{(\oldtheequation)}
}{%
  \renewcommand\tagform@[1]{\maketag@@@{\ignorespaces#1\unskip\@@italiccorr}}
  \renewcommand\theequation{(\oldtheequation)}
}
\makeatother

\newcommand{\Autoref}[1]{%
	\begingroup%
	\def\chapterautorefname{Chapter}%
	\def\sectionautorefname{Section}%
	\def\subsectionautorefname{Subsection}%
	\autoref{#1}%
	\endgroup%
}
\newcommand{\Autorefs}[1]{%
	\begingroup%
	\def\figureautorefname{Figures}%
	\autoref{#1}%
	\endgroup%
}

\begin{document}

\title{Comparative study of different quadrature methods for cut elements}

\author*[2]{\fnm{Michael} \sur{Loibl}\,\orcid{https://orcid.org/0009-0006-2146-3727}}
\email{michael.loibl@unibw.de}
\equalcont{These authors contributed equally to this work.}
\author*[1]{\fnm{Guilherme H.} \sur{Teixeira}\,\orcid{https://orcid.org/0000-0003-3857-5291}}
\email{teixeira@tugraz.at}
\equalcont{These authors contributed equally to this work.}
\author*[3]{\fnm{Teoman} \sur{Toprak}\,\orcid{https://orcid.org/0009-0005-8817-5044}}
\email{toprak@fdy.tu-darmstadt.de}
\equalcont{These authors contributed equally to this work.}

\author[3]{\fnm{Irina} \sur{Shishkina}\,\orcid{https://orcid.org/0000-0002-1094-8386}}
\author[3]{\fnm{Chen} \sur{Miao}\,\orcid{https://orcid.org/0009-0000-9429-529X}}
\author[2]{\fnm{Josef} \sur{Kiendl}\,\orcid{https://orcid.org/0000-0002-3025-1084}}
\author[3]{\fnm{Florian} \sur{Kummer}\,\orcid{https://orcid.org/0000-0002-2827-7576}}
\author[1]{\fnm{Benjamin} \sur{Marussig}\,\orcid{https://orcid.org/0000-0001-6076-4326}}
\affil[1]{
	\orgdiv{Institute of Applied Mechanics, Graz Center of Computational Engineering (GCCE)},
	\orgname{Graz University of Technology},
	\orgaddress{\street{Technikerstraße 4/II}, \city{Graz}, \postcode{8010}, \country{Austria}}
}
\affil[2]{
	\orgdiv{Institute of Engineering Mechanics and Structural Analysis}, 
	\orgname{University of the Bundeswehr Munich}, 
	\orgaddress{\street{Werner-Heisenberg-Weg 39}, \city{Neubiberg}, \postcode{85577}, \country{Germany}}
	}
\affil[3]{
	\orgdiv{Chair of Fluid Dynamics},
	\orgname{Technical University of Darmstadt},
	\orgaddress{\street{Otto-Berndt-Str. 2},
	\city{Darmstadt}, \postcode{64287}, \country{Germany}}
}

\abstract{
The quadrature of cut elements is crucial for all Finite Element Methods  that do not apply boundary-fitted meshes. It should be efficient, accurate, and robust. Various approaches balancing these requirements have been published, with some available as open-source implementations.
This work reviews these open-sources codes and the methods used. Furthermore, benchmarking examples are developed for 2D and 3D geometries. Implicit and explicit boundary descriptions are available for all models. The different examples test the efficiency, accuracy, versatility, and robustness of the codes. Special focus is set on the influence of the input parameter, which controls the desired quadrature order, on the actual integration error. A detailed comparison of the discussed codes is carried out. The benchmarking allows a conclusive comparison and presents a valuable tool for future code development. All tests are published in an accompanying open-source repository.
}


\maketitle




\section{Introduction}
The Finite Element Method (FEM), which is a widely used tool in research and engineering to perform various simulations, requires the robust meshing of a geometry as an important step in the preprocessing.
This might become a cumbersome and tedious task for arbitrarily complex geometries since manual intervention is often demanded in the meshing process, e.g., geometry repair, de-featuring, and repeated meshing. 
Immersed methods are developed to circumvent this operation,
where the general idea is to embed a (intricate) geometry in a simple background mesh without involving such mesh manipulations. 
The governing equations are then discretized on the simpler background mesh, which is often chosen to be uniform and Cartesian. The mesh is split by the domain boundaries into active and inactive regions. In consequence, some elements might be intersected by the domain boundaries -- such elements are referred to as cut elements -- and, therefore, such approaches necessitate special considerations.
There are numerous methods that implement this described fundamental idea: CutFEM, Embedded Domain Method, Extended FEM (XFEM), Fictitious Domain Method, Finite Cell Method, Unfitted Finite Elements, etc. 
This general concept appears also in many CAD models which are constructed by trimming operations.
Although they may differ in name or in the formulations such as stabilization, boundary-condition enforcement and geometry handling,
we broadly adopt the terminology cut elements to underscore their common problems and tasks on such elements. 

A conclusive overview of the historical development of these methods can be found in \cite{de_prenter_stability_2023}. 
Immersed formulations streamline model preparation, but they introduce distinct numerical tasks compared with standard boundary-fitted FEM. 
The review \cite{de_prenter_stability_2023} lists three core challenges and focuses on the latter one:
\begin{itemize}
	\item the numerical
	evaluation of integrals over cut elements
	\item the imposition
	of (essential) boundary conditions over the immersed
	or unfitted boundaries
	\item the stability of the formulation
	in relation to small cut elements
\end{itemize}
Further comprehensive reviews of this topic can be found in \cite{schillinger_finite_2015} with focus on the Finite Cell Method (FCM), in \cite{marussig_review_2018} with focus on the Isogeometric Analysis (IGA) and trimming, and in \cite{burman_cut_2025} with focus on CutFEM.

The herein presented work deals with the first listed challenge -- the quadrature over cut elements. 
The tensor-product structure of the background mesh is disturbed by arbitrary cutting boundaries. 
This necessitates the application of tailored quadrature rules on these intersected elements. 
Massive research was conducted to address this task.
Possible strategies include quadtree/octree subdivision, moment fitting, element reparametrization, polynomial integration or dimension-reduction of integrals \cite{de_prenter_stability_2023}. 
The vast amount of methods makes the choice of an appropriate tool complicated. 
Fortunately, some of the developed methods were published as open-source codes making them directly accessible and faster applicable. 

In this paper, we review available codes and focus on their applied methods for the quadrature over cut elements. To be precise, thirteen
open-source repositories were found in total, of which ten are discussed and tested in detail. 
A Matlab benchmark environment was implemented which allows the execution of these codes from a common base. 
This interface is also available open-source in \cite{CutElementIntegration2025LatestVersion} and 
its structure was previously presented in \cite{Toprak2026}. 
Here, a number of benchmarks is presented for a comprehensive comparison of the different codes and their applied methods. 
The focus lies on testing their accuracy, robustness, versatility and efficiency.

The benchmarks consider Cartesian background meshes and are designed to simultaneously cover both implicit and parametric boundary descriptions. 
Level set functions are considered as implicit, and B-spline, respectively, Non-Uniform Rational B-Splines (NURBS) as parametric descriptions. 
Implicit boundaries also include voxel representations (cf., e.g., \cite{zander_fcmlab_2014} and \cite{elhaddad_multilevel_2018}), which are, for example, supported by the FCMLab (one of the open-source tools).
However, this is not further investigated herein. 2D and 3D examples are covered.
These benchmarks not only enable an objective comparison of existing methods for this review but also provide the community with a standardized test set to evaluate future implementations, addressing the widely recognized challenge of 'test case design' in research software \cite{eisty_testing_2025}.

The paper is organized as follows. In \autoref{sec:theory}, a general problem description and a corresponding notation are introduced. \Autoref{sec:review} provides a conclusive review of all existing open-source codes. For each code, the applied method and some code details are at first discussed followed by a small study of the quadrature order for three basic 2D geometries. In the subsequent \autoref{sec:numerical_examples}, several numerical benchmarks are computed. Thereby, the codes are compared in detail and particular characteristics are identified. The subsections \ref{sec:numerical_examples_2d} and \ref{sec:numerical_examples_3d} discuss 2D and 3D examples, respectively. The final \autoref{sec:conclusion} concludes the paper and
provides an outlook to further research.

\section{Problem description and notation}
\label{sec:theory}

\begin{standalone}

Our principal interest is the quadrature of cut elements, which specify arbitrary domains $\elemDomain \subset \EuclideanSpace^{\dimElem}$ in $\EuclideanSpace^\dimEuclideanSpace$ with a smooth parametrization $\elemParametrization: \elemDomain\rightarrow\EuclideanSpace^\dimEuclideanSpace$ with $\dimElem \leq \dimEuclideanSpace$.
In this section, we recap some approximation theory mainly based on the seminal works \cite{Stroud1971,Chien1995,Atkinson1978,Atkinson1997,Trefethen2013,Ern2004}.
First, we introduce the basic notation and concepts of numerical quadrature formulas. Subsequently, we summarize how domain transformations affect integration accuracy, particularly when geometric approximations are necessary, providing the theoretical foundation for assessing the accuracy of cut element quadrature methods. Finally, we discuss the determination of the number of quadrature points required.

\subsection{The basic quadrature problem}

The term quadrature denotes the numerical calculation of integrals.
In its simplest form, the quadrature problem\footnote{\autoref{eq:basicQuadratureProblem} may include a known weight function $w(\xQuadDomain)$, which can be chosen to remove integrable singularities. If no singularities are present, it is often set to $w(\xQuadDomain)=1$ and, thus, it will not be considered further herein.} is to compute 
\begin{equation}
	\label{eq:basicQuadratureProblem}
	\integral\left(\integrandQuadDomain\right) = \int_{\quadDomain}\integrandQuadDomain(\xQuadDomain) \diffOf{\xQuadDomain}
\end{equation}
on the reference domain $\quadDomain$ 
for a given continuous function, $\integrandQuadDomain\in C(\quadDomain)$. 
Quadrature methods aim to provide formulas to approximate $\integral$ by a set of $\nQuadPoints$ points $\{\xQuadDomain_1, \ldots, \xQuadDomain_{\nQuadPoints}\}$ in $\quadDomain$ called quadrature points, which are associated with real numbers $\{\wQuadDomain_1, \ldots, \wQuadDomain_{\nQuadPoints}\}$ called quadrature weights.

The quadrature order, $\quadDegree$, is the largest degree $\quadDegreeLower$ of the function space of polynomials $\polynomialSpace_{\quadDegreeLower}$ such that
\begin{align} \label{eq:quadrature_exactness}
\int_{\quadDomain} \polynomial(\xQuadDomain) \diffOf{\xQuadDomain} = \sum_{i=1}^{\nQuadPoints} \wQuadDomain_i \polynomial(\xQuadDomain_i), && \forall \, \polynomial \in \polynomialSpace_{\quadDegreeLower}.
\end{align}
In other words, a quadrature formula introduces an approximation family of polynomials\footnote{As noted by \cite{Trefethen2013}, polynomial interpolants have been the standard idea in numerical quadrature since the 18th century, but other interpolants may be applied as well, see for example \cite[Chapter 22]{Trefethen2013}.}
for the integrand $\integrandQuadDomain$\footnote{Most quadrature methods can be viewed in this framework, but some are better analyzed from another perspective, see e.g., \cite[Section 5.4]{Atkinson1978}.}, which results in
\begin{align}
	\numIntegral^{\quadDegree}\left(\integrandQuadDomain\right)&= \sum_{i=1}^{\nQuadPoints} \wQuadDomain_i  \integrandQuadDomain(\xQuadDomain_i) \nonumber \\
  &=\int_{\quadDomain} \quadFunc(\xQuadDomain) \diffOf{\xQuadDomain}=\integral\left(\quadFunc\right).
\end{align}
where  $\numIntegral^{\quadDegree}$ is the numerical approximation of the integral of order $\quadDegree$, and $\quadFunc$ denotes the result of applying 
the polynomial interpolation operator over $\quadDomain$ to $\integrandQuadDomain$ utilizing polynomials in $\polynomialSpace_{\quadDegree}$.
Thus, we can write the integration error following \cite{Atkinson1997,Ern2004} as
\begin{align}
	\intErrorQuadDomain_{\quadDegree} &=  
	\integral\left(\integrandQuadDomain\right) - \numIntegral^{\quadDegree}\left(\integrandQuadDomain\right) \nonumber \\
	&= \int_{\quadDomain}\left[\integrandQuadDomain(\xQuadDomain)-\quadFunc(\xQuadDomain)\right] \diffOf{\xQuadDomain},
	\\
  |\intErrorQuadDomain_{\quadDegree}| 
  &\leq h \|\integrandQuadDomain(\xQuadDomain)-\quadFunc(\xQuadDomain)\|_\infty \leq c \elemSize_{\quadDomain}^{\quadDegree+1}
    \semiNormWspk{\integrandQuadDomain}{\quadDegree+1,\infty,\quadDomain}
\end{align}
where $\elemSize_{\quadDomain}$ is the diameter of $\quadDomain$ and $\semiNormWspk{\cdot}{\quadDegree+1,\infty,\quadDomain}$ is the seminorm in $W^{\quadDegree+1,\infty}(\quadDomain)$, i.e., the Sobolev space of functions on domain $\quadDomain$ whose weak derivatives of order $\quadDegree+1$ exist and are essentially bounded.
While there is usually no need to set up the approximation polynomials
explicitly, it is worth knowing that we usually require
\begin{align}
	\lim_{\quadDegree\rightarrow\infty} \|\integrandQuadDomain(\xQuadDomain)-\quadFunc(\xQuadDomain)\|_\infty = 0
\end{align}
which also highlights the influence of the integrand $\integrandQuadDomain(\xQuadDomain)$ and its derivatives \cite{Atkinson1978}.

\subsection{Domain transformations}
\label{sec:domaintransformations}

Let $\elemDomain$ be a non-empty, Lipschitz, compact, 
connected subset of $\EuclideanSpace^{\dimEuclideanSpace}$, defined by the mapping $\TK:\quadDomain\rightarrow\elemDomain$ and $\integrand$ a function in $W^{\ell+1,p}(\elemDomain)$.
The change of variables $\xElemDomain=\TK\left(\xQuadDomain\right)$ yields
$\integrand\left(\TK\left(\xQuadDomain\right)\right)=\integrandQuadDomain\left(\xQuadDomain\right)=\left(\integrand \circ \TK\right)\left(\xQuadDomain\right)$ and the integration of a function $\integrand$ over $\elemDomain$ reduces to
\begin{align}
I\left(\integrand\right) &= \int_{\elemDomain} \integrand(\xElemDomain) \diffOf{\xElemDomain} \nonumber\\
    &= \int_{\quadDomain}   \integrand(\TK(\xQuadDomain)) \det\left(\JK\right)  \diffOf{\xQuadDomain}
\end{align}
where $\JK(\xQuadDomain)=\frac{\partial \TK(\xQuadDomain)}{\partial \xQuadDomain}$ is the Jacobian matrix of $\TK$ at $\xQuadDomain$.
The quadrature formula for $\elemDomain$ is 
\begin{align}
	\label{eq:mappedQuadratureProblem}
  \numIntegral\left(\integrand\right) \equiv
	 \sum_{i=1}^{\nQuadPoints}  \wQuadDomain_i    \integrand(\TK(\xQuadDomain_i)) \detOf{\JK(\xQuadDomain_i)}.
\end{align}
Assuming the quadrature order $\quadDegree \geq 0$, $\TK \in C^{\quadDegree+2}(\quadDomain)$, and $\integrand \in C^{\quadDegree+1}(\elemDomain)$ the integration error is \cite[Theorem 5.3.2.]{Atkinson1997}
\begin{equation}
|\integral\left(\integrand\right) - \numIntegral\left(\integrand\right)| \leq c \, \elemSize_{\quadDomain}^{\quadDegree+1} 
\semiNormWspk{\gamma}{\quadDegree+1,\infty,\quadDomain}
\end{equation}
with
\begin{equation}
    \gamma(\xQuadDomain) \equiv \integrand(\TK(\xQuadDomain)) \det\left(\JK(\xQuadDomain)\right),\quad \xQuadDomain\in \quadDomain.
\end{equation}

Note that in the case of cut elements, it is not straightforward to enforce the continuity conditions on  $\TK$. 
Moreover, the integration domain $\elemDomain$ is usually represented by an approximation $\elemDomainApprox$ given by a mapping $\TKApprox: \quadDomain \rightarrow\elemDomainApprox$, as indicated in \autoref{fig:domain_transformation}.
\begin{figure}
	\centering
	\def\plotwidthfactor{1.1} 
	\includegraphics{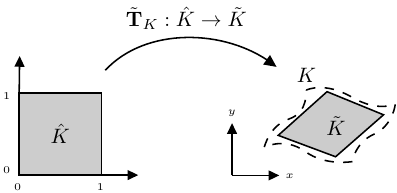}
	\caption{Approximation $\elemDomainApprox$ of an integration domain $\elemDomain$ enclosed by the dashed lines.}
	\label{fig:domain_transformation}	
\end{figure}
The integral over $\elemDomainApprox$ reads
\begin{equation}
  \integralApprox\left(\integrand\right)= \int_{\quadDomain}  \gApprox \detJApprox \diffOf{\xQuadDomain}
\end{equation}
and the quadrature formula for $\elemDomainApprox$ is 
\begin{equation}
	\label{eq:mappedQuadratureProblemApprox}
  \numIntegralApprox\left(\integrand\right)\equiv
	 \sum_{i=1}^{\nQuadPoints}  \wQuadDomain_i    \integrand\left(\TKApprox(\xQuadDomain_i)\right)
   \det\left(\JKApprox(\xQuadDomain_i)\right). 
\end{equation}
With $\gApproxNumInt$ being the numerical quadrature approximation of $\gApprox$,
the integration error becomes
\newcommand{\degreeEvenOdd}{\tilde{\degreeApprox}}
\begin{align}
  \label{eq:integrationErrorApproxDomain}
  E &=
  \integral\left(\integrand\right) - \numIntegralApprox\left(\integrand\right) \nonumber \\
  &= 
  \int_{\quadDomain}
  \gExact \detJExact\,
  \diffOf{\xQuadDomain} \nonumber \\
  &\quad - 
  \int_{\quadDomain} \gApproxNumInt
  \detJApprox \,
  \diffOf{\xQuadDomain} \nonumber \\
  &
  = E_1 + E_2 + E_3 
  \end{align}
with
  \begin{align}
  E_1 &= 
  \int_{\quadDomain} \gExact \nonumber\\
  & \cdot
  \left[\detJExact - \detJApprox\right] \, \diffOf{\xQuadDomain}\\
  E_2 &= 
  \int_{\quadDomain} \left[\gExact -
  \gApproxNumInt\right] \nonumber\\
  &\quad \cdot \detJExact \, \diffOf{\xQuadDomain}\\
  E_3 &= 
  \int_{\quadDomain} \left[\gExact -
  \gApproxNumInt\right] \nonumber\\
  &\cdot \left[\detJApprox - \detJExact\right]     
  \, \diffOf{\xQuadDomain}
  \end{align}
  where $E_1$ compares the difference between exact and approximate Jacobians, $E_2$ compares the errors during the function evaluations, and $E_3$ captures the interaction between both types of errors.
  Suppose $\elemDomainApprox$ is specified by an interpolation of degree $\degreeApprox\geq1$, the error estimate\footnote{Even-degree polynomial interpolation on symmetric meshes leads to an extra order of convergence, i.e., $\degreeApprox + 2$, see \cite{Atkinson1997} for details. We have omitted these superconvergence effects in \autoref{eq:integrationErrorApprox} since cut elements usually do not yield symmetric meshes.} can be expressed as
\begin{align}
  \label{eq:integrationErrorApprox}
  |\integral\left(\integrand\right) - \numIntegralApprox\left(\integrand\right)| 
  \leq c\,\elemSize_{\quadDomain}^{\minOrder} 
\end{align}
with $\minOrder=\min\{\quadDegree+1,\degreeApprox + 1\}$
where $\TK \in C^{\minOrder+1}(\quadDomain)$, $\integrand \in C^{\minOrder}(\elemDomain)$, and
$c$ is dependent on $\integrand$ and the parametrization functions $\TK$, and it is a multiple of the maximum of the norms of all
derivatives of $\gExact$ of order $\leq \minOrder$ and derivatives of $\TK$ of order $\leq \degreeApprox + 1$, including products of these norms, see \cite[Theorem 5.3.3.]{Atkinson1997}. 

In cut element integration, different methods produce different approximations $\elemDomainApprox$, ranging from low-order tessellations to higher-order domain representations. The accuracy of the former is dominated by the interpolation of degree $\degreeApprox$. Higher-order methods, on the other hand, involve more complex and essentially arbitrary domain transformations $\TKApprox$, which directly impact the integration error \ref{eq:integrationErrorApprox}. In addition, the complexity of $\TKApprox$ determines the number of quadrature points required, which we discuss next.


\subsection{Number of quadrature points}
\label{sec:Number-of-quadrature-points}
\subsubsection{The quadrature order of a quadrature formula}

The quadrature order of a quadrature formula, $\quadDegree$, indicates that it is exact for all $\integrandQuadDomain$ within the function space 
 of polynomials of degree $\leq \quadDegree$, $\integrandQuadDomain\in\polynomialSpace_{\quadDegree}$, for any $\quadDegree\ge0$.
The quadrature order, $\quadDegree$, is encoded in the number of quadrature points $\nQuadPoints$.
For instance, the closed Newton-Cotes formulas yield $\quadDegree=\nQuadPoints$ for odd $\nQuadPoints$, but $\quadDegree=\nQuadPoints-1$ for even $\nQuadPoints$, see \cite[Theorem 5.1]{Atkinson1978}.
Furthermore, Gauss~formulas achieve $\quadDegree=2\nQuadPoints-1$.
The standard derivation introduced in \cite{Jacobi1826} is based on orthogonal polynomials. 
This high quadrature order made the Gauss~formulas the most prominent quadrature method used. At the same time, it is worth mentioning that the
quadrature order is not the only aspect determining the rate of convergence of a quadrature method. 
For instance, although Gauss formulas have higher algebraic order ($\quadDegree=2\nQuadPoints-1$) than Clenshaw--Curtis ($\quadDegree=\nQuadPoints-1$), both converge exponentially fast for analytic integrands, see \cite[Chapter 19]{Trefethen2013}.

\subsubsection{The influence of domain transformations}
\label{sec:influence_domain_transformations}

In case of a domain transformation as in \autoref{eq:mappedQuadratureProblemApprox}, the integrand $\integrand\left(\TKApprox\left(\xQuadDomain\right)\right)=\integrandQuadDomain\left(\xQuadDomain\right)=\left(\integrand \circ \TKApprox\right)\left(\xQuadDomain\right)$
is the composition of $\integrand$ and $\TKApprox$.
In the following, we discuss the influence of this composition on the algebraic degree of the integrand to gain insight into the required number of quadrature points to evaluate \autoref{eq:mappedQuadratureProblemApprox} precisely.  
For the sake of clarity, we will focus on the two-dimensional case and consider a bi-variate polynomial tensor product space with bi-degree $(a,b)$ denoted as $\Q{a,b}$.

\paragraph*{General mapping}
Suppose $\integrand \in \Q{q_1,q_2}$ and $\TKApprox \in \Q{p_1,p_2}$. 
We emphasize that $\Q{q_1,q_2}$ contains all polynomials with $\deg_x \leq q_1$ and $\deg_y \leq q_2$ simultaneously, which includes polynomials of total degree up to $q_1+q_2$.
The composition is of bi-degree $\integrand \circ \TKApprox\in\Q{q_1p_1+q_2p_1,q_1p_2+q_2p_2}$.
Furthermore, the scaling factor of the quadrature weight is $ \det\left(\JKApprox\right)  \in \Q{2p_1-1,2p_2-1}$.
Thus, we obtain for equal degrees $p=p_1=p_2$ and $q=q_1=q_2$
\begin{align}
  \integrand \circ \TKApprox \det\left(\JKApprox\right) \in\Q{2p(q+1)-1,2p(q+1)-1}.
\end{align}
For the extension to the three-dimensional case, we refer the interested reader to \cite{Antolin2022}.

\paragraph*{Quadrilateral mapping with one curved edge}
Let's consider a quadrilateral element $\elemDomainApproxOne$ with one curved edge of degree $\pApproxOne$. 
To obtain the corresponding algebraic degree, we can use the general setting with  $p_1=1$, $p_2=\pApproxOne$, and $q=q_1=q_2$, yielding
\begin{align}
  \integrand \circ \TKApproxOne \det\left(\JKApproxOne\right) \in \Q{2q+1,2\pApproxOne(q+1)-1}.
\end{align}
Note the degree reduction in the straight direction, reducing the determinant degree to $\det\left(\JKApproxOne\right)  \in \Q{1,2\pApproxOne-1}$. 

\paragraph*{Axis-aligned quadrilateral mapping with one curved edge}
For cut elements on Cartesian grids, the straight edges are typically aligned with the coordinate axes, which introduces additional structure that significantly reduces the overall degree.
Consider the blended mapping
\begin{align}
  \TKApproxOneB(u,v)=(1-u)\,A(v) + u\,B(v),
\end{align}
where $A(v)=(0,v)$ is the left axis-aligned vertical edge and $B(v)=(v,B_y(v))$ is the curved right edge with $B_y$ of degree $\pApproxOneB$. 
This gives the explicit form
\begin{align}
  \TKApproxOneB(u,v) = (uv,\, (1-u)v + uB_y(v)).
\end{align}
The partial derivatives are
\begin{align}
\partial_u \TKApproxOneB &= (v, B_y(v)-v )&\in\,& \Q{0,\pApproxOneB}, \\
\partial_v \TKApproxOneB &= (u, (1-u) + uB'_y(v))&\in\,& \Q{1,\pApproxOneB-1},
\end{align}
giving the determinant
\begin{align}
\det(\JKApproxOneB) &= v[(1-u) + uB'_y] - u[B_y - v] \nonumber\\
&= v + u[vB'_y - B_y] \quad\in \Q{1,\pApproxOneB}.
\end{align}
For the integrand composition, $\integrand(x,y) = x^i y^j$ with $i,j \leq q$ gives
\begin{align}
(\integrand \circ \TKApproxOneB)(u,v) = (uv)^i \cdot [(1-u)v + uB_y(v)]^j,
\end{align}
i.e., $\integrand \circ \TKApproxOneB\in \Q{2q,q+q\pApproxOneB}$.
Note the degree reduction in the $v$-direction compared to the generic case $\integrand \circ \TKApprox$ discussed above due to the axis-alignment constraint $B_x = v$.
Finally, we obtain
\begin{align}
\integrand \circ \TKApproxOneB \det(\JKApproxOneB) \in \Q{2q+1,\, q\pApproxOneB+q+\pApproxOneB}.
\end{align}

\paragraph*{Degenerate mapping}
Another frequent geometric configuration is that a cut element $\elemDomainApproxDeg$ forms a triangle with two straight edges and a curved one. This can be represented by a degenerate mapping
\begin{align}
  \TKApproxDeg(u,v)&=(\TKApproxDeg^x,\TKApproxDeg^y)\nonumber\\
  &=(u,\,(1-u)\,A(v)+u\,B(v)),
\end{align}
where the $x$-component is constrained to $\TKApproxDeg^x(u,v) = u$, forcing a coordinate-separated structure.
$\TKApproxDeg^y$ describes a blended mapping where $A(v)$ and $B(v)$ denote two opposite edges in the $u$--direction. 
When $A(v) = A_0$ (fixed point), the edge collapses to a vertex and a significant degree reduction occurs.
The partial derivatives simplify to
\begin{align}
\partial_u \TKApproxDeg &= (1,\, B(v)-A_0)&\in~& \Q{0,\pApproxDeg}, \\
\partial_v \TKApproxDeg &= (0,\, u B'(v)) &\in~& \Q{1,\pApproxDeg-1},
\end{align}
giving $\det(\JKApproxDeg) = u\, B'(v) \in \Q{1,\pApproxDeg-1}$.
The degree in $v$ reduces from $2\pApproxDeg - 1$ to merely $\pApproxDeg - 1$ because $A'(v) = 0$ eliminates the corresponding polynomial term from $\partial_v \TKApproxDeg$.
The explicit factor $u$ reflects the geometric singularity at the collapsed vertex ($u=0$).
The algebraic degree of the overall integrand is 
\begin{align}
\integrand \circ \TKApproxDeg \det(\JKApproxDeg) \in \Q{2q+1,\, \pApproxDeg(q+1)-1}.
\end{align}

%
%

%
\begin{table*}
\centering
\caption{Algebraic degree of $\integrand \circ \TK \det(\JK)$ for an integrand of $\integrand \in \Q{q,q}$ and different domain transformations}\label{tab:overviewDegree}
\begin{tabular}{lll}
\toprule
Mapping type & Determinant & Transformed integrand \\ 
\midrule
General mapping $\TKApprox$ & $\Q{2p-1,2p-1}$ & $\Q{2p(q+1)-1,2p(q+1)-1}$  \\
One curved edge $\TKApproxOne$ (generic) & $\Q{1,2p-1}$ & $\Q{2q+1,2p(q+1)-1}$  \\
One curved edge $\TKApproxOneB$ (axis-aligned) & $\Q{1,p}$ & $\Q{2q+1,qp+q+p}$  \\
Degenerate mapping $\TKApproxDeg$ & $\Q{1,p-1}$ & $\Q{2q+1,p(q+1)-1}$  \\
Affine transformations $\TKa$ & $\Q{0,0}$ & $\Q{q,q}$ \\
\bottomrule
\end{tabular}
\end{table*}
\paragraph*{Affine transformations}
An affine transformation, $\TKa : \quadDomain \rightarrow \elemDomain$, takes the form
\begin{align}
  \TKa(\xQuadDomain) = \JKa \xQuadDomain + b_{\elemDomain}
\end{align}
where $b_{\elemDomain}\in \EuclideanSpace^\dimEuclideanSpace$. The Jacobian matrix $\JKa$ is invertible since $\TKa$ is bijective and we further get that its determinant is constant 
\begin{align}
	\left\lvert \detOf{\JKa} \right\rvert=\frac{\text{meas}(\elemDomain)}{\text{meas}(\quadDomain)},
\end{align}
and thus, $\detOf{\JKa} \in \Q{0,0}$.

\paragraph*{Summary of the overall algebraic degrees}
\autoref{tab:overviewDegree} summarizes the polynomial degrees of the tensor-product integrand $\integrand \in \Q{q,q}$ composed with the domain transformations discussed above with maximal degree $p$.
Note that the dominant degree $\dMax=\max(d_x,d_y)$ of $\Q{d_x,d_y}$ of the transformed integrand depends significantly on the domain transformation. 
\autoref{fig:integrand_complexity_surface} illustrates the different complexities of $\dMax$ for combinations of moderate degrees $q$ and $p$. 
\begin{figure}[ht]
	\centering
	\def\plotwidthfactor{0.9} 
	\includegraphics{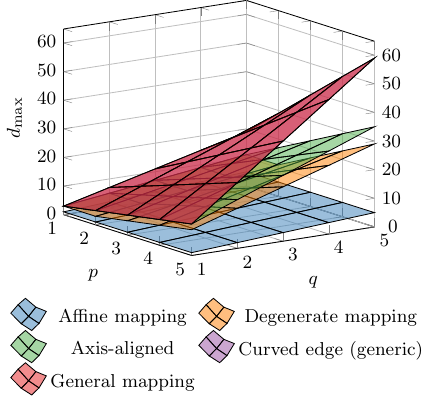}
	\caption{Dominant degree $\dMax$ of $\integrand \circ \TK \det(\JK)$ for $f \in \Q{q,q}$ and different domain transformations of maximal degree $p$. The case \emph{Curve edge (generic)} is not visible since its dominant degree coincides with the \emph{General mapping}.}
	\label{fig:integrand_complexity_surface}	
\end{figure}
It is apparent that the general mapping $\TKApprox$ has the highest $\dMax$, which is significantly reduced in the case of the structured domain transformations, especially for higher degrees $p$ and $q$.
Considering $q=p=5$ for instance, the general mapping $\TKApprox$ results in $\dMax=59$, which drops to around $\dMax\approx30$ for the axis-aligned transformation with one curved edge $\TKApproxOneB$ and the degenerate mapping $\TKApproxDeg$, or to the minimum of $\dMax=5$ in the affine case.

The key takeaway is that the polynomial degree $p$ of the geometric mapping $\TK$ is not sufficient to get a sharp estimate of the dominant degree $\dMax$. 
At the same time, knowing $\dMax$ is relevant to set the proper number of quadrature points.
For tensor-product Gauss rules with $\nQuadPoints$ points per direction, the requirement $2\nQuadPoints-1 \geq \dMax$ gives $\nQuadPoints = \lceil(\dMax+1)/2\rceil$.
Using the general mapping $\TKApprox$ provides the worst case scenario which is often not required in cut element integration.   
We will continue these considerations in \autoref{sec:three_geometries}, where we discuss the set-up to numerically investigate the number of required quadrature points for the investigated integration schemes.

\end{standalone}

\section{Review of open-source codes}\label{sec:review}

This section introduces all open-source codes considered. 
At first, an overview is provided in \autoref{sec:overview}.
The different codes are discussed in \autoref{sec:quadtree} to \autoref{sec:further_codes}.
For first basic studies,
\autoref{sec:three_geometries} presents three 2D test case geometries which are subsequently used to test each code and study how the input parameters discussed in \autoref{sec:quadtree} to \autoref{sec:further_codes} affect the integration accuracy.

\subsection{Overview}
\label{sec:overview}

\autoref{tab:overview} documents all 13 open-source codes dealing with immersed methods which we identified in our search. They are grouped by their method for cut element integration.
The `reparametrization' category is intentionally broad to provide organizational structure for the following subsections. 
The second column lists the code names used throughout this paper with hyperlinks providing direct access to their open-source repositories; complete URLs are given in the respective subsections (last column).
In the numerical examples in \autoref{sec:numerical_examples}, we also introduce the additional name QuaHOGPE to label a special version of the QuaHOG code, as outlined in \autoref{sec:QuaHOG}. 
 
Most codes handle implicit boundary representations, while two support parametric descriptions. 
QuESo is a special case, using STL files for 3D boundary representations.
All codes can handle 2D as well as 3D geometries except for QUGaR and QuaHOG.
It is worth noting that the application domains and programming languages are diverse.
The operating system (OS) column indicates on which platform we integrate a code into our Matlab benchmark environment \cite{CutElementIntegration2025LatestVersion}.
In other words, the code itself may still be compatible with an OS, even if it is not listed.
Codes marked `n/a' are not included in the comparison, hence we did not test their compatibility.

As marked in the last column, ten codes are reviewed and compared in detail, while three are only briefly discussed in \autoref{sec:further_codes}, including a rationale for their exclusion. 

\begin{sidewaystable*}
	\caption{Overview of open source codes}\label{tab:overview}
	\renewcommand{\arraystretch}{1.1} 
	\begin{tabular*}{\textheight}{@{\extracolsep\fill}p{1.7cm}|p{2.4cm}|p{1.5cm}|p{1.8cm}|p{3.3cm}|p{1.2cm}|p{1.7cm}|p{1.2cm}|p{1.3cm}|p{1.5cm}}%
		
		\toprule%
		\multicolumn{2}{l|}{\textbf{Quadrature method}} & \textbf{Name} & \textbf{Boundary rep.} & \textbf{Application} & \textbf{Dimen-sions} & \textbf{Language} & \textbf{Tested OS} & \textbf{License} & \textbf{Included} \\ \midrule \midrule
		
		\multicolumn{2}{l|}{\multirow{6}{*}{\parbox{1.5cm}{Quadtree/Octree}}} & \href{https://gitlab.lrz.de/cie_sam_public/fcmlab}{FCMLab} & Implicit & p-FEM, FCM & \raggedright 2D \& 3D & Matlab & Linux \& Win & GPL-3.0 & \raggedright \raggedright Yes (Sec. \ref{sec:fcmlab}) \\ \cmidrule{3-10}
		
		\multicolumn{2}{l|}{} & \href{https://github.com/evalf/nutils}{Nutils} & Implicit & \raggedright FEM, IGA,
		FCM, multi-physics & \raggedright 2D \& 3D & Python & Linux \& Win & MIT & \raggedright Yes (Sec. \ref{sec:nutils}) \\ \cmidrule{3-10}
		
		\multicolumn{2}{l|}{} & \href{https://github.com//tpmc}{TPMC} & Implicit & Marching cubes algorithm & \raggedright 2D \& 3D & C++, Python UI & n/a & GPL-2, LGPL-3 & \raggedright No (Sec. \ref{sec:further_codes})  \\ \cmidrule{1-10}
		
		\multirow{6}{*}{\parbox{2cm}{Reparame- trization by}} & Background mesh mapping & \href{https://github.com/ngsxfem}{ngsxfem} & Implicit & Active-mesh XFEM,	space-time FEM & \raggedright 2D \& 3D & C++, Python UI & Win & LGPL-3.0 & \raggedright Yes (Sec. \ref{sec:ngsxfem})\\ \cmidrule{2-10}
		
		& Blending Function Method & \href{https://github.com/pantolin/qugar}{QUGaR} & Parametric & Solution of sparse linear systems & 2D & Matlab & Win & BSD-3 & \raggedright Yes (Sec. \ref{sec:QUGaR}) \\ \cmidrule{2-10}
		
		& Triangulation of cut elements & \href{https://github.com/gridap/Gridap.jl}{Gridap} & Implicit & non-linear, multi-field PDEs & \raggedright 2D \& 3D & Julia & Win & MIT & \raggedright Yes (Sec. \ref{sec:gridap}) \\ \cmidrule{1-10}
		
		\multirow{6}{*}{\parbox{1.7cm}{Moment fitting with reference from}} & \multirow{4}{*}{\parbox{2.5cm}{Divergence theorem}} & \href{https://github.com/FDYdarmstadt/BoSSS}{BoSSS} & Implicit & High-order DG for multi-physics flows & \raggedright 2D \& 3D & C\# & Win & Apache 2.0 & \raggedright Yes (Sec. \ref{sec:BoSSS}) \\ \cmidrule{3-10}
		
		& & \href{https://github.com/manuelmessmer/QuESo}{QuESo} & STL & FEM, IGA, multi-physics & 3D & C++, Python UI & Linux \& Win & BSD-4 & \raggedright Yes (Sec. \ref{sec:queso}) \\ \cmidrule{2-10}
		
		& Quadtree/ Octree & \href{https://gitlab.com/hpfem/code/mlhp}{mlhp} & Implicit & hp-FEM & \raggedright 2D \& 3D & C++, Python UI & Linux \& Win & MIT & \raggedright Yes (Sec. \ref{sec:mlhp}) \\ \cmidrule{1-10}
		
		\multirow{9}{*}{\parbox{2cm}{Dimension-reduction through}} & \multirow{7}{*}{\parbox{2.4cm}{\raggedright Recasting as the graph of height functions}} & \href{https://github.com/algoim/algoim}{Algoim} & Implicit & High-order multi-scale	multi-physics & \raggedright 2D \& 3D  & C++ & Linux & BSD-3 & \raggedright Yes (Sec. \ref{sec:algoim}) \\ \cmidrule{3-10}
		
		&& \href{https://github.com/CutFEM/CutFEM-Library}{CutFEM-Library} & Implicit & CutFEM algorithms & \raggedright 2D \& 3D & C++ & n/a & GPL-3.0 & \raggedright No (Sec. \ref{sec:further_codes}) \\ \cmidrule{3-10}
		
		&& \href{https://github.com/dealii/dealii}{deal.II} & Implicit & FEM, multi-physics & \raggedright 2D \& 3D & C++ & n/a & LGPL-2.1 & \raggedright No (Sec. \ref{sec:further_codes}) \\ \cmidrule{2-10}
		
		& Green's theorem & \href{https://github.com/davidgunderman/QuaHOG}{QuaHOG} & Parametric & Higher-order quadrature algorithms & 2D & Matlab & Linux \& Win & BSD-3 & \raggedright Yes (Sec. \ref{sec:QuaHOG}) \\
		\botrule
	\end{tabular*}
\end{sidewaystable*}

\subsection{Three 2D test case geometries}
\label{sec:three_geometries}

\begin{figure}
	\centering
	\begin{subfigure}{0.2\textwidth}
		\centering
		\includegraphics[width=\textwidth]{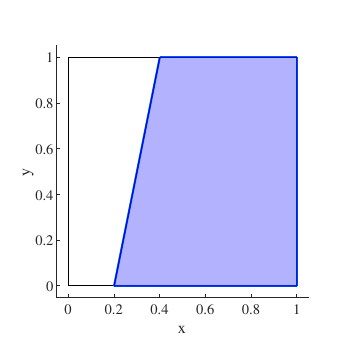}
		\caption{Case 1, $\curveDegree=1$}
		\label{fig:geo_area_single_q1}
	\end{subfigure}
	\hfill
	\begin{subfigure}{0.2\textwidth}
		\centering
		\includegraphics[width=\textwidth]{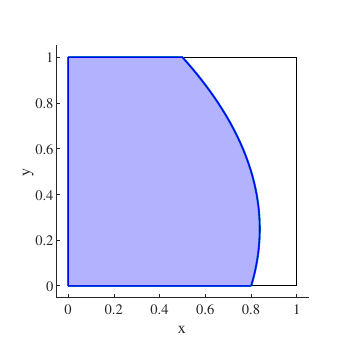}
		\caption{Case 2, $\curveDegree=2$}
		\label{fig:geo_area_single_q2}
	\end{subfigure}
	\hfill
	\begin{subfigure}{0.2\textwidth}
		\centering
		\includegraphics[width=\textwidth]{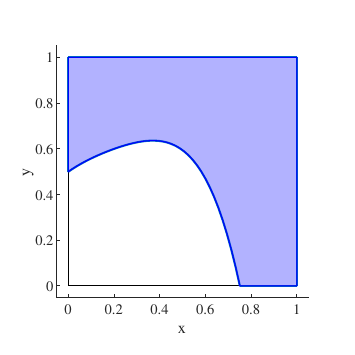} 
		\caption{Case 3, $\curveDegree=5$}
		\label{fig:geo_area_single_q5}
	\end{subfigure}
	\caption{Three 2D test case geometries with different curve degree $\curveDegree$. The active domain is indicated by the thicker blue line and the shaded area. The background domain is indicated by the black line and consists of $\Omega=\{\bm{x}\in [0,1]\times [0,1]\}$.}
	\label{fig:geo_area_single}
\end{figure}

A first numerical study is carried out when discussing the single codes in the following. For this purpose, three distinct 2D geometries with different curve degree (linear, quadratic, quintic) are introduced at this point. They are 
illustrated in \autoref{fig:geo_area_single}. 
The boundary curves are explicitly described  
in these three cases which enables an analytical computation of reference integrals.

The three geometries resemble cutting situations of single elements in order to investigate the ability of the different codes/methods to deal with such features.
The first test case (\autoref{fig:geo_area_single_q1}) presents a cut by a straight inclined line resulting in a trapezium, the second test case (\autoref{fig:geo_area_single_q2}) shows a cut by a quadratic curve resulting in a convex shape, and the third test case (\autoref{fig:geo_area_single_q5}) depicts a cut by a quintic curve resulting in a concave, five-sided shape. The last geometry is of particular interest since it requires subdivision of the domain for some quadrature methods because it is five-sided.

For any quadrature rule, an important parameter, which is subsequently studied, is the quadrature order $\quadDegree$ respectively the number of quadrature points $\numQuadPointsSetting$.
In all following studies, $\numQuadPointsSetting$ is chosen as unified input parameter to control the desired quadrature order in each code. $\numQuadPointsSetting$ defines the number of quadrature points per element and direction. All codes, that directly apply $\numQuadPointsSetting$, use a rule based on a 1D Gaussian quadrature. However, many codes rather require a quadrature order $\quadDegree$ as input. Therefore, the desired quadrature order is internally computed to be $\quadDegree = 2\cdot\numQuadPointsSetting-1$ (Gaussian relation). This input parameter is kept flexible in all codes because it depends on the specific application case (i.e., the order of the considered integrand).
Be aware that $\numQuadPointsSetting$ denotes an input parameter/\underline{set}ting whereas $\numQuadPoints$ refers to an actual number of points in this work. Some codes are further controlled by other input values, whereby the most important settings are stated in the specific code descriptions.
The study is performed based on three questions:
\begin{enumerate}
	\item How does an increasing number of quadrature points $\numQuadPointsSetting$ improve the accuracy of an area computation?
	\item How does $h$-refinement improve the accuracy of an area computation?
	\item How does an increasing number of quadrature points $\numQuadPointsSetting$ improve the accuracy of an integral computation with a monomial as integrand?
\end{enumerate}

These questions cover different aspects of quadrature errors discussed in \autoref{sec:theory}.
Question 1 investigates the error due to the approximate Jacobians of a quadrature's domain transformation (cf.~error $E_1$ from \autoref{eq:integrationErrorApproxDomain}).
Question 2 measures the convergence order $\minOrder$ of the integration error specified in \autoref{eq:integrationErrorApprox}.
Question 3 adds the errors during the function evaluation to the assessment (cf.~errors $E_2$ and $E_3$ from \autoref{eq:integrationErrorApproxDomain}).

For the last investigation, the considered monomials $f$ have the same degree $\integrandDegree$ in both directions: $f(x,y)=x^{\integrandDegree} y^{\integrandDegree}$.
Since the actual integrals are computed over the active and not the complete domain, the functions $f$ are shifted and scaled to resemble monomials on the bounding box of the active domain being defined by $\{\bm{x}\in [x_{min},x_{max}] \times [y_{min},y_{max}]\}$.
I.e., the actual integrands are consequently 
\begin{align}
	\tilde{f}(x,y)&=((x-x_{min})/(x_{max}-x_{min}))^{\integrandDegree} \nonumber\\
	&\cdot ((y-y_{min})/(y_{max}-y_{min}))^{\integrandDegree} .
\end{align}
All investigated codes are designed to correctly integrate polynomials, which appear, e.g., as basis functions in FEM.

Not all three study questions are investigated for all codes. Particular codes are limited in resolving the considered test geometries with a single element. In this case, only an $h$-refinement study is performed. On the other hand, some codes are able to exactly compute the integrals on a single element with a sufficiently high $\numQuadPointsSetting$. Therefore, no $h$-refinement study is performed for them.

For the higher-order quadrature schemes Algoim, BoSSS, ngsxfem, QuaHOG and QUGaR, the expected value for an optimal $\numQuadPointsSetting$ for an exact integral depends on the domain transformation, as discussed in \autoref{sec:Number-of-quadrature-points}.
Considering a general tensor-product mapping of degree $\curveDegree$ (i.e., the degree of the boundary curve), the number of required Gauss quadrature points becomes
\begin{equation}\label{eq:nQP_exact}
	n_{QP,set,general} = \lceil\curveDegree(\integrandDegree+1)\rceil.
\end{equation}
This number provides a conservative choice.
For an axis-aligned quadrilateral mapping with one curved edge of degree $\curveDegree$, the number of Gauss quadrature points reduces to
\begin{equation}\label{eq:nQP_exact_single}
	\numQuadPointsSetting = \left\lceil \frac{\integrandDegree\curveDegree + \integrandDegree + \curveDegree + 1}{2}\right\rceil.
\end{equation}
Note that this mapping situation applies to our three test case geometries, in particular to cases 1 and 2. Thus, we expect that these numerical experiments will be more aligned with \autoref{eq:nQP_exact_single}. 

Therefore, all studies regarding $\numQuadPointsSetting$ are carried out up to this value plus one additional point to confirm convergence. This means, for example, that for a curve degree $\curveDegree=2$ (second test case) and an integrand of degree $\integrandDegree=6$ $\numQuadPointsSetting$ is studied up to $\numQuadPointsSetting=12$.
The same parameter range is chosen for FCMLab and mlhp, although they employ the quadtree method for which \autoref{eq:nQP_exact_single} does not apply, since $\curveDegree$ controls quadrature point density rather than domain interpolation degree in this context. The same holds for Nutils where, however, the maximum quadrature order is limited by the program to a maximum $\quadDegree=6$ which corresponds to a maximum $\numQuadPointsSetting=4$.

A comparison of the different codes with respect to these studies is provided in \autoref{sec:monomials} for the monomial integrands. The input parameter `reparametrization degree', which is used by some codes as stated in the following subsections, is chosen to be the respective curve degree. Please note that $2.22\cdot 10^{-15}$ was chosen to define machine precision which varied in the actual simulations due to the usage of different computers and OS. Values below this machine precision are plotted as this number in all graphs. A higher decimal precision is not considered.


\begin{figure*}[h!]
	\centering
	{\includegraphics[width=\textwidth,trim={0 20 0 10},clip]{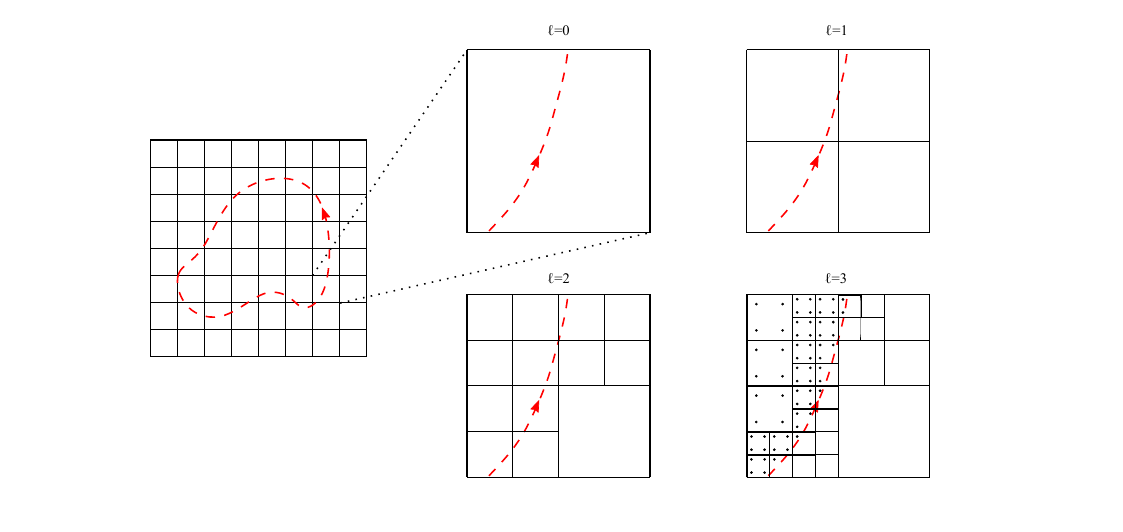}}
	\caption{Quadtree concept, $\numSubLevel$ is the subdivision level}
	\label{fig:fcmlab_concept}
\end{figure*}

\subsection{Quadtree/Octree}
\label{sec:quadtree}

The quadtree approach (in 3D the corresponding octree version) is a well established concept in the context of immersed methods and is very popular due to its robustness and simplicity. \autoref{fig:fcmlab_concept} illustrates the general concept for a 2D problem. Each cut element is uniformly subdivided (bisected in each direction). For each subdivided element, it is recursively checked whether the subelements are intersected by the domain's boundary curve. If a subelement is cut, it is further subdivided until a predefined level of subdivisions $k$ is reached. In the end, (Gaussian) quadrature points are placed within all active subelements. In a next step, an inside/outside check is performed for all quadrature points in active, cut subelements. The inside/outside check is easily performed with implicit boundary descriptions. Therefore, the method is less often used in combination with parametric boundary descriptions.

\subsubsection{FCMLab}
\label{sec:fcmlab}

\paragraph{Applied method and code details}

FCMLab\footnote{\url{https://gitlab.lrz.de/cie_sam_public/fcmlab}, accessed: 25.3.2025} (FCM = Finite Cell Method) is the earliest published software we identified. It deals with the integration of cut elements in the context of FEM. The open-source code also contains IGA modules which are however not fully supported\footnote{Private communication with S. Kollmannsberger, 2023}.
FCMLab applies the quadtree and octree method for 2D and 3D, respectively, which were explained above. The required inside/outside check is easily performed since FCMLab uses an implicit boundary description. Two ways of computing boundary integrals are implemented. The more inaccurate approach also uses a quadtree to track the boundary by means of cut elements. Then, the boundary integration is performed by simply integrating over all cut elements. The more accurate approach builds a linear approximation of the implicit boundary. The most important parameters to control the integration settings are:
\begin{itemize}
	\item number of Gauss points
	\item number of subdivision levels $\numSubLevel$; three levels are used in our studies since this value was mostly chosen in \cite{zander_fcmlab_2014}
	\item number of points per element to generate linear boundary approximation (called SeedPoints in \cite{zander_fcmlab_2014})
\end{itemize}
The computation of normal vectors along the boundaries does not seem to be directly provided in the form which would be required for flux computations. The code is ready to run and requires no installation. A helpful Wiki is published\footnote{\url{https://gitlab.lrz.de/cie_sam_public/fcmlab/-/wikis/home}, accessed: 25.3.2025}. The interfacing of FCMLab in our benchmark environment is eased due to the fact that it is written in Matlab which is also the scripting language we used. In detail, multiple inherited objects are added to overload existing class methods.

\paragraph{Study of quadrature order for the three 2D test case geometries}

The general setup of the following study is provided in \autoref{sec:three_geometries}.

\begin{figure}
	\centering
	\def\plotwidthfactor{0.8} 
	\includegraphics{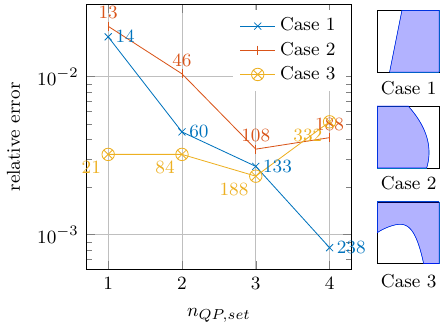}
	\caption{FCMLab: Error of area computation with different $\numQuadPointsSetting$ for the three test case geometries. The numbers next to the markers note $\numQuadPoints$.}
	\label{fig:fcmlab_area}
\end{figure}

\begin{figure*}
	\centering
	\def\plotwidthfactor{0.35} 
	\includegraphics{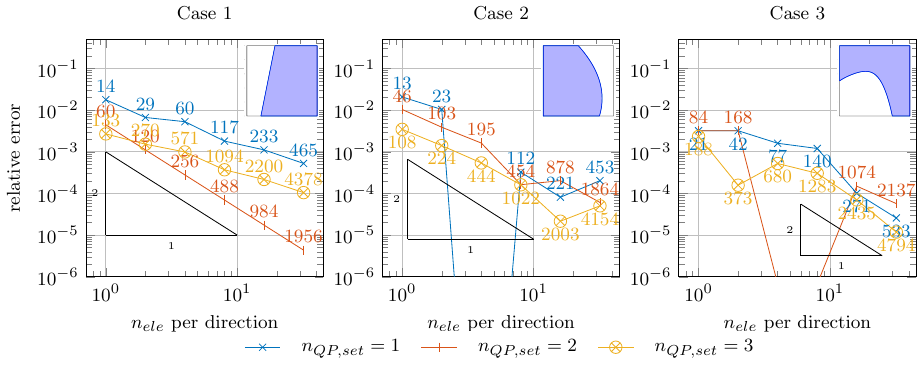}
	\caption{FCMLab: Error of area computation with $h$-refinement for the three test case geometries and different number of quadrature points $\numQuadPointsSetting$. The numbers next to the markers note $\numQuadPoints$.}
	\label{fig:fcmlab_href}	
\end{figure*}

\begin{figure*}
	\centering
	\def\plotwidthfactor{0.35} 
	\includegraphics{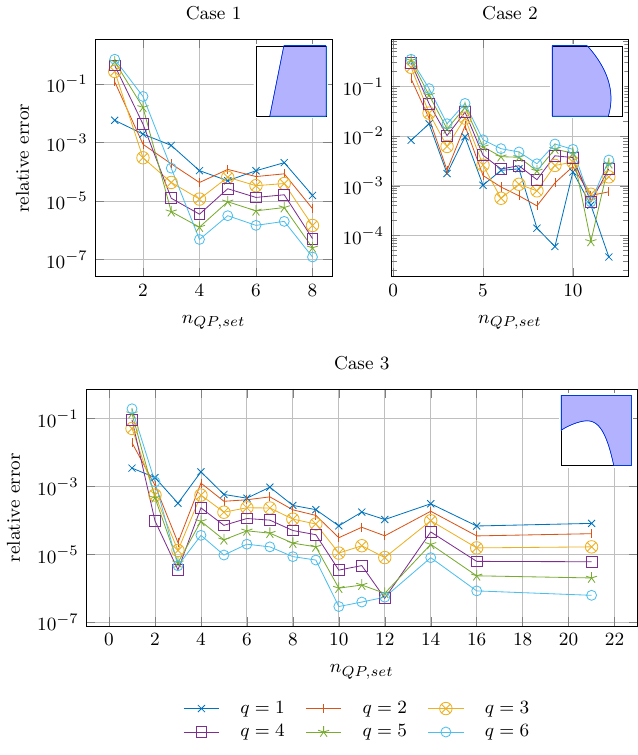}
	\caption{FCMLab: Error of integral computation with different $\numQuadPointsSetting$ for the three test case geometries and different integrands of degree $\integrandDegree$.} 
	\label{fig:fcmlab_integrand}	
\end{figure*}

\autoref{fig:fcmlab_area} and \autoref{fig:fcmlab_href} show the influence of $\numQuadPointsSetting$ and of $h$-refinement on the accuracy for an area computation, respectively. \autoref{fig:fcmlab_integrand} depicts the influence of $\numQuadPointsSetting$ on the accuracy for an integral computation with a monomial of degree $\integrandDegree$ as integrand.
It has to be mentioned that arbitrarily high values of $\numQuadPointsSetting$ cannot be computed with FCMLab since it uses hard-coded Gaussian quadrature rules which are not provided for all orders (cf.~\autoref{fig:fcmlab_integrand}).

In \autoref{fig:fcmlab_area}, FCMLab shows a reduction of the error for case 1 and 2. The error for case 3 is already lower from the beginning but does not show any decrease. General problems to correctly resolve the geometries with a single element are observed.

Three different values of $\numQuadPointsSetting$ are investigated in the $h$-refinement study (\autoref{fig:fcmlab_href}) since no clear dependence was observed in \autoref{fig:fcmlab_area}.
However, a strong trend regarding the relative errors for increasing $\numQuadPointsSetting$ also cannot be detected by this study.
At the same time, the total number of quadrature points $\numQuadPoints$ significantly grows with higher $\numQuadPointsSetting$. The convergence rate is observed to be roughly of second order for all graphs except for case 1 with $\numQuadPointsSetting=1$ and $\numQuadPointsSetting=3$ where it is lower. For case 3, the graphs start with a plateau and the rate becomes visible starting from $\numElem\approx 10^1$.
Quadtree methods often yield a trend of decreasing geometrical error but also introduce randomness disturbing the convergence behavior as observed in the cases 2 and 3.

In the study with monomials as integrand (\autoref{fig:fcmlab_integrand}), FCMLab clearly demonstrates an increasing accuracy with higher $\numQuadPointsSetting$. Nevertheless, the error starts to level off far from machine precision for $\numQuadPointsSetting$ greater than four.

\subsubsection{Nutils}
\label{sec:nutils}

\begin{figure*}[h]
	\centering
	\includegraphics[width=\textwidth]{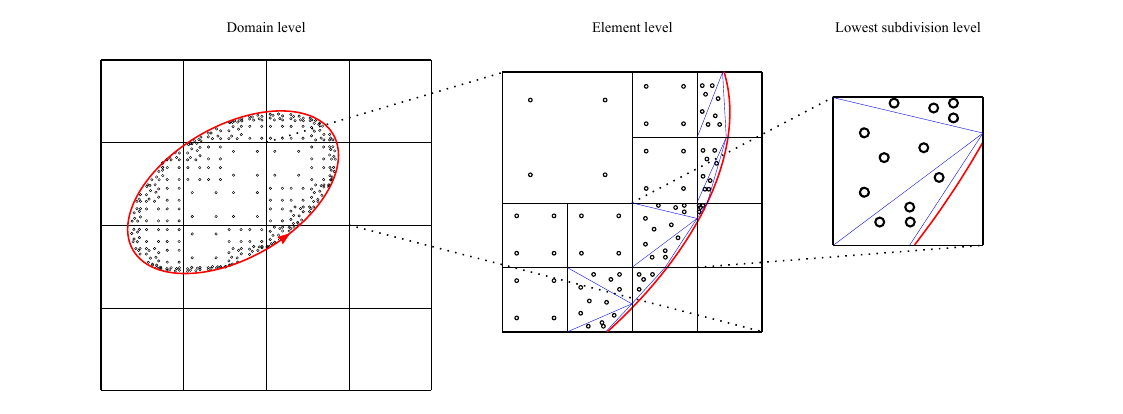}
	\caption{Quadtree concept with tessellation on lowest subdivision level. Tessellation is identified by blue lines.}
	\label{fig:nutils_concept}
\end{figure*}

\paragraph{Applied method and code details}
Nutils\footnote{\url{https://github.com/evalf/nutils}, accessed: 25.3.2025} is a Python library with a broad field of applications including FEM and IGA. For the simulation of immersed models, the software uses the quadtree and the octree method for 2D and 3D, respectively, which were explained above. In contrast to FCMLab, a tessellation is additionally applied on the lowest subdivision level. 
A simplex quadrature rule is used in the resulting triangles or tetrahedra which improves accuracy in contrast to a masked Gauss rule \cite{Verhoosel2015}. Nutils only supports implicit boundary descriptions. Therefore, it is advantageous that the tessellation directly provides a boundary parametrization, e.g., for the imposition of boundary conditions. The concept is illustrated in \autoref{fig:nutils_concept}.
It can be observed that the tessellation does not exactly catch the intersection points between boundary curve and subdivision grid since a linear interpolation of the level set function is used. Nutils also supports the definition of different quadrature orders on each subdivision level which increases the efficiency of the underlying method \cite{divi_error-estimate-based_2020}. However, we did not employ this feature in our benchmark environment for simplicity reasons. The most important parameters to control the integration settings are:
\begin{itemize}
	\item quadrature order
	\item number of subdivision levels $k$; the tessellation is not counted as separate level; three levels are used in our studies analogically to the setting for the FCMLab (cf. \autoref{sec:fcmlab})
\end{itemize}
The code is available as Python package. A short tutorial\footnote{\url{https://nutils.org/}, accessed: 25.3.2025} and several example files\footnote{\url{https://examples.nutils.org/}, accessed: 25.3.2025} including immersed models are provided. Furthermore, Nutils maintains a community forum\footnote{\url{https://github.com/evalf/nutils/discussions}, accessed: 25.3.2025}. 
The interfacing of Nutils in our benchmark environment is
done by creating a Python file with a generic  model which can be called from Matlab by \texttt{pyrunfile}. Unfortunately, a small change to the library has to be carried out because the Matlab-Python interface redirects the command \texttt{sys.stdout} to \texttt{MexPrintf} which is missing a property Nutils tries to call (see \cite{CutElementIntegration2025LatestVersion} for details). Furthermore, special care is required when transferring large integers w.r.t.~the possibly differing invoked integer precision in Matlab versus the underlying Numpy library\footnote{\url{https://numpy.org/}, accessed: 25.3.2025}. In addition, quadrature orders higher or equal to seven are not supported. The authors of Nutils already investigated the influence of their integration settings on the geometry and integration error of the described method. Furthermore, they successfully performed a robustness test for an example with an elliptic exclusion where they varied the ratio and the angle of the exclusion \cite{divi_error-estimate-based_2020}.

\paragraph{Study of quadrature order for the three 2D test case geometries}
The general setup of the following study is provided in \autoref{sec:three_geometries}.

\begin{figure}[b]
	\centering
	\def\plotwidthfactor{0.8} 
	\includegraphics{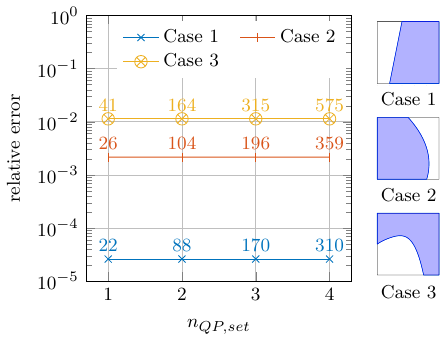}
	\caption{Nutils: Error of area computation with different $\numQuadPointsSetting$ for the three test case geometries. The numbers next to the markers note $\numQuadPoints$.}
	\label{fig:nutils_area}
\end{figure}

\autoref{fig:nutils_area} shows the influence of $\numQuadPointsSetting$ on the accuracy for an area computation. No change of accuracy is observed for increasing $\numQuadPointsSetting$. Furthermore, the error is lower for lower degree of the boundary curve $\curveDegree$. The reason for both observations is that Nutils applies a triangulation on the lowest subdivision level which yields identical results for an unchanged mesh. The triangulation obviously works better for lower $\curveDegree$. At first glance, it might be surprising that Nutils does not demonstrate the machine precision for the test case 1 with the linear boundary curve. The reason is that Nutils uses a certain tolerance when computing intersections. The developers argue that this tolerance in general helps to match intersections between neighbouring elements and to discard nearly vanishing elements\footnote{\url{https://github.com/evalf/nutils/discussions/920\#discussioncomment-13693233}, accessed: 8.7.2025}.
This tolerance can be actually changed by the user in the program. However, the default value was used herein.

\begin{figure}
	\centering
	\def\plotwidthfactor{0.8} 
	\includegraphics{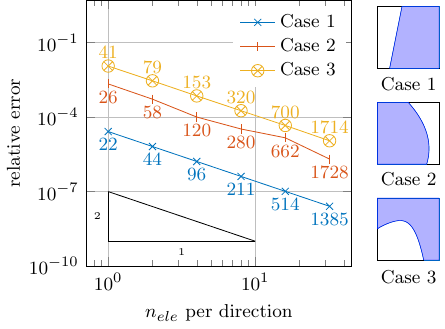}
	\caption{Nutils: Error of area computation with $h$-refinement for the three test case geometries. $\numQuadPointsSetting=1$. The numbers next to the markers note $\numQuadPoints$.}
	\label{fig:nutils_href}
\end{figure}

\autoref{fig:nutils_href} illustrates a corresponding $h$-refinement study. $\numQuadPointsSetting$ is chosen to be one since this is sufficient to compute the area of a triangulation as was observed in \autoref{fig:nutils_area}. The triangulation stably results in a second-order convergence. Compared to the FCMLab results shown in \autoref{fig:fcmlab_area}, the graphs have the same maximal convergence rate but do not possess any oscillations. Thus, we can conclude that the tessellation significantly increased the robustness of the scheme.

\begin{figure*}
	\centering
	\def\plotwidthfactor{0.35} 
	\includegraphics{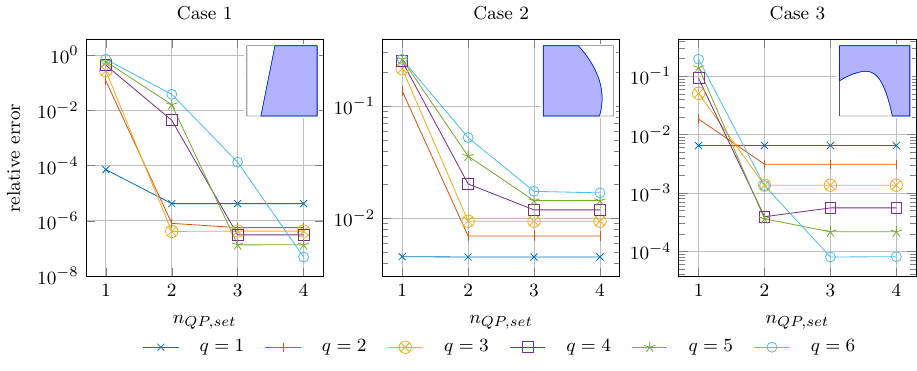}
	\caption{Nutils: Error of integral computation with different $\numQuadPointsSetting$ for the three test case geometries and different integrands of degree $\integrandDegree$.} 
	\label{fig:nutils_integrand}	
\end{figure*}

\autoref{fig:nutils_integrand} depicts the influence of $\numQuadPointsSetting$ on the accuracy for an integral computation with a monomial of degree $\integrandDegree$ as integrand. The investigated values of $\numQuadPointsSetting$ are limited to a maximum of four which corresponds to a quadrature order $\quadDegree$ of six/seven -- the relation between $\quadDegree$ and $\numQuadPointsSetting$ always matches two values to one -- because Nutils does not support higher order rules\footnote{Message from Nutils: 'UserWarning: inexact integration for polynomial of degree 7'; same for higher degrees. Code still runs with a fall-back value of a single quadrature point.}. 
The graphs in \autoref{fig:nutils_integrand} clearly demonstrate an increasing accuracy with higher $\numQuadPointsSetting$. Nevertheless, the error starts to level off far from machine precision. Due to its triangulation, Nutils is limited by its geometrical error and, therefore, is less influenced by $\numQuadPointsSetting$ for the cases 2 and 3.

\subsection{Reparametrization}

This subsection collects a broad group of methods that share the idea that one could map a standard quadrature rule onto a well-parametrized cut element.

\subsubsection{ngsxfem}
\label{sec:ngsxfem}

\begin{figure*}[h]
	\centering
	{\includegraphics[width=\textwidth,trim={0 25 0 25},clip]{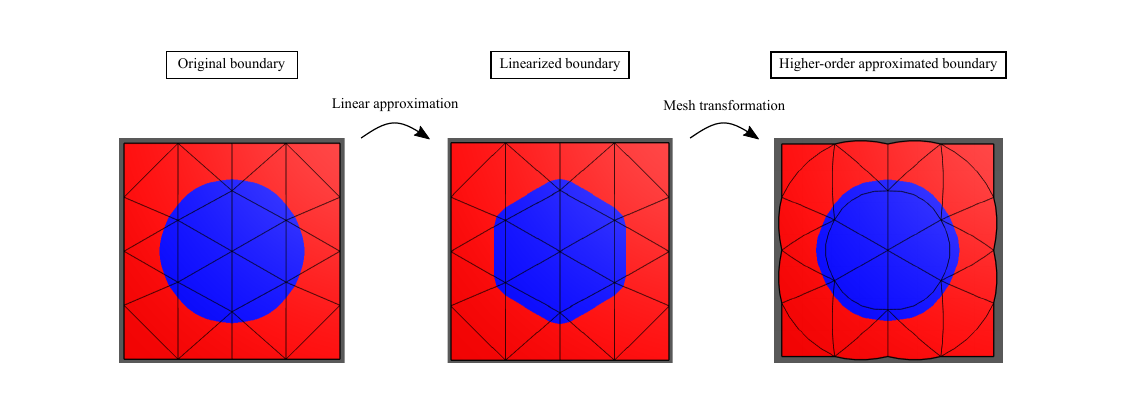}}
	\caption{ngsxfem: Concept of linear and higher-order approximation. The pictures are generated using \texttt{intlset.ipynb} from \cite{ngsxfem}.}
	\label{fig:ngsxfem_concept}
\end{figure*}

\paragraph{Applied method and code details}
ngsxfem\footnote{\url{https://github.com/ngsxfem}, accessed: 22.4.2025} is an add-on library to the FEM-software NGSolve \cite{ngsolve}. It was successfully used in a variety of mechanical problems such as scalar and Stokes interface problems, space-time finite elements, fluid structure interaction or model order reduction (a publication list is provided on \url{https://github.com/ngsxfem/ngsxfem/blob/master/doc/literature.md}). ngsxfem uses an integration procedure identical to the Gridap code (see \autoref{sec:gridap}). In order to increase accuracy, they use a higher-order mesh transformation that maps the cut elements such that the boundary is better approximated. This procedure is illustrated in \autoref{fig:ngsxfem_concept}.
Furthermore, ngsxfem preferably applies triangle and tetrahedron meshes. Even though quadrilateral meshes are also supported, they are limited to single level set functions.
The same limitation seems to hold for the current open-source implementation of the mesh transformation even though the method was studied for more general cases in \cite{Lehrenfeld2017}.
Mapped Gauss integration is used for all element types. 
The most important parameters to control the integration settings are:
\begin{itemize}
	\item quadrature order
	\item reparametrization degree; it is chosen equal to the curve degree $\curveDegree$ in our studies
\end{itemize}
\begin{remark}
	The term `reparametrization degree', which is also used in the discussion of other open-source codes within this work as well as in our benchmark environment, is used here because it serves the same goal of a higher-order approximation of the boundary. However, a mesh transformation instead of a direct reparametrization is used in ngsxfem.
\end{remark}
The code is available as Python package. A helpful Jupyter tutorial and several example files are published\footnote{\url{https://ngsuite.pages.gwdg.de/ngsxfem/}, accessed: 25.3.2025}. Furthermore, NGSolve maintains an active community forum\footnote{\url{https://forum.ngsolve.org/}, accessed: 25.3.2025}. The interfacing of ngsxfem in our benchmark environment is done by creating a Python file with a generic ngsxfem model that can be called from Matlab by \texttt{pyrunfile}. For all 2D examples with one level set function, we apply quadrilateral meshes for a better comparison to other codes. The ngsxfem authors already investigated the geometrical error of the described method and successfully tested the robustness by moving a complex 2D geometry through the domain in \cite{Lehrenfeld2016}. They observed the expected convergence rates associated with the used reparametrization degree.

\paragraph{Study of quadrature order for the three 2D test case geometries}

\begin{figure}
	\centering
	\def\plotwidthfactor{0.8} 
	\includegraphics{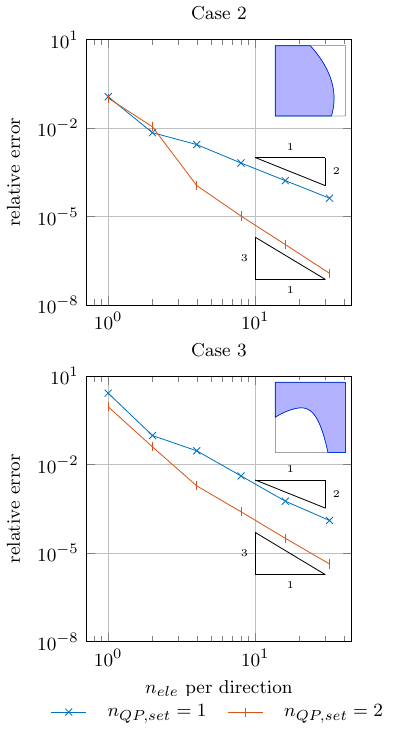}
	\caption{ngsxfem: Error of area computation with $h$-refinement for the three test case geometries and different number of quadrature points $\numQuadPointsSetting$.} 
	\label{fig:ngsxfem_href}	
\end{figure}

The general setup of the following study is provided in \autoref{sec:three_geometries}.
ngsxfem exactly resolves test case 1 with the linear boundary curve since it is applying a triangulation as a first step -- no extra plot in \autoref{fig:ngsxfem_href}.
Furthermore, ngsxfem exactly resolves test case 1 with the linear boundary curve since it applies a triangulation as a fist step. Therefore, we only present an $h$-refinement study for the cases 2 and 3 in \autoref{fig:ngsxfem_href}.

The other two cases cannot be resolved well with a single element. Therefore, only an $h$-refinement study is presented for these two geometries in \autoref{fig:ngsxfem_href}. Only $\numQuadPointsSetting=1$ and $\numQuadPointsSetting=2$ are depicted because higher values resulted in the same graphs as for $\numQuadPointsSetting=2$.
It can be observed that higher-order convergence is achievable due to the background mesh mapping. The order of the mapping is chosen to be equal to the curve degree. It is interesting to note that also for case 3 only second- and third-order convergence were achieved. The reason might be that the curve degree does not necessarily resemble its complexity.
\begin{remark}
	Numerically unstable results -- relevant differences when running on different machines -- were observed in case that the reparametrization degree was chosen to be higher than an appropriate degree resembling its complexity which could be lower than the actual curve degree. This especially holds in combination with coarse meshes.
\end{remark}
Furthermore, it is observed that $\numQuadPointsSetting$ must be sufficiently high to reach higher-order convergence, whereby the convergence order is defined by \autoref{eq:integrationErrorApprox}. The influence of the order of the background mesh mapping was studied in more detail in \cite{Lehrenfeld2017}. In contrast to other codes, it is not directly possible to retrieve the quadrature points for ngsxfem in our Matlab interface and, therefore, they are not reported herein. 

\begin{figure}
	\centering
	\def\plotwidthfactor{.8} 
	\includegraphics{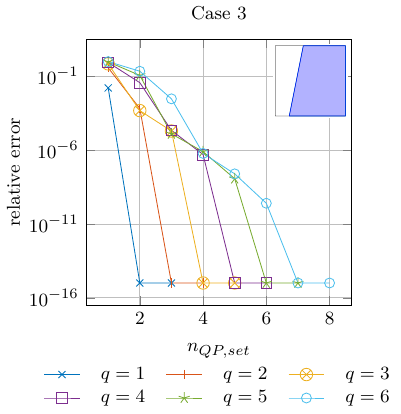}
	\caption{ngsxfem: Error of area computation with different $\numQuadPointsSetting$ for test case geometry 1 and different integrands of degree $\integrandDegree$.}
	\label{fig:ngsxfem_integrand}
\end{figure}
\Autoref{fig:ngsxfem_integrand} depicts the influence of $\numQuadPointsSetting$ on the accuracy for an integral computation with a monomial of degree $\integrandDegree$ as integrand. Only the case 1 with the linear boundary curve is investigated because the other two are not resolved well with a single element as already mentioned. ngsxfem is able to reach machine precision with the expected $\numQuadPointsSetting$ defined by \autoref{eq:nQP_exact_single}.

\subsubsection{QUGaR}
\label{sec:QUGaR}

\paragraph{Applied method and code details}

\begin{figure}[h]
	\centering
	\includegraphics[width=0.47\textwidth,trim={0 500 0 0},clip]{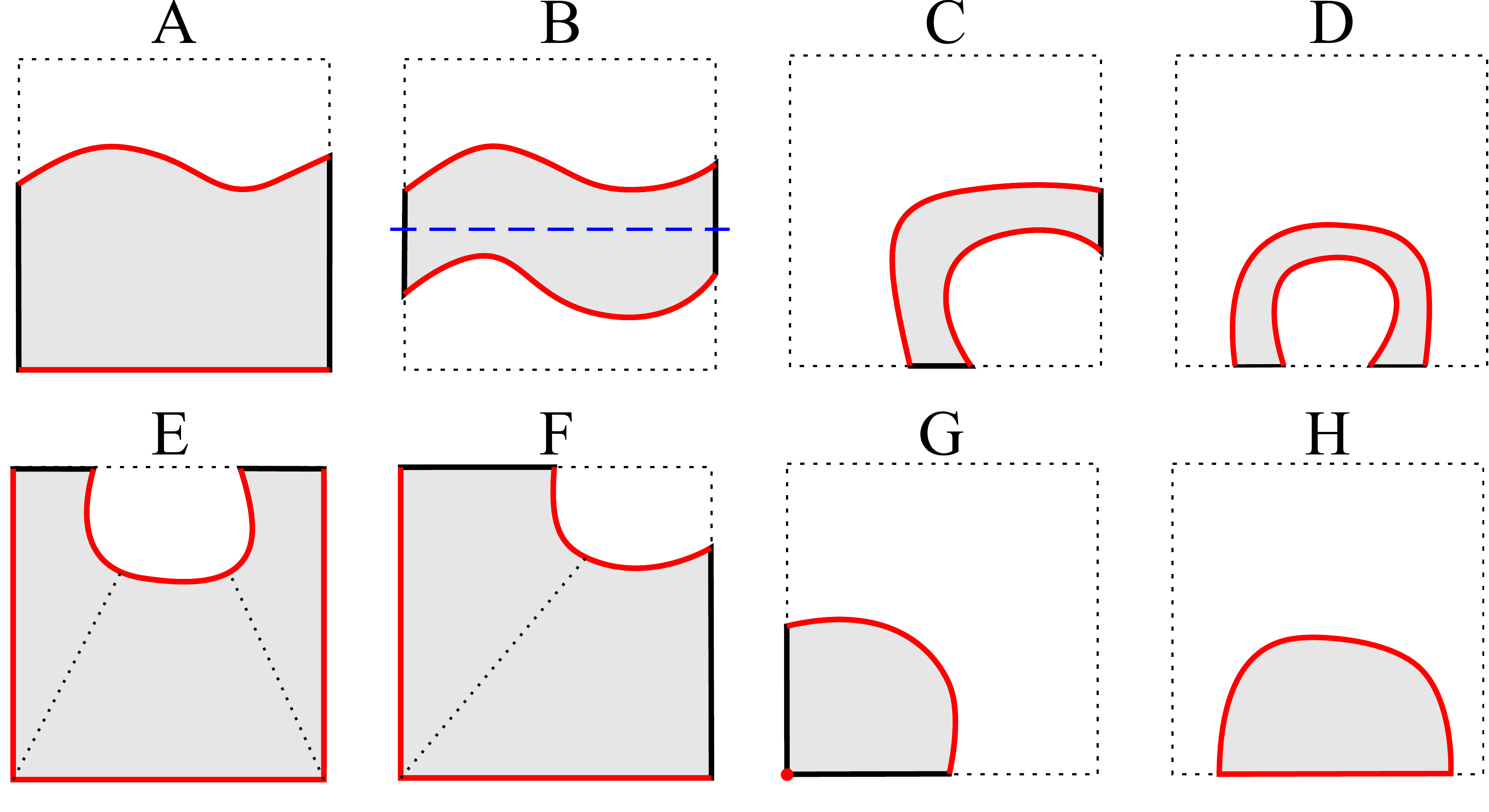}
	\caption{QUGaR: Reparameterization base categories of cut elements using quadrilateral cells as described in \cite{wei_immersed_2021}.} 
	\label{fig:QUGaRreparametrization}
\end{figure}

QUGaR\footnote{\url{https://github.com/pantolin/qugar}, accessed 7.5.2025} (Quadratures for Unfitted GeometRies) is an open-source C++ library. It is dedicated to generate quadrature rules for cut high-order elements for finite element analysis with cut elements such as unfitted, immersed boundary, or CutFEM methods. 
While it is a standalone library, it provides out-of-the-box interoperability with the Python interfaces of the FEniCSx computing platform\footnote{\url{https://fenicsproject.org/}, accessed: 2.1.2026}.
For implicit boundaries, QUGaR employs a fork of the Algoim library (see \autoref{sec:algoim}). At the same time, it offers functionalities for generating unfitted domain reparameterizations. They are mainly for visualization but can also be employed to obtain a proper distribution of standard quadrature points (e.g., Gauss points) within cut elements.
This option holds for elements cut by parametric interfaces as well. 
In fact, while there is no publication dedicated to this concept, this reparametrization was utilized in several publications on isogeometric simulations with cut elements such as \cite{antolin_isogeometric_2019,wei_immersed_2021} and QUGaR developer announced future updates for supporting 2D and 3D parametric boundary representations. 
The key challenge in reparametrizing cut elements is the robust detection of a proper partition into regular regions. 
QUGaR's construction aims for quadrilateral regions and utilizes templates based on the topology of the cut element, as shown in \autoref{fig:QUGaRreparametrization}.
The key is to find a pair of curves (one being the interface) and create ruled surfaces between them.
This construction may require splitting the element (case B,E,F). 
If the cut element does not fall into one of the base categories, a divide-and-conquer strategy is applied.
A robust detection of such patterns is far from trivial and gets significantly more complex when considering 3D models.
Thus, the team of QUGaR proposed so-called folded decompositions to treat cut trivariate domains \cite{Antolin2022}. 

For this paper, the developer Pablo Antolin provided us with a MATLAB interface to the 2D reparameterization routines for parametric interfaces. 
In other words, we enrich our study by a further integration concept by employing QUGaR as an example for treating cut elements by a reparameterization concept, despite its current primary focus being on implicit interfaces.

The most important parameters to control the integration settings are:
\begin{itemize}
	\item number of Gauss points
	\item reparametrization degree
\end{itemize}

\paragraph{Study of quadrature order for the three 2D test case geometries}

The general setup of the following study is provided in \autoref{sec:three_geometries}.

\begin{figure}[b]
	\centering
	\def\plotwidthfactor{0.8} 
	\includegraphics{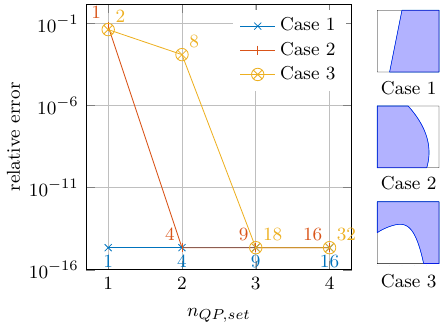}
	\caption{QUGaR: Error of area computation with different $\numQuadPointsSetting$ for the three test case geometries. The numbers next to the markers note $\numQuadPoints$.}
	\label{fig:qugar_area}
\end{figure}

\begin{figure*}[th]
	\centering
	\def\plotwidth{0.384\textwidth} 
	\def\plotheight{0.32\textwidth} 
	\includegraphics{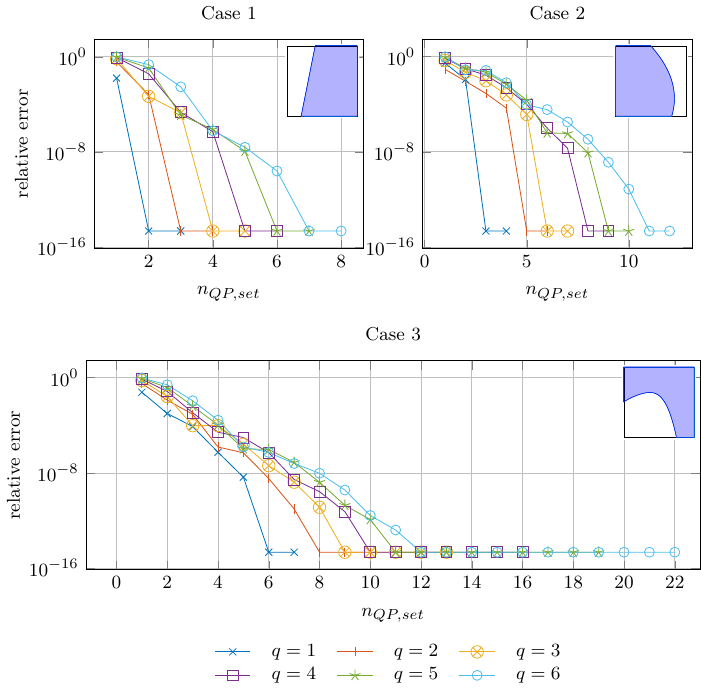}
	\caption{QUGaR: Error of integral computation with different $\numQuadPointsSetting$ for the three test case geometries and different integrands of degree $\integrandDegree$.} 
	\label{fig:qugar_integrand}	
\end{figure*}

\autoref{fig:qugar_area} shows the influence of $\numQuadPointsSetting$ on the accuracy for an area computation. QUGaR is able to reach machine precision with the expected $\numQuadPointsSetting$ defined by Eq. \ref{eq:nQP_exact_single}. For case 3, a subdivision is required because it is a five-sided element which leads to double as many quadrature points as for the other two geometries which is indicated by the numbers next to the markers. The subdivision is also visualized in \autoref{fig:example_unibw16_QP}.
No $h$-refinement study is performed for the area computation since machine precision can be already reached by increasing $\numQuadPointsSetting$.

\autoref{fig:qugar_integrand} illustrates the influence of $\numQuadPointsSetting$ on the accuracy for an integral computation with a monomial of degree $\integrandDegree$ as integrand. QUGaR is able to reach machine precision with the expected $\numQuadPointsSetting$ defined by Eq. \ref{eq:nQP_exact_single}. Whereby, Eq. \ref{eq:nQP_exact_single} slightly overestimates the required $\numQuadPointsSetting$ for case 3 with the quintic boundary curve for all integrands $\integrandDegree$ greater than two. The reason is that the curve degree does not resemble the curve's complexity.

%
\begin{figure*}
	\centering
	{\includegraphics[width=\textwidth,trim={0 15 0 20},clip]{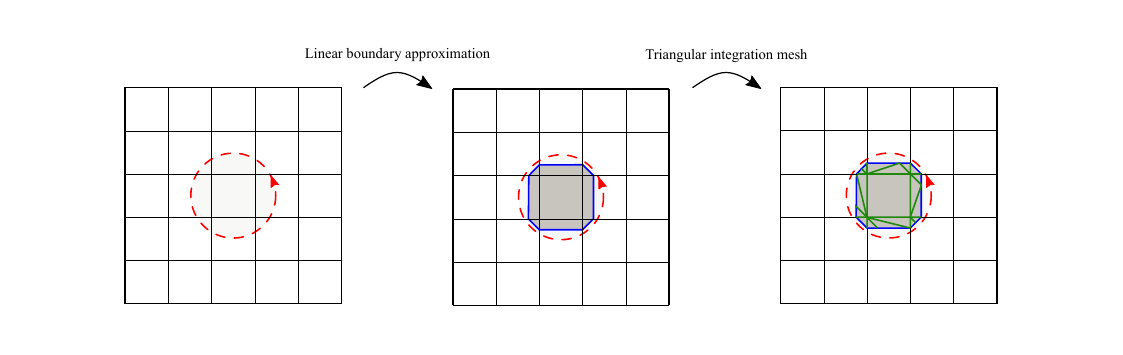}}
	\caption{Gridap: Applied immersed concept, boundary curve as red dashed line, linear boundary approximation as blue lines, triangulated integration mesh as green lines}
	\label{fig:gridap_concept}
\end{figure*}

\subsubsection{Gridap}
\label{sec:gridap}

\paragraph{Applied method and code details}
Gridap\footnote{\url{https://github.com/gridap/Gridap.jl}, accessed: 22.4.2025} is a Julia library with an expressive API using mathematical notations. Its module GridapEmbedded\footnote{\url{https://github.com/gridap/GridapEmbedded.jl}, accessed: 22.4.2025} provides functionalities to solve immersed problems. A linear approximation of the implicitly described boundary is used. Thereby, the level set function of the boundary is evaluated at the element corners and these values are then used for a linear interpolation within the cut elements.
Subsequently, the linearly bounded cut elements are triangulated. This procedure is illustrated in \autoref{fig:gridap_concept}.
A mapped Gauss integration is used for the triangles in 2D and tetrahedron quadrature rules are used in 3D (cf.~e.g., \cite{zienkiewicz_finite_2013} for tetrahedron rules).
The most important parameter to control the integration settings is:
\begin{itemize}
	\item quadrature order
\end{itemize}
In their tutorial\footnote{\url{https://gridap.github.io/Tutorials/dev/pages/t020\_poisson\_unfitted/\#poisson\_unfitted.jl-1}, accessed: 22.04.2025}, they explicitly recommend to choose the quadrature degree $\quadDegree$ for Galerkin problem as:
\begin{equation}
	\quadDegree = 2 \cdot q \cdot \dimEuclideanSpace
\end{equation}
where $q$ is the order of the basis functions and $\dimEuclideanSpace$ is the spatial dimension of the problem. The first product `$2\cdot q$' comes from the bilinear form of the Galerkin method. The reason for the additional multiplication with the dimension is that Fubini's theorem, which would allow computing the integral as sequence of one-dimensional integrals, is not applicable to arbitrary cut elements. This is in agreement with the conclusive derivation of the exact quadrature degree for cut elements in \cite[Appendix]{Antolin2022} under the assumption of a linear boundary and constant mappings. The derivations of the required quadrature orders in \autoref{sec:influence_domain_transformations} focus on two-dimensional geometries, but a derivation for a different dimensionality could be analogically performed and is presented for the case of a general mapping in 3D by \cite[Appendix]{Antolin2022}.

A tutorial\footnote{\url{https://gridap.github.io/Tutorials/dev/}, accessed: 22.04.2025} and a community forum\footnote{\url{https://github.com/gridap/Gridap.jl/discussions}, accessed: 22.04.2025} are available for Gridap. The interfacing in our benchmark environment is done by creating a Julia file with a generic Gridap model. Unfortunately, a direct interaction between Matlab and Julia is not available in a robust and comprehensive way. Therefore, Python with its package julia\footnote{\url{https://pypi.org/project/julia/}, accessed: 22.04.2025} is used as connecting tool. For the translation of the boundary definition from Matlab to Julia, the \textit{eval} command of Julia is used. In contrast to Matlab, it is thereby important that Julia is a high-level, optimized language which pre-compiles parts of the program for faster runtime. The consequence is that a dynamic use of \textit{eval} would prevent this pre-compilation leading to Julia's world age problem \cite{belyakova_world_2020}. Therefore, the translation of the boundary definition should take place before the main Julia file is called.
Regarding the model setup, processing embedded geometries which are provided as STL files is additionally supported. This functionality is not further investigated in the present work (see the STLCutters\footnote{\url{https://github.com/gridap/STLCutters.jl}, accessed: 22.04.2025} module for further information).
The Gridap authors already tested the robustness of their method for 2D and 3D by moving a geometry through the domain \cite{badia_aggregated_2018} (the publication refers to FEMPAR which is the predecessor of Gridap\footnote{Private communication with A. Martin, 2024}).

\paragraph{Study of quadrature order for the three 2D test case geometries}

The general setup of the following study is provided in \autoref{sec:three_geometries}.

\begin{figure}
	\centering
	\def\plotwidthfactor{0.8} 
	\includegraphics{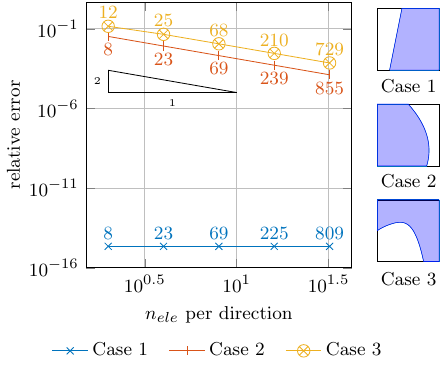}
	\caption{Gridap: Error of area computation with $h$-refinement for the three test case geometries. The numbers next to the markers note $\numQuadPoints$. $\numQuadPointsSetting=1$.}
	\label{fig:gridap_href}
\end{figure}

Gridap is not able to resolve the geometry for a single element. Therefore, the studies of $\numQuadPointsSetting$ as shown for other codes are omitted. \autoref{fig:gridap_href} illustrates an $h$-refinement study on the accuracy for an area computation. $\numQuadPointsSetting$ is chosen to be one since this is sufficient to compute the area of a triangulation as used by Gridap -- this was already discussed and observed in \autoref{sec:nutils}. The triangulation results in a second-order convergence. The linear boundary is exactly resolved and, therefore, results in machine precision for any mesh.

\subsection{Moment fitting}
\label{sec:moment_fitting}

A moment fitting method defines a tailored quadrature rule on a cut element by solving the following system of equations:
\begin{align}
	&
	\left[\begin{array}{cccc}
		f_1(\bm{x}_1) &f_1(\bm{x}_2) &\hdots &f_1(\bm{x}_{\numQuadPoints}) \\
		f_2(\bm{x}_1) &f_2(\bm{x}_2) &\hdots &f_2(\bm{x}_{\numQuadPoints}) \\
		\vdots & &\ddots &\vdots \\
		f_n(\bm{x}_1) &f_n(\bm{x}_2) &\hdots &f_n(\bm{x}_{\numQuadPoints})
	\end{array}\right] \left[\begin{array}{c}
		w_1 \\
		w_2 \\
		\vdots \\
		w_{\numQuadPoints}
	\end{array}\right] = \nonumber \\ 	
	&= \left(\begin{array}{c}
		\int_{\elemDomain} f_1(\bm{x})d\bm{x} \\
		\int_{\elemDomain} f_2(\bm{x})d\bm{x} \\
		\vdots \\
		\int_{\elemDomain} f_n(\bm{x})d\bm{x}
	\end{array}\right)
	\label{eq:moment_fitting}
\end{align}
where $\bm{x}_j$ is the position of the $j^{th}$ quadrature point, $w_j$ is the integration weight of the $j^{th}$ quadrature point, $\numQuadPoints$ is the number of quadrature points, $f_i$ is the $i^{th}$ function of the target space, $n$ is the number of moments (considered functions in the target space), and $\elemDomain$ is the active domain of the cut element. Thereby, the right-hand side must be computed by a sufficiently accurate quadrature rule as a reference.

The three codes discussed in the following subsections apply the idea of the moment fitting method. However, their concepts differ regarding the choice of the positions of the quadrature points $\bm{x}_j$, the considered functions $f_i$, and the computation of the right-hand side. All three codes fix the positions of the quadrature points $\bm{x}_j$, whereby \autoref{eq:moment_fitting} becomes linear in the quadrature weights $w_j$. This simplifies the solving of the equation system. Since it is a linear system, the number of unknowns, i.e., the quadrature weights, 
must in general be equal to or greater 
than the number of moment conditions to guarantee the existence of a solution.
It is worth to note that optimal quadrature rules with less points could be derived by additionally optimizing the positions as presented in \cite{Nagy2015}.

\subsubsection{BoSSS}
\label{sec:BoSSS}

\paragraph{Applied method and code details}
\label{sec:bosss_method}
		
BoSSS\footnote{\url{https://github.com/FDYdarmstadt/BoSSS}, accessed: 27.11.2025} (Bounded Support Spectral Solver) is an open-source package, mainly for the simulation of flows with interfaces,
e.g., multiphase flows \cite{kummerEtAl_bosss_2024}.
The package is based on the extended discontinuous Galerkin (XDG) method, 
which introduces cut elements to capture these interfaces, 
similar to Extended Finite Element Methods; Therefore, one might also classify it as DG-XFEM.
While it also can use Algoim (see \autoref{sec:algoim}) to compute the respective quadrature rules,
it alternatively provides its own method, the so-called hierarchical moment fitting method (HMF)
\cite{Mueller2013}.

The above described moment fitting method is applied.
The nodes are predefined by a Gauss-Legendre quadrature rule and only the quadrature weights $w_j$ are optimized, resp., 
computed from \autoref{eq:moment_fitting}.
Obviously, this might lead to nodes in a `void' region, i.e., outside of the active domain of the cut element and hence, a negative weight.
This might lead to numerical problems for certain applications, e.g., the discrete norm of certain functions might be negative.

In contrast to approaches such as Algoim, which are obtained by applying a transformation onto a Gauss-Legendre rule,
the relation between the number of nodes and the quadrature order is subject to a tuning parameter.
To guarantee a certain quadrature order, the weights must fulfill a set of moment conditions.
Since this is a linear system, the number of unknowns, i.e., the number of weights, 
must in general be equal or greater 
to the number of moment conditions as a prerequisite for the existence of a solution.
It has been observed during the preparation of the original works \cite{Mueller2013,Kummer2017}, that, 
if the number of nodes and weights is equal or only sightly larger than the number of moment conditions,
the weights show an oscillatory behavior, i.e., one encounters very large positive and negative weights.
This is likely caused by the placement of quadrature nodes, which are predefined and do not adapt to the actual geometry of the cut element.
Therefore, a safety factor $\varpi$ regarding the number of quadrature nodes is introduced,
i.e., the number of nodes is chosen to be about $\varpi$ times the number of moment conditions (values given below).



The original HMF \cite{Mueller2013}, involves two steps: 
First, the surface integral is computed using divergence-free test functions
and second, the volume integral is computed using the surface integral. 
In both steps moment fitting equations are defined.
Variations of the original HMF have been proposed which address specific issues,
e.g., construction in one step \cite{Kummer2017} or
avoiding placing the quadrature in the void region \cite{geisenhofer_discontinuous_2019},
which are beyond the scope of this review.

The divergence theorem (Gauss' theorem) allows constructing the integral 
over a implicitly given surface $I=\{\bm{x} \in \partial \elemDomain | \ \phi(\bm{x}) = 0\}$ 
for a divergence-free basis with functions $\bm{g}_i$ (i.e., $\text{div}(\bm{g}_i)=0$) as
\begin{equation}
	\int_I \bm{g}_i \cdot \bm{n}_I d\Gamma = -\int_{\partial\elemDomain\setminus I} \bm{g}_i \cdot \bm{n} d\Gamma
	\label{eq:surface_integral_div_free}
\end{equation}
where $\bm{n}_I = \nabla \phi / \| \nabla \phi \|$ is the normal on the interface $I$, $\bm{n}$ is the outward unit normal on the element boundary and $\phi$ is the level set function defining the interface. 
A moment fitting can be applied to determine a quadrature rule for the right-hand side:
\begin{align}
	& \left[\begin{array}{ccc}
		\bm{g}_1(\bm{x}_1) \cdot \bm{n}_I(\bm{x}_1) &\hdots & \bm{g}_1(\bm{x}_{\numQuadPoints}) \cdot \bm{n}_I(\bm{x}_{\numQuadPoints}) \\
		\vdots &\ddots &\vdots \\
		\bm{g}_n(\bm{x}_1) \cdot \bm{n_I}(\bm{x}_1) &\hdots &\bm{g}_n(\bm{x}_{\numQuadPoints}) \cdot \bm{n}_I(\bm{x}_{\numQuadPoints})
	\end{array}\right] \cdot \nonumber \\
	&\cdot \left[\begin{array}{c}
		w_1 \\
		w_2 \\
		\vdots \\
		w_{\numQuadPoints}
	\end{array}\right] = 	
	\left(\begin{array}{c}
		-\int_{\partial\elemDomain\setminus I} \bm{g}_1(\bm{x}) \cdot \bm{n} d\Gamma \\
		\vdots \\
		-\int_{\partial\elemDomain\setminus I} \bm{g}_n(\bm{x}) \cdot \bm{n} d\Gamma
	\end{array}\right) \label{eq:moment_fitting_div_free}
\end{align}
Note, that it is crucial to use divergence-free test functions $\bm{g}_i$ in order to obtain the surface integral.
Otherwise, \autoref{eq:surface_integral_div_free} would contain an additional, unknown volume integral over $\elemDomain$.
Unfortunately, this can lead to a reduced accuracy for integrands which are not well represented in the divergence-free space.
A fix (not considered here) has been proposed \cite{Kummer2017}, at the cost of increased computational effort.
Furthermore, one assumes that the 
right-hand-side of \autoref{eq:moment_fitting_div_free} can be computed with sufficient accuracy.
In a 2D-setting, this is typically an open polygon.
One has to find the zero-set of $\phi$, aka., the roots of $\phi$, 
on the straight edges of the polygonal element, usually by a Newton method.
Then, 1D Gauss rules are mapped onto the respective edges.
In a 3D-setting, one first has to carry out the 2D HMF procedure 
for each face, hence the name \emph{hierarchical} moment fitting.

Once the surface integral is available, 
one can compute the weights for the volume quadrature defined in \autoref{eq:moment_fitting} 
using the divergence theorem to compute the respective right-hand side:
\begin{align}\label{eq:divergence_theorem}
	\int_{\elemDomain}f_i(\bm{x})d\bm{x} &= \int_{\elemDomain} \nabla \cdot \bm{F}_i(\bm{x}) d\bm{x} = \nonumber \\
	&= \int_{\partial \elemDomain} \bm{F}_i(\bm{x}) \cdot \bm{n}(\bm{x}) d\Gamma
\end{align}
where $\bm{F}_i$ are the anti-derivatives of $f_i$. 
The anti-derivatives can be analytically computed as:
\begin{align}
	\dimEuclideanSpace &= 2: 
	\bm{F}_i(\bm{x})=\dfrac{1}{2} 
	\left[\begin{array}{c}
		\int f_i(\bm{x})dx \\
		\int f_i(\bm{x})dy
	\end{array}\right] \\
	\dimEuclideanSpace &= 3:
	\bm{F}_i(\bm{x})=\dfrac{1}{3} 
	\left[\begin{array}{c}
		\int f_i(\bm{x})dx \\
		\int f_i(\bm{x})dy \\
		\int f_i(\bm{x})dz
	\end{array}\right]
\end{align}
The linear systems Eq. \ref{eq:moment_fitting} and \ref{eq:moment_fitting_div_free} are under-determined,
therefore a least-squares solution is selected.
%
%
%
%
The most important parameters to control the integration settings are:
\begin{itemize}
	\item quadrature order, which determines the number of moment-equations, i.e., 
	      the number of rows in \autoref{eq:moment_fitting} and \autoref{eq:moment_fitting_div_free}
	\item the safety factor $\varpi$, which determines the number of nodes and weights in relation to the number of moment conditions,
	      and thereby the number of columns in \autoref{eq:moment_fitting} and \autoref{eq:moment_fitting_div_free};
		  In this study, the default values for $\varpi$ were used
		  (surface integrals: 1.6 for 2D, 3.0 for 3D; volume integrals: 1.0).
\end{itemize}

The code is written in C\#. An installation that does not require any compilation is available online. A helpful Jupyter tutorial is published\footnote{\url{
https://git.rwth-aachen.de/kummer/bosss-public/-/wikis/Getting-Started/Jupyter-Tutorials}, accessed: 27.11.2025}.
By the time at which this study started, interfacing C\# from Matlab was only possible on Windows.
Recently, this also is supported for Linux\footnote{\url{
https://www.mathworks.com/help/matlab/call-net-from-matlab.html}, accessed 11.12.2025}, but this was not tested in this study.


\paragraph{Study of quadrature order for the three 2D test case geometries}

\begin{figure}[h]
	\centering
	\def\plotwidthfactor{0.8} 
	\includegraphics{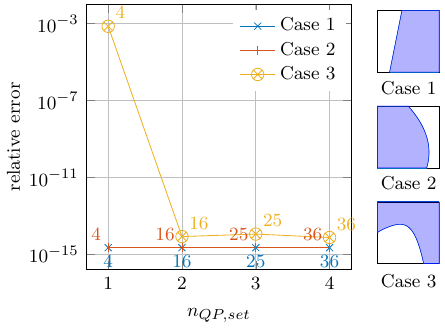}
	\caption{BoSSS: Area computation. Study of property $\numQuadPointsSetting$ for the three test case geometries. The numbers next to the markers note $\numQuadPoints$.}
	\label{fig:bosss_area}
\end{figure}

The general setup of the following study is provided in \autoref{sec:three_geometries}.
Figure \ref{fig:bosss_area} shows the influence of $\numQuadPointsSetting$ on the accuracy for an area computation. 
BoSSS is able to reach machine precision for all three test cases with $\numQuadPointsSetting \geq 2$.
While this threshold seems to be exceptionally low, 
in comparison to libraries like QUGaR,
it should be noted that due to the safety factor $\varpi$, BoSSS already produces 16 nodes for $\numQuadPointsSetting=2$,
while QUGaR only produces 4, c.f. Figure \ref{fig:qugar_area}.


\begin{figure*}
	\centering
	\def\plotwidthfactor{0.35} 
	\includegraphics{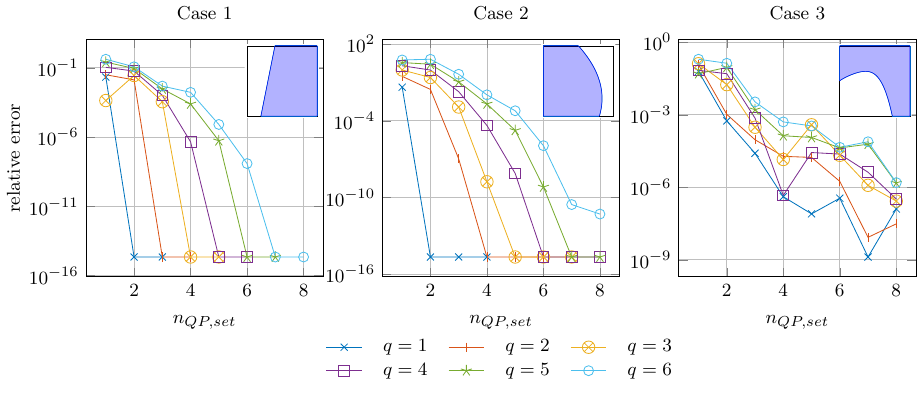}
	\caption{BoSSS: Study of property $\numQuadPointsSetting$ for the three test case geometries and integrands of degree $\integrandDegree$.} 
	\label{fig:bosss_integrand}
\end{figure*}

\autoref{fig:bosss_integrand} illustrates the influence of $\numQuadPointsSetting$ 
on the accuracy for an integral computation with a monomial of degree $\integrandDegree$ as integrand. 
BoSSS converges again with $\numQuadPointsSetting$ being lower than defined by \autoref{eq:nQP_exact_single},
due to the afore-mentioned reason.
The investigated values of $\numQuadPointsSetting$ are limited to a maximum of eight which corresponds to a quadrature order $\quadDegree$ of 15/16 -- the relation between $\quadDegree$ and $\numQuadPointsSetting$ always matches two values to one -- because BoSSS does not support higher order rules. 
If one were to request higher orders, BoSSS would fall back to the highest supported rule.

Note that HMF rules are typically not embedded,
i.e., a rule designed for order $k_1$ typically does not exactly give the same result as one for order $k_2 > k_1$,
for some integrand of order $\integrandDegree < k_1$.
The reason for this is the non-triangular shape of the moment fitting system.
Furthermore, surface integrals are indeed constructed to be exact for integrands as $\bm{g}_i \cdot \bm{n}_I$,
which are typically not polynomial.
The consequence is that all integrals in test case 3 cannot reach machine precision. 
The same holds for the integrand of degree $\integrandDegree=6$ in combination with test case 2. 
Furthermore, the ordering of the convergence curves becomes chaotic in case 3 indicating a pre-asymptotic regime. 
Nevertheless, all integrals reach errors below $10^{-5}$.

\subsubsection{mlhp}
\label{sec:mlhp}

\begin{figure*}
	\centering
	{\includegraphics[width=\textwidth,trim={0 22 0 10},clip]{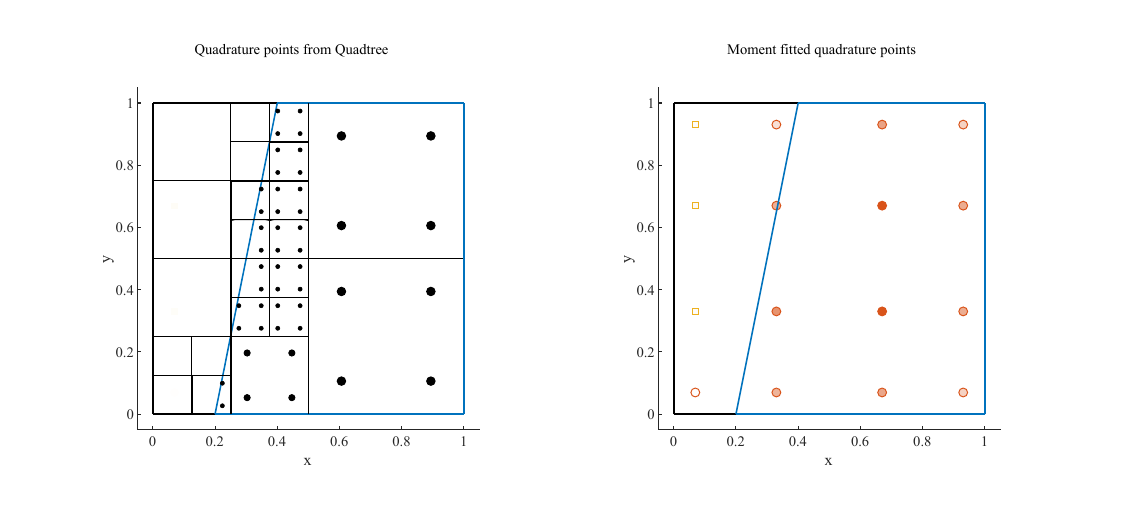}}
	\caption{mlhp: Quadtree and resulting moment fitted quadrature points. The geometry is illustrated by the blue line. Circular black or red markers indicate positive weights. Golden square markers indicate negative weights. The transparency of the marker faces in the right plot implies the magnitude of each weight compared to the maximum absolute value of all weights.}
	\label{fig:mlhp_concept}
\end{figure*}

\paragraph{Applied method and code details}
mlhp\footnote{\url{https://gitlab.com/hpfem/code/mlhp}} is a software successor of the FCMLab which was described in \autoref{sec:fcmlab}. This code applies \textit{hp}-finite elements \cite{kopp_efficient_2022} and provides quadtree/octree quadrature rules. More interestingly, mlhp supports a second quadrature concept that overcomes the fundamental problem of high numbers of quadrature points inherent to quadtree/octree methods 
by applying a moment fitting approach as defined in \autoref{eq:moment_fitting}.
The positions $\bm{x}_j$ are chosen as the Gauss-Legendre positions which results in a linear system of equations with respect to the integration weights $w_j$. The functions $f_i$ considered for the derivation of the quadrature rule are Lagrange polynomials of order $q$ where the supporting points are chosen to be the positions $\bm{x}_j$ respectively the Gauss-Legendre positions. Thus, the matrix in \autoref{eq:moment_fitting} becomes the identity matrix. In consequence, the system of equations can be solved accurately without any iterative procedure. The exact integrals (right-hand side of \autoref{eq:moment_fitting}) are computed by means of the quadtree/octree rule. The quadrature points for the quadtree rule and the resulting moment fitted quadrature points are illustrated in \autoref{fig:mlhp_concept}.
It can be observed that the final points might lie outside of the active domain and, therefore, might have negative weights. A possible influence of negative weights was already discussed in \autoref{sec:bosss_method}.
Boundary integrals are computed by means of the marching cubes algorithm \cite{engwer_geometric_2018} and are so far only supported in 3D.
The most important parameters to control the integration settings are:
\begin{itemize}
	\item number of Gauss points for the final moment fitted rule $\numQuadPoints$
	\item number of Gauss points for the quadtree/octree rule to compute the right-hand side $\numQuadPointsSetting$; this determines also the requested quadrature order
	\item number of subdivision levels $k$
	\item number of equidistant points to check the cutting state of each element (called seedpoints within the code - this shouldn't be confused with the meaning of `SeedPoints' within the FCMLab code (cf.~\autoref{sec:fcmlab}))
\end{itemize}
The number of moment fitted points $\numQuadPoints$ and the number of Gauss points $\numQuadPointsSetting$ used for regular elements as well as the quadtree/octree references are related as (cf.~also \autoref{fig:mlhp_concept})\footnote{Private communication with P. Kopp, 2025}:
\begin{equation}\label{eq:nQP_MF}
	\numQuadPoints = \left(2 \cdot \numQuadPointsSetting\right)^{\dimEuclideanSpace} = \left(2 \cdot \left\lceil\dfrac{q+1}{2}\right\rceil \right)^{\dimEuclideanSpace} \geq \left(q+1\right)^{\dimEuclideanSpace}
\end{equation}
 In contrast to the BoSSS code, no safety factor is introduced (cf.~\autoref{sec:bosss_method}). The set of Lagrange polynomials of order $q$ enables a quadrature rule of order $q$ due to their linear independence \cite{ron_lecture_nodate}.
 
 The mlhp code is written in C++ with Python bindings. A precompiled version is available as Python library. The interfacing of mlhp in our benchmark environment is done by creating a Python file with a generic mlhp model which can be called from Matlab by \texttt{pyrunfile}. The developers highlight their data oriented design (DOD) of the software contrasting its predecessor FCMLab which is strongly based on object oriented programming (OOP). Thus, they are aiming for a better performance due to faster memory access and better branch prediction\footnote{\url{https://gitlab.com/hpfem/code/mlhp/-/wikis/Core concepts}, accessed: 14.7.2025}. The implementation was examined in \cite{kopp_efficient_2022}, but did not consider cut elements.

\paragraph{Study of quadrature order for the three 2D test case geometries}
The general setup of the following study is provided in \autoref{sec:three_geometries}.

\begin{figure}[h]
	\centering
	\def\plotwidthfactor{0.8} 
	\includegraphics{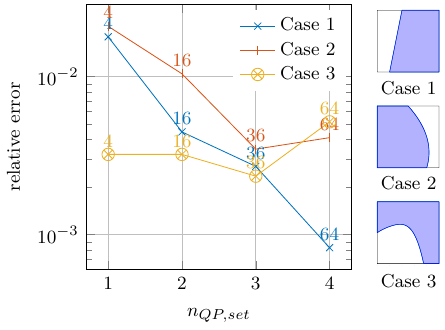}
	\caption{mlhp: Error of area computation with different $\numQuadPointsSetting$ for the three test case geometries. The numbers next to the markers note $\numQuadPoints$.}
	\label{fig:mlhp_area}
\end{figure}

\autoref{fig:mlhp_area} shows the influence of $\numQuadPointsSetting$ on the accuracy for an area computation. 
mlhp shows a certain reduction of the error for cases 1 and 2, but not for case 3. General problems in correctly resolving the geometries with a single element are observed. The values for $\numQuadPointsSetting$ and $\numQuadPoints$ reported in \autoref{fig:mlhp_area} agree with the relation given in \autoref{eq:nQP_MF}.

\begin{figure*}
	\centering
	\def\plotwidthfactor{0.35} 
	\includegraphics{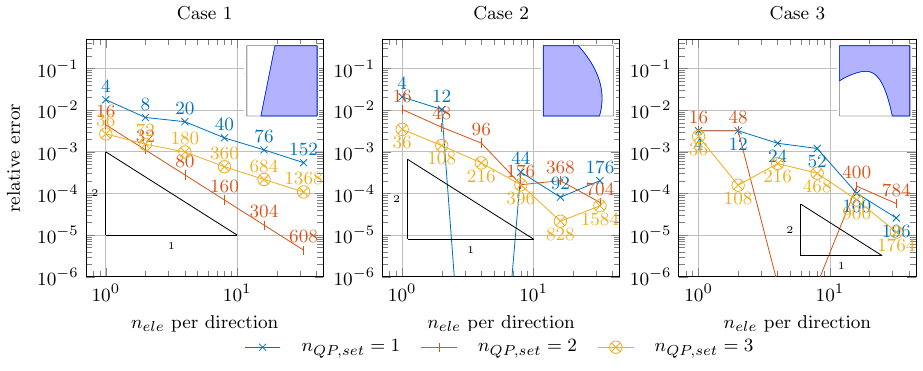}
	\caption{mlhp: Error of area computation with $h$-refinement for the three test case geometries and different number of quadrature points $\numQuadPointsSetting$. The numbers next to the markers note $\numQuadPoints$.}
	\label{fig:mlhp_href}
\end{figure*}

\autoref{fig:mlhp_href} illustrates a corresponding $h$-refinement study, whereby three different values of $\numQuadPointsSetting$ are investigated since no clear dependence was observed in \autoref{fig:mlhp_area}. However, a strong trend for increasing $\numQuadPointsSetting$ can also not be detected by this study.
The convergence rate is observed to be roughly of second order for all graphs except for case 1 with $\numQuadPointsSetting=1$ and $\numQuadPointsSetting=3$ where it is lower. For case 3, the graphs start with a plateau and the rate becomes visible starting from $\numElem\approx 10^1$.
The mlhp code uses a quadtree method to compute the reference for its moment fitting equation which clearly influences its accuracy. Quadtree methods yield often a trend of decreasing geometrical error but also introduce randomness disturbing the convergence behaviour as observed in the cases 2 and 3.

\begin{figure*}
	\centering
	\def\plotwidthfactor{0.35} 
	\includegraphics{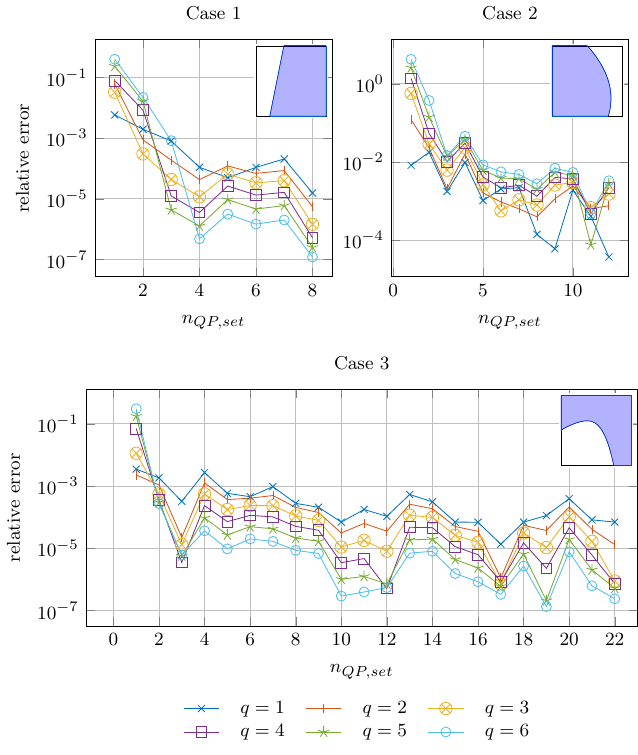}
	\caption{mlhp: Error of integral computation with different $\numQuadPointsSetting$ for the three test case geometries and different integrands of degree $\integrandDegree$.} 
	\label{fig:mlhp_integrand}	
\end{figure*}
\autoref{fig:mlhp_integrand} depicts the influence of $\numQuadPointsSetting$ on the accuracy for an integral computation with a monomial of degree $\integrandDegree$ as integrand. mlhp clearly demonstrates an increasing accuracy with higher $\numQuadPointsSetting$. Nevertheless, the error starts to levels off far from machine precision for $\numQuadPointsSetting$ greater than four.

The results in \autoref{fig:mlhp_area} and \autoref{fig:mlhp_href} exactly and in \autoref{fig:mlhp_integrand} almost coincide with the ones by the FCMLab (cf.~\autoref{sec:fcmlab}). The reason is that mlhp uses a quadtree method as reference in its moment fitting equations. The major difference is that mlhp clearly reduces the number of quadrature points $\numQuadPoints$ as can be seen by comparing the numbers next to the markers of \autoref{fig:fcmlab_area} with \autoref{fig:mlhp_area} and of \autoref{fig:fcmlab_href} with \autoref{fig:mlhp_href}.

A comparison of the results from mlhp with the ones from BoSSS reveals the importance of an accurate computation of the right-hand side of \autoref{eq:moment_fitting}. This can be observed in particular from \autoref{fig:bosss_area} compared to \autoref{fig:mlhp_area} and from \autoref{fig:bosss_integrand} to \autoref{fig:mlhp_href} showing that the less accurate quadtree method employed by mlhp increases the relative errors. 
\subsubsection{QuESo}\label{sec:queso}

\begin{figure}
	\centering
	{\includegraphics[width=0.45\textwidth,trim={140 20 140 10},clip]{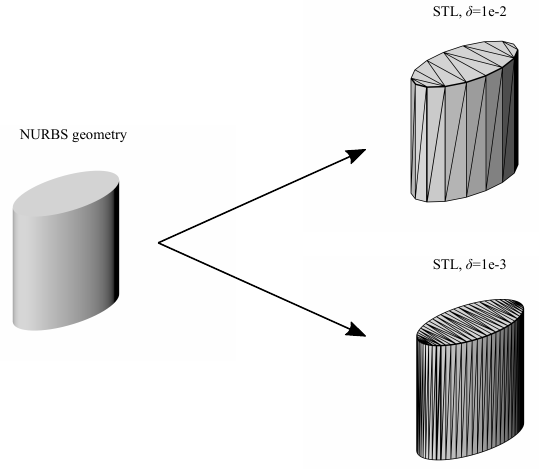}}
	\caption{STL of an elliptic cylinder with different chordal tolerances $\delta$ generated with Rhino \cite{Mcneel}}
	\label{fig:queso_concept}
\end{figure}

QuESo\footnote{\url{https://github.com/manuelmessmer/QuESo}, accessed: 8.4.2025} (Quadrature for Embedded Solids) is a preprocessor for embedded 3D geometries that are provided as STL files. The STL format can be understood as a low-order parametric boundary description. In our context, it is a triangular approximation of a NURBS geometry where the chordal tolerance is typically the main parameter controlling the accuracy of the approximation (see \autoref{fig:queso_concept}). 
An interface to the open-source simulation software Kratos Multiphysics\footnote{\url{https://github.com/KratosMultiphysics/Kratos}, accessed: 8.4.2025} is provided. QuESo uses a tailored quadrature rule derived by solving a system of moment fitting equations as defined in \autoref{eq:moment_fitting}.
The functions $f_i$ are chosen as the 3D Legendre polynomial of degree $q$. Thus, $n=(q+1)^3$ is the number of Legendre polynomials respectively the number of moments in \autoref{eq:moment_fitting}. The optimal number of quadrature points for a complex cutting situation is unknown. Therefore, a point elimination algorithm is applied to find the ideally optimal quadrature points \cite{Nagy2015}. The positions of the quadrature points are fixed in this algorithm, and thereby, the system of equations (\autoref{eq:moment_fitting}) becomes linear in the weights. A correct computation of the exact integrals (right-hand side of \autoref{eq:moment_fitting}) is crucial to obtain an accurate quadrature rule. The divergence theorem as defined in \autoref{eq:divergence_theorem} is used. For the particular case of triangles defining the boundary, it can be rewritten as:
\begin{align}
	\int_{\elemDomain}f_i(\bm{x})d\bm{x} &= \int_{\elemDomain} \nabla \cdot \bm{F}(\bm{x}) d\bm{x} = \nonumber \\
	&= \int_{\partial\elemDomain} \bm{F}(\bm{x}) \cdot \bm{n}(\bm{x}) d\Gamma = \nonumber \\
	&= \sum_{a=1}^{n_t}\int_{\partial\elemDomain_a} \bm{F}(\bm{x}) \cdot \bm{n}(\bm{x}) \det(\bm{J}_{t_a})d\Gamma_a \label{eq:divergence_theorem_triangle}
\end{align}
where $\Gamma_a$ is the $a^{th}$ of $n_t$ triangles building the boundary $\Gamma$ and $\det(\bm{J}_{t_a})$ is the mapping between the physical space of $\Gamma$ and the parametric space $\Gamma_a$ of each boundary triangle $t_a$.

QuESo provides enhanced methods needed in this context for the computation of intersections, the element classification, the clipping of the boundary and the enclosing of domains \cite{mesmer_robust_2024}. The most important parameters to control the integration settings are:
\begin{itemize}
	\item quadrature order
	\item tolerance on the residual of the moment fitting equation
\end{itemize}
In addition, the settings for the generation of the STL files are decisive. Further settings are described in the provided Wiki\footnote{\url{https://github.com/manuelmessmer/QuESo/wiki}, accessed: 8.4.2025}. In particular, Generalized Gaussian Quadrature rules (resp.~patch-wise quadrature rules) are supported for smooth spline basis as used in IGA. This is not further considered in our comparison since the focus of the present work should be on cut element integration (see \cite{Messmer2022} for further details). 
The code is designed for polynomial orders of basis functions in FEM up to degree four, with degree two being recommended, which corresponds to quadrature orders of nine and five, respectively. It is possible to select higher degrees, but this significantly slows down the code. QuESo is written in C++ coming with a Python interface, which enables the call of a generic Python QuESo model from Matlab using \texttt{pyrunfile}. Model settings are defined in separate json-files. Unfortunately, Matlab does not directly provide a powerful STL generation tool. The \texttt{delaunayTriangulation}\footnote{\url{https://de.mathworks.com/help/matlab/ref/delaunaytriangulation.html}, accessed: 24.07.2025}
of Matlab was tested but resulted sometimes in very flawed geometries which could not be processed correctly by QuESo. Therefore, the geometries investigated in \autoref{sec:numerical_examples_3d} were transferred to a STL representation with Rhino\cite{Mcneel}. In \cite{mesmer_robust_2024}, the QuESo authors already tested the accuracy of their approach with the Thingi10K database \cite{zhou_thingi10k_2016} which contains 4948 STL geometries. Further, they assessed the robustness using (artificially) flawed geometries mimicking gaps, overlaps, and incorrect orientations.

A study of the quadrature order with the three geometries from \autoref{sec:three_geometries}, as performed for all other codes, is not applicable at this point since QuESo is the only code that is solely available for 3D geometries.

\subsection{Dimension-reduction}

The two codes discussed in the following subsection apply methods that allow reducing the dimension of a considered integral. This dimension-reduction eases the computation of the integral.

\subsubsection{Algoim}
\label{sec:algoim}

\paragraph{Applied method and code details}
Algoim\footnote{\url{https://github.com/algoim/algoim}, accessed: 27.11.2025} is an open-source package developed using C++.
It is based on dimension-reduction by recasting the problem as the graph of height functions and has two main algorithms following the works by \cite{Saye2015,Saye2022}, with the latter extending the former to support multiple level sets, automatic topology changes, and simplex elements.
In this work, the newer algorithm is employed.

The problem is fundamentally described by an implicit boundary representation, defined through the sufficiently smooth level‐set function $\phi:\mathbb{R}^d\to\mathbb{R}$ 
within the hyperrectangle $H = [a_1,b_1]\times\cdots\times[a_d,b_d]$. 
The volume and the surface of the active cut element are given by
$\elemDomain = \{\bm{x}\in H:\phi(\bm{x})\le0\}$
and 
$\Gamma = \{\bm{x}\in H:\phi(\bm{x})=0\}$, respectively.
Hence, the algorithm seeks to construct quadrature rules with high-order convergence rates for the integrals over $\elemDomain$ and $\Gamma$ in the form given as in \autoref{eq:quadrature_exactness}. 

\begin{figure}
	\centering
	\begin{subfigure}{0.45\textwidth}
	\centering
		\includegraphics{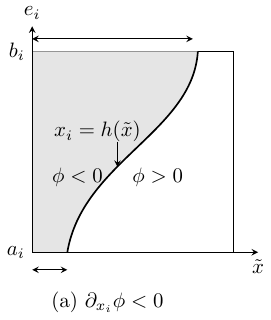}
	\end{subfigure}
	\hfill
	\begin{subfigure}{0.45\textwidth}
	\centering
		\includegraphics{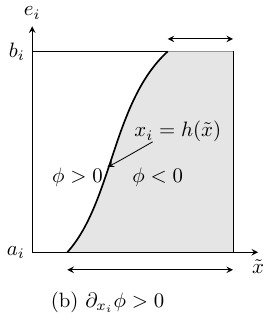}
	\end{subfigure}
	\caption{Algoim:
		Concept of dimension-reduction with height functions $h(\bm{\tilde x})$ from \cite{Saye2015}.
	 Depending on the sign of $\partial_{x_i}\phi$, the active region is defined either above (a) or beneath (b) of the graph of $h(\bm{\tilde x})$, 
	 and the limits $x_i^L$ and $x_i^U$ are accordingly defined between $h(\bm{\tilde x})$ and either of $a_i$ or $b_i$. 
	\label{fig:algoim_height_functions}
	 }
\end{figure}

Initially, the direction $i$ along which the level-set function changes most at the center of $H$, denoted by $\bm{x_c}$, is determined by:
\begin{equation}
	i = \argmax_{1 \le j \le d} \bigl|\partial_{x_j}\phi(\bm{x_c})\bigr|.
\end{equation}
Subsequently, a height function is defined as $x_i=h(\bm{\tilde x})$ with $\bm{\tilde x} \in \mathbb{R}^{d-1}$ and $\bm{x}=(\tilde x_1,...,h(\bm{\tilde x}),...,\tilde x_d)$, and with the use of its graph, the integral is collapsed into two nested integrals in height and tangential directions as:
\begin{equation}
	\int_{\Omega} f(\bm{x}) d \bm{x} = \int_{\bm{\tilde x}}  \int_{x_i=x_i^\mathrm{L}}^{x_i=x_i^\mathrm{U}} f(\bm{x}) d x_i d \bm{\tilde x}.
\end{equation}
\Autoref{fig:algoim_height_functions} illustrates the height function for two monotonic examples of the level-set function $\phi$.
For each fixed $\bm{\tilde x}$, 
the height function is defined between the limits of $H$ in the height direction and $\Gamma$, depending on the sign of $\phi$.
Accordingly, the integral is discretized
between the newly defined lower ($x_i^\mathrm{L}$) and upper ($x_i^\mathrm{U}$) limits with the Gauss–Legendre (or tanh–sinh) quadrature rule, thereby inheriting their convergence behavior. Herein, only the Gauss-Legendre quadrature rule is employed.
Consequently, the same dimension-reduction procedure is called recursively for the tangential direction, while the final $d$-dimensional quadrature nodes and weights are obtained by a tensor product of 1D rules. 
Surface integrals are handled analogously. 

However, if the level-set function is not monotonic in the chosen height direction (i.e., its derivative along that axis changes sign within the element), the height function becomes multi-valued. 
In that case, the element is subdivided into two parts, and the algorithm is re-executed on each part separately.

In the later work~\cite{Saye2022}, the initial approach is generalized to multi‐component domains defined by arbitrary Boolean combinations of multivariate polynomials, including singular and self‐intersecting features.
Moreover, subdivision of elements is performed in a topology-aware fashion, e.g., considering tunnels, corners, cusps and multiple components, and the approach is extended to handle simplex elements with the ultimate aim of providing a black box quadrature algorithm.


The most important parameters to control the integration settings are:
\begin{itemize}
	\item number of quadrature points
	\item reparametrization degree
\end{itemize}

A small Wiki is provided\footnote{\url{https://algoim.github.io/}, accessed: 27.11.2025}. 
We implemented a Matlab wrapper\footnote{\url{https://gitlab.tugraz.at/84445DADE7215DA7/matlabalgoimwrapper}, accessed: 14.1.2026} providing a C++ MEX interface to Algoim, which we employ in our benchmark environment.
The wrapper passes level set functions to Algoim and retrieves the corresponding quadrature points. As the level set data is provided at runtime rather than compile time, this approach is slower than using Algoim directly.

\paragraph{Study of quadrature order for three geometries with different curve degrees}

The general setup of the following study is provided in \autoref{sec:three_geometries}.

\begin{figure}
	\centering
	\def\plotwidthfactor{0.8} 
	\includegraphics{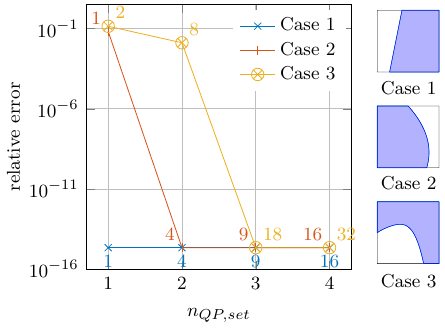}
	\caption{Algoim: Error of area computation with different $\numQuadPointsSetting$ for the three test case geometries. The numbers next to the markers note $\numQuadPoints$.}
	\label{fig:algoim_area}
\end{figure}
\begin{figure*}
	\centering
	\def\plotwidth{0.384\textwidth} 
	\def\plotheight{0.32\textwidth} 
	\includegraphics{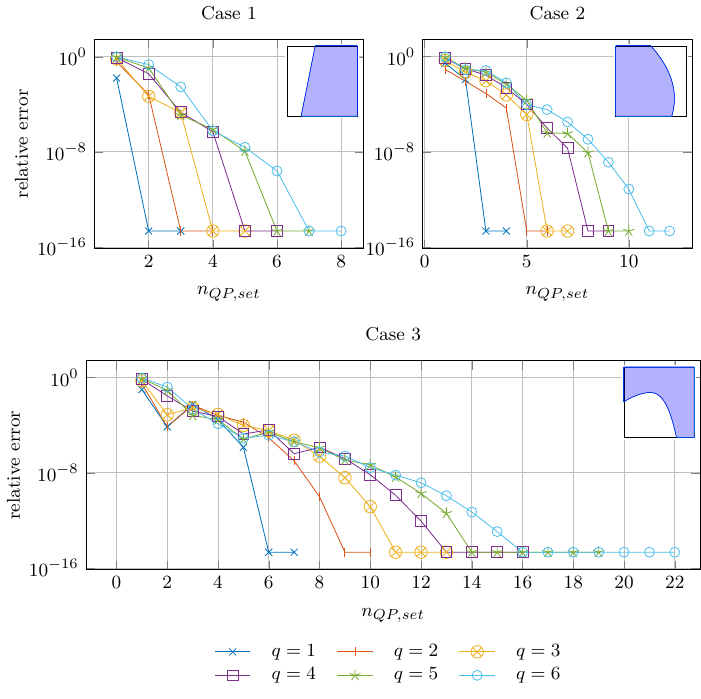}
	\caption{Algoim: Error of integral computation with different $\numQuadPointsSetting$ for the three test case geometries and different integrands of degree $\integrandDegree$.} 
	\label{fig:algoim_integrand}	
\end{figure*}
\autoref{fig:algoim_area} shows the influence of $\numQuadPointsSetting$ on the accuracy for an area computation. Algoim is able to reach machine precision with the expected $\numQuadPointsSetting$ defined by \autoref{eq:nQP_exact_single}. For test case 3, a subdivision is required because it is a five-sided element, as indicated by the numbers next to the markers, this leads to twice as many quadrature points as for the other two geometries. The subdivision is also visualized in \autoref{fig:example_unibw16_QP}.
No $h$-refinement study is performed for the area computation since machine precision can be already reached by increasing $\numQuadPointsSetting$.

\autoref{fig:algoim_integrand} illustrates the influence of $\numQuadPointsSetting$ on the accuracy for an integral computation with a monomial of degree $\integrandDegree$ as integrand. Algoim is able to reach machine precision with the expected $\numQuadPointsSetting$ defined by \autoref{eq:nQP_exact_single}. Whereby, \autoref{eq:nQP_exact_single} slightly overestimates the required $\numQuadPointsSetting$ for case 3 with the quintic boundary curve for all integrands $\integrandDegree$ greater than two. The reason is that the curve degree does not resemble the curve's complexity.

\subsubsection{QuaHOG}\label{sec:QuaHOG}

\begin{figure*}
	\centering
	\includegraphics[width=\textwidth]{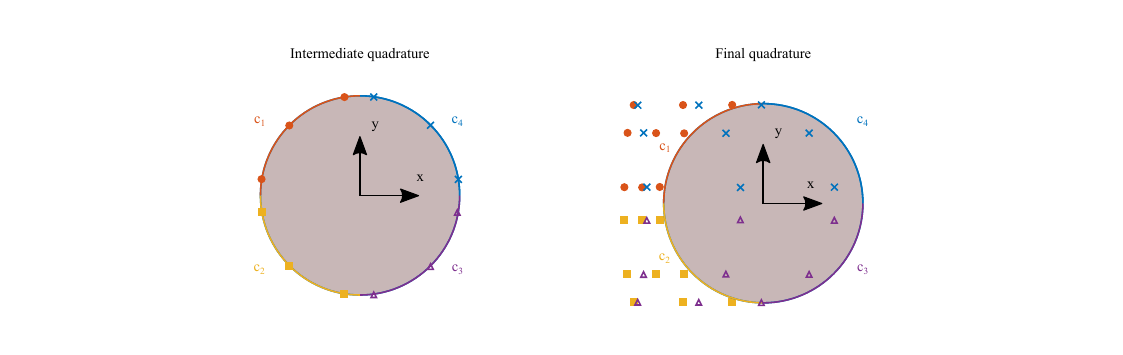}
	\caption{QuaHOG: Positioning of quadrature points applying Green's Theorem on a circular disk. The boundary is described by four curves $c_i$. The quadrature points corresponding to same curves are illustrated with same markers. The number of quadrature points is $Q_i=P=3$. The transition from the intermediate to the final quadrature can be understood such that a quadrature point ($x_{i,q}$,$y_{i,q}$) is mapped to three points with coordinates ($x_{i,q,\zeta}$,$y_{i,q}$) where $x_{i,q,\zeta}$ is distributed between $C$ and $x_{i,q}$ according to a Gaussian quadrature.}
	\label{fig:quahog_concept}
\end{figure*}

\paragraph{Applied method and code details}
QuaHOG\footnote{\url{https://github.com/davidgunderman/QuaHOG}, accessed: 26.3.2025} (Quadrature for High-Order Geometries)  produces
higher-order quadrature rules for intersected 2D geometries in combination with
arbitrary integrands. It features no physical application but is a pure quadrature tool. 
In contrast to the other codes discussed herein, QuaHOG applies
a mesh-free approach. Nevertheless, it is included in the present work since it can
be adapted for any immersed problem dealing with cut elements and proves to enable
high-order accuracy for any high-order polynomial or rational boundary description.
Parametric curves are considered. The procedure is based on Green's theorem
\begin{align}
	\iint_{\Omega}f(x,y)dxdy &= \int_{\partial\Omega} A_f(x,y)dy 
\end{align}
where $\Omega$ is the active domain and $A_f$ is the $x$-antiderivative of the integrand $f$. Green's theorem reduces
the integral by one dimension, whereby the choice of the direction is arbitrary. 
Furthermore, the boundary is split up into $n_c$ rational Bernstein-Bézier curves $c_i$
\begin{align}
	\int_{\partial\Omega} A_f(x,y)dy = \sum_{i=1}^{n_c}\int_{c_i} A_f(x,y)dy.
\end{align}

An intermediate quadrature is generated to compute the line integral
\begin{align}
	\int_{c_i}A_f(x,y)dy = \int_{0}^{1} A_f(x_i(s),y_i(s)) \dfrac{dy_i(s)}{ds} ds
\end{align}
where $s\in[0,1]$ is the curve parameter. 
This 1D-integral can be evaluated by Gaussian quadrature. Thereby, intermediate quadrature points ${s_{i,q}}$ and weights ${\gamma_{i,q}}$ are obtained. The
points in $\mathbb{R}^2$ read
\begin{align}
	(x_{i,q},y_{i,q}) = (x_i(s_{i,q}),y_i(s_{i,q})).
\end{align}
The antiderivative $A_f$ can also be computed by means of Gaussian quadrature
\begin{align}
	A_f(x_{i,q},y_{i,q}) &= \int_{C}^{x_{i,q}} f(x,y_{i,q})dx \approx \nonumber \\
	&\approx \sum_{\zeta=1}^{P} \gamma_{i,q,\zeta}f(x_{i,q,\zeta},y_{i,q})
\end{align}
where $C$ is an integration constant and $\gamma$ an integration weight. $C$ can be chosen freely, but it affects the locations and weights of the quadrature points as 
well as the floating point stability if it is chosen to be very far away from the integration domain. Therefore, it is defined 
according to \cite{Gunderman2021} as the minimum $x$-coordinate of any control point defining the boundary
\begin{align}
	C=\text{min}\{x_j\} \quad \text{for} \quad j=1,...,n_{CP}
\end{align}
where $n_{CP}$ is the total number of control points of the boundary.
The final quadrature reads
\begin{align}
	\iint_{\Omega}f(x,y)dxdy \approx \sum_{i=1}^{n_c}\sum_{q=1}^{Q_i}\sum_{\zeta=1}^{P}w_{i,q,\zeta}f(x_{i,q,\zeta},y_{i,q})
\end{align}
where $Q_i$ is the number of intermediate quadrature points and $P$ is the number of antiderivative quadrature points. The quadrature weight is
\begin{equation}
	w_{i,q,\zeta} = \gamma_{i,q}\gamma_{i,q,\zeta}\dfrac{dy_i}{ds}(s_{i,q}).
\end{equation}
An example for the resulting intermediate and final quadrature for a circular disk is illustrated in \autoref{fig:quahog_concept}.
QuaHOG supports an additional variation of this method where a rational quadrature rule is 
used for the intermediate quadrature instead of a Gaussian rule. Herein, this alternative is called 
QuaHOGPE since it is exact for any polynomial integrand 
(PE = \underline{p}olynomial \underline{e}xact).
The most important parameters for QuaHOG to control the integration settings are:
\begin{itemize}
	\item number of intermediate quadrature points $Q_i$
	\item number of antiderivative quadrature points $P$
\end{itemize}
The choice of these numbers is not trivial. For simplicity, \cite{Gunderman2021} sets $Q_i=P$, which we adopt.
The only integration setting in case of QuaHOGPE is:
\begin{itemize}
	\item quadrature order
\end{itemize}
The open-source repository only supports 2D, but a similar procedure for 3D, based on the generalized Stoke's theorem, is published in \cite{Gunderman2021a}. The interfacing of QuaHOG in our benchmark environment is straightforward due to the fact that both are written in Matlab. 
In detail, overloading allowed to overwrite problematic functions. The QuaHOG authors carried out a comparison of their code to existing methods such as quadtree and tessellation concluding a superior performance \cite{Gunderman2021}.

\paragraph{Study of quadrature order for the three 2D test case geometries}

The general setup of the following study is provided in \autoref{sec:three_geometries}.

\begin{figure}
	\centering
	\def\plotwidthfactor{0.8} 
	\includegraphics{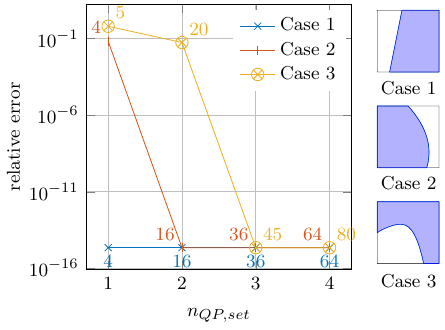}
	\caption{QuaHOG: Error of area computation with different $\numQuadPointsSetting$ for the three test case geometries. The numbers next to the markers note $\numQuadPoints$.}
	\label{fig:quahog_area}
\end{figure}

\autoref{fig:quahog_area} shows the influence of $\numQuadPointsSetting$ on the accuracy for an area computation. QuaHOG is able to reach machine precision with the expected $\numQuadPointsSetting$ defined by \autoref{eq:nQP_exact_single}. The relation between $\numQuadPointsSetting$ and $\numQuadPoints$ is given as
\begin{equation}
	\numQuadPoints=n_c \cdot Q_i \cdot P = n_c \cdot (\numQuadPointsSetting)^2
\end{equation}
where $Q_i=P$ as mentioned above.
The results for QuaHOGPE are omitted because it is able to reach machine precision for all geometries with $\numQuadPointsSetting=1$. The reason is that the geometrical
error is only influenced by its intermediate quadrature rule (in the context of a pure area computation) for which the method determines the appropriate
number of quadrature automatically. $\numQuadPointsSetting$ solely influences the antiderivative quadrature rule for which
a single quadrature point is sufficient since the computation of the area equals a constant integrand.

\begin{figure*}
	\centering
	\def\plotwidth{0.384\textwidth} 
	\def\plotheight{0.32\textwidth} 
	\includegraphics{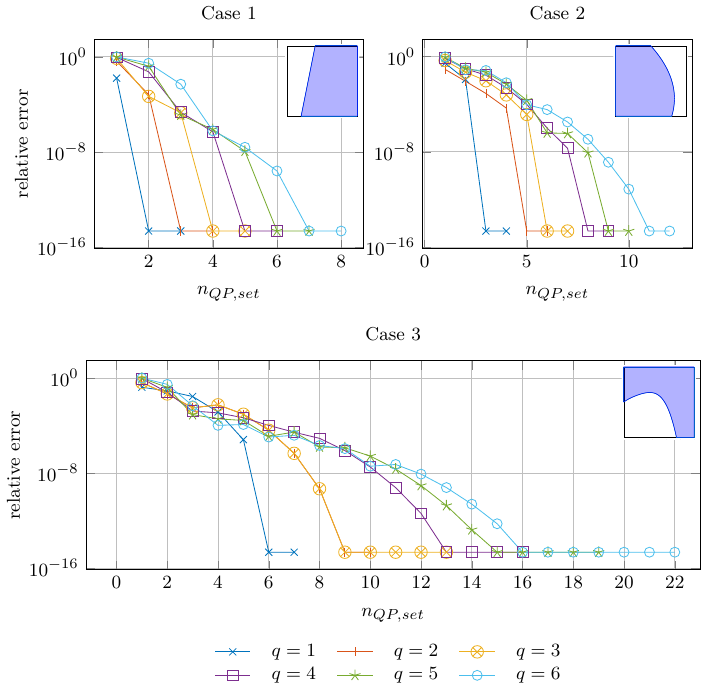}
	\caption{QuaHOG: Error of integral computation with different $\numQuadPointsSetting$ for the three test case geometries and different integrands of degree $\integrandDegree$.} 
	\label{fig:quahog_integrand}	
\end{figure*}

\begin{figure*}
	\centering
	\def\plotwidth{0.35\textwidth} 
	\def\plotheight{0.3\textwidth} 
	\includegraphics{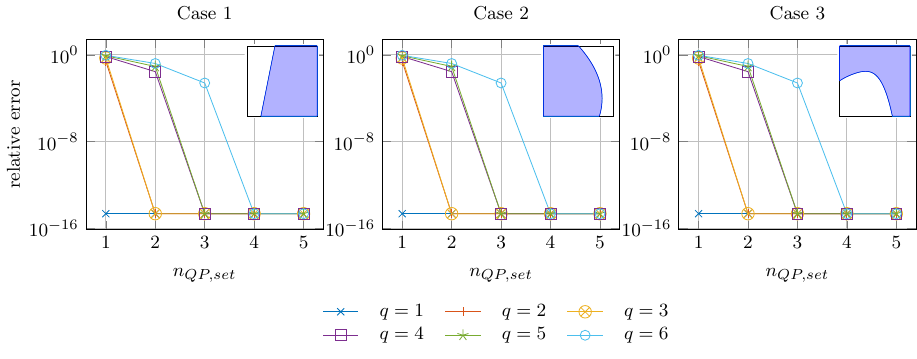}
	\caption{QuaHOGPE: Error of integral computation with different $\numQuadPointsSetting$ for the three test case geometries and different integrands of degree $\integrandDegree$.} 
	\label{fig:quahogpe_integrand}	
\end{figure*}

\autoref{fig:quahog_integrand} and \autoref{fig:quahogpe_integrand} illustrate the influence of $\numQuadPointsSetting$ on the accuracy for an integral computation with a monomial of degree $\integrandDegree$ as integrand for QuaHOG respectively QuaHOGPE. QuaHOG is able to reach machine precision with the expected $\numQuadPointsSetting$ defined by \autoref{eq:nQP_exact_single}. However, \autoref{eq:nQP_exact_single} slightly overestimates the required $\numQuadPointsSetting$ for test case 3 with the quintic boundary curve. The reason is that the curve degree does not resemble the curve's complexity.
QuaHOGPE is also able to reach machine precision. The required $\numQuadPointsSetting$ resembles the standard number of Gauss points needed for the antiderivative quadrature rule. The convergence curves for all three test cases look alike because the geometrical error is negligible resulting in a pure integration error depending on the integrand function.

\subsection{Further codes}\label{sec:further_codes}
The following codes are not included in our repository and the subsequent comparison: CutFEM-Library, deal.II and TPMC.

The CutFEM-Library\footnote{\url{https://github.com/CutFEM/CutFEM-Library}, accessed: 24.7.2025} is a relatively small code used, e.g., for the simulation of Darcy flow \cite{frachon_divergence_2024} and time-dependent convection-diffusion equations \cite{myrback_high-order_2024}. It directly applies the algorithms provided by the code Algoim, which is included in our discussion and explained in more detail in \autoref{sec:algoim}. 

deal.II\footnote{\url{https://github.com/dealii/dealii}, accessed: 24.7.2025}, which is an extensive FEM library, also directly applies the algorithms provided by Algoim. Its immersed module is for example used for the solution of the wave equation in \cite{sticko_higher_2019}. Two small additional repositories based on deal.II and working with immersed problems are found online: fem-cut-cell-3D\footnote{\url{https://github.com/afonsoal/fem-cut-cell-3D}, accessed: 24.7.2025, applied e.g. in \cite{londero_cut-cell_2015}} and ans-ifem\footnote{\url{https://github.com/luca-heltai/ans-ifem}, accessed: 24.7.2025, applied e.g. in \cite{heltai_fully_2012}}. They are not further investigated since they are directly based on deal.II.

TPMC\footnote{\url{https://github.com//tpmc}, accessed: 24.7.2025} is a module for immersed computations within DUNE\footnote{\url{https://dune-project.org/}, accessed: 24.7.2025}, which is an extensive library for the Finite Element and the Finite Volume Method. TPMC provides a topology preserving marching cubes algorithm that robustly determines cutting patterns on element level for level-set functions.
The authors present comprehensive accuracy and robustness tests for their method \cite{engwer_geometric_2018}.
A marching cubes algorithm is for example also used in the mlhp code (cf.~\autoref{sec:mlhp}). The respective code is not considered because the corresponding Github repository is lacking sufficient documentation and examples to be incorporated with a reasonable workload. For future work, it might be worth to check the Zenodo version of TPMC\footnote{\url{https://doi.org/10.5281/zenodo.252153}, accessed: 12.02.2026} which was published separately accompanying \cite{engwer_geometric_2018}.


\section{Comparative numerical examples}\label{sec:numerical_examples}

In this section, multiple numerical examples are investigated in order to assess and compare the different open source 
codes and their respective quadrature methods, which were discussed in the preceding sections. Since the included codes
are so diverse, the considered problems involve purely numerical integration without a specific physical application.
Thereby, the methods can be analyzed w.r.t.~geometrical and integration errors. Two-dimensional and three-dimensional
geometries are considered where certain codes might only support either of them (cf. \autoref{tab:overview}). All models
computed herein can be found in the folder `publications/Paper\_CutElemComp'  in our open-source repository \cite{CutElementIntegration2025LatestVersion}. The presented examples are only a fraction of all benchmarks available in \cite{CutElementIntegration2025LatestVersion} whereby all models directly run with the discussed codes.


Some general remarks and limitations regarding the following studies should be mentioned here, so that they do not need to be repeated in each example:
\begin{itemize}
	\item The results from FCMLab and mlhp almost coincide because mlhp applies the quadtree/octree method, which is also used by FCMLab, for the reference integral in its moment fitting.
	
	\item It is not possible to compute single element meshes with level set functions in Gridap.
	
	\item The ngsxfem code supports quad-meshes and its mesh transformation method for single level-set functions only. If applicable, a quad-mesh with a mesh transformation is employed, whereby the degree of the mesh transformation (reparametrization degree) equals the degree of the boundary curve. Furthermore, it is not possible to retrieve quadrature points for that code in our Matlab interface. In the following examples, the same mesh size $h$ is chosen as input in case of triangular meshes, which is the same input as for the other codes with quadrilateral meshes, making aware that a direct comparison between a triangle and a quadrilateral mesh is limited.
	
	\item It is not possible to sort the quadrature points in points coming from cut and uncut elements for Nutils.
	
	\item QuaHOG and QuaHOGPE are not considered if an $h$-refinement is performed, as they employ a mesh-free method.
	
	\item Values below $2.22\cdot 10^{-15}$ are defined to be machine precision and are plotted as such values. A higher decimal precision is not considered.
\end{itemize}

\subsection{Two-dimensional geometries}
\label{sec:numerical_examples_2d}

At first, \autoref{sec:QP_positioning} demonstrates the variety of approaches for cut element integration by a comparison of the final quadrature points layout.
Afterwards, \autoref{sec:testsuite_unibw} investigates the
capability of the discussed codes to deal with a number of cutting situations including different boundary curve degrees, rational 
boundary curves or boundaries with kinks. Subsequently, the robustness of the codes is tested by step-wise shifting a geometry
in \autoref{sec:robustness}. Afterwards, a comparison of the efficacy of the codes w.r.t. monomials as integrands is studied in \autoref{sec:monomials} which was already separately discussed in the previous \autoref{sec:review} for each single code.
The following nine 
codes support two-dimensional computations: Algoim, BoSSS, FCMLab, QUGaR, mlhp, ngsxfem, Nutils, QuaHOG, QuaHOGPE.

\subsubsection{Quadrature point layout}
\label{sec:QP_positioning}

\begin{figure*}
	\centering
	\begin{subfigure}{0.32\textwidth}
		\centering
		\includegraphics[width=\textwidth]{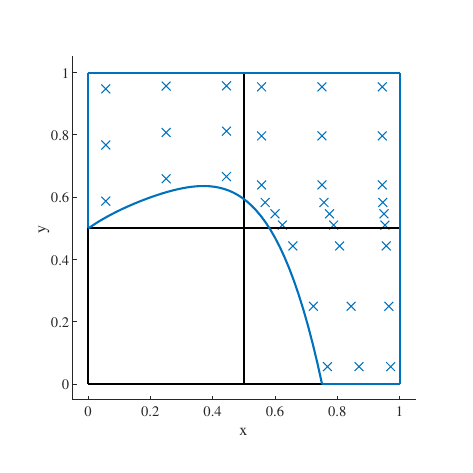}
		\caption{Algoim}
	\end{subfigure}
	\hfill
	\begin{subfigure}{0.32\textwidth}
		\centering
		\includegraphics[width=\textwidth]{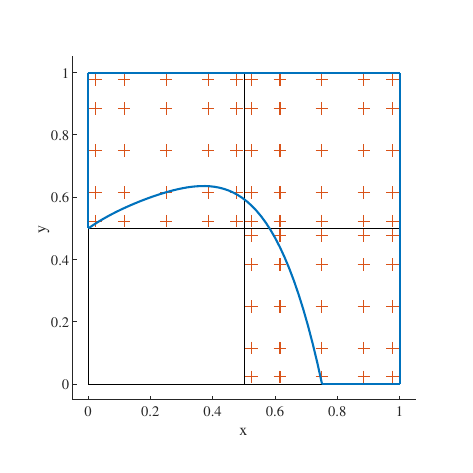}
		\caption{BoSSS}
	\end{subfigure}
	\hfill
	\begin{subfigure}{0.32\textwidth}
		\centering
		\includegraphics[width=\textwidth]{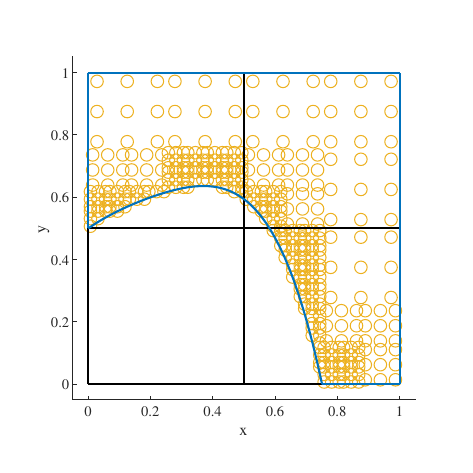}
		\caption{FCMLab}
	\end{subfigure}
	
	\begin{subfigure}{0.32\textwidth}
		\centering
		\includegraphics[width=\textwidth]{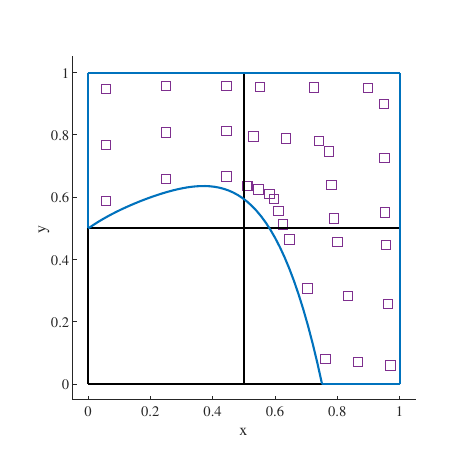}
		\caption{QUGaR}
	\end{subfigure}
	\hfill
	\begin{subfigure}{0.32\textwidth}
		\centering
		\includegraphics[width=\textwidth]{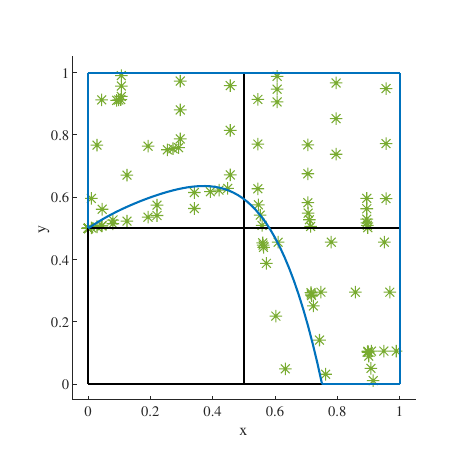}
		\caption{Gridap}
	\end{subfigure}
	\hfill
	\begin{subfigure}{0.32\textwidth}
		\centering
		\includegraphics[width=\textwidth]{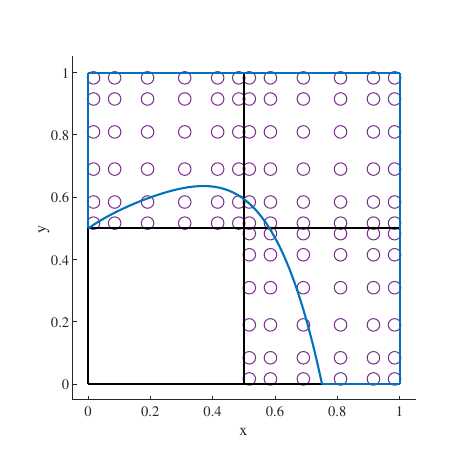}
		\caption{mlhp}
	\end{subfigure}
	
	\begin{subfigure}{0.32\textwidth}
		\centering
		\includegraphics[width=\textwidth]{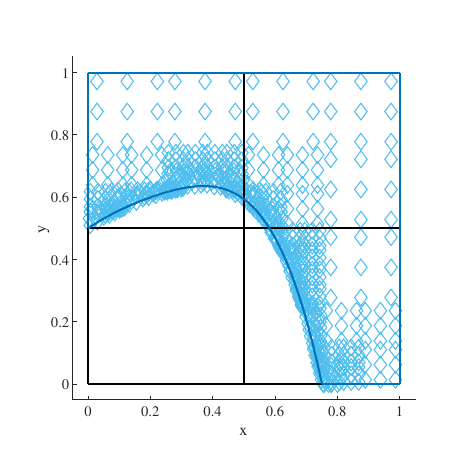}
		\caption{Nutils}
	\end{subfigure}
	\hfill
	\begin{subfigure}{0.32\textwidth}
		\centering
		\includegraphics[width=\textwidth]{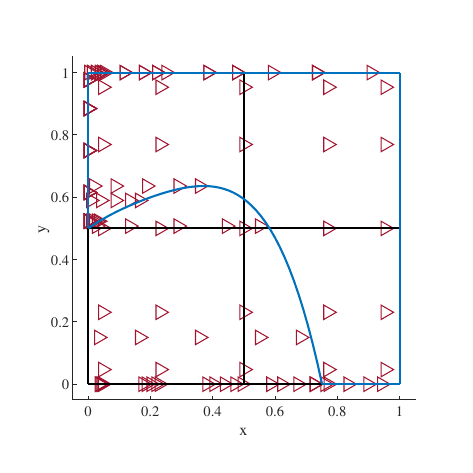}
		\caption{QuaHOG}
	\end{subfigure}
	\hfill
	\begin{subfigure}{0.32\textwidth}
		\centering
		\includegraphics[width=\textwidth]{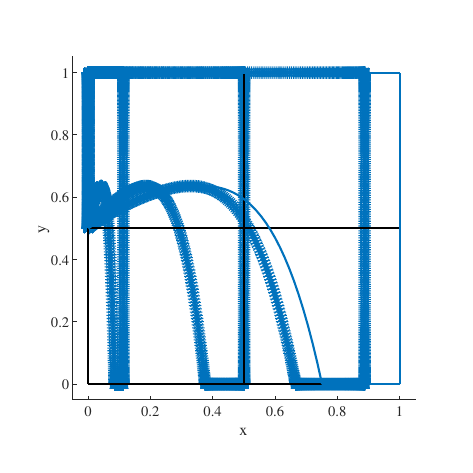}
		\caption{QuaHOGPE}
		\label{fig:example_unibw16_QP_QuaHOGPE}
	\end{subfigure}
	\caption{Quadrature points for the test case geometry 3 (cf.~\autoref{fig:geo_area_single_q5}). Active domain indicated by blue line. Background mesh indicated by black line. $n_{ele}=2$ per direction. Settings for illustration purposes only: $\numQuadPointsSetting=3$, $\reparamDegree=5$, $\numSubLevel=3$ (number of subdivision levels in quadtree), $Q_i=P=5$ (number of intermediate and antiderivative quadrature points employed by QuaHOG and QuaHOGPE).}
	\label{fig:example_unibw16_QP}
\end{figure*}

\autoref{fig:example_unibw16_QP} presents plots of the quadrature points for the test case geometry 3 which was introduced in \autoref{sec:three_geometries}. A refinement of two elements per direction is used leading to three cut and one inactive element. The point layouts for Algoim and QUGaR illustrate the case where a subdivision of the five-sided cut element is necessary -- cf.~upper, right element. Thereby, Algoim employs a horizontal subdivision, whereas QUGaR employs a diagonal subdivision.
FCMLab and Nutils clearly illustrate the quadtree scheme with its high number of quadrature points along the cutting curve. BoSSS, Gridap, mlhp, QuaHOG, and QuaHOGPE lead to quadrature points that are also outside of the active domain, which might be disadvantageous for certain applications. In case of QuaHOG and QuaHOGPE, the points often lay even in inactive elements as can be seen here. BoSSS and mlhp choose a uniform distribution across a cut element which often results in points outside of the active domain of the element in combination with negative quadrature weights. For Gridap, it depends on the geometry being concave or convex and the resulting boundary approximation whether points occur in the inactive domain of a cut element. However, they are never outside of the approximated geometry. A similar behaviour is expected for ngsxfem. Yet, it is not possible to retrieve the quadrature points for that code in our Matlab interface. In case of BoSSS, the bottom left element is at first detected as cut due to the intersection of the boundary curve at the top left vertex of this element. This results in a single quadrature point with a zero weight being assigned to this element. This point is omitted in \autoref{fig:example_unibw16_QP} because it does not contribute in the integration due to its zero weight.

\subsubsection{Diverse cutting scenarios}\label{sec:testsuite_unibw}

The following example tests the versatility of the different codes by running a sequence of 26 test geometries -- three of them were already discussed in detail in \autoref{sec:review}, whereby the three cases from \autoref{sec:three_geometries} are associated with the geometries 1, 2, and 13 in this larger testsuite, respectively. 

All the different geometries are visualized in \autoref{sec:app_testsuite_unibw}. Their goal is to cover typical cutting situations including varying curve degrees, rational curve functions, kinks, inner knots, small features, etc. 
They mimic cutting patterns occurring on an element level. Therefore, all geometries intersect with the domain boundary and no boundary curves are completely encapsulated within the domain. This incorporates the assumption that no features which are smaller than one element are of practical interest -- respectively mesh refinement would be suggested for a proper resolution. In consequence, reasonable results are already expected for the considered geometries with a single element ($\numElem=1$).

\begin{figure*}
	\centering
	\def\plotwidthfactor{0.38} 
	\includegraphics{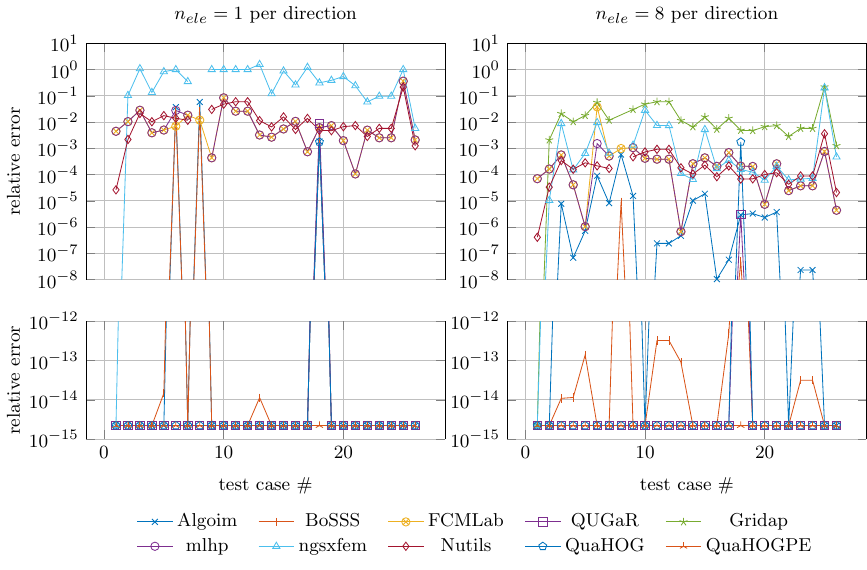}
	\caption{Error of area computation for 26 different geometries (test cases). Note that Gridap is missing for $n_{ele}=1$ is not applicable to a single element.}
	\label{fig:testsuite_unibw}
\end{figure*}
An area computation is carried out with two different refinements, namely $\numElem=1$ and $\numElem=8$ per direction. 
The following values are chosen for $\numQuadPointsSetting$: $\numQuadPointsSetting=2$ for FCMLab and mlhp, $\numQuadPointsSetting=1$ for all codes with a linear boundary approximation and $\numQuadPointsSetting=\lceil(\curveDegree+1)/2\rceil=2$ for all remaining codes, whereby $\curveDegree$ is the degree of the boundary curve. The chosen values are motivated by the studies performed in the particular code parts in \autoref{sec:review}.
The results are presented as relative error in \autoref{fig:testsuite_unibw}.

It can be observed that the codes can be distinguished into two groups: the first group applies some sort of boundary approximation leading to a geometrical error and the second group tries to resolve the boundary as accurately as possible. FCMLab, Gridap, mlhp, ngsxfem, and Nutils count to the first group, whereas Algoim, BoSSS, QUGaR, QuaHOG, and QuaHOGPE count to the second group. However, the distinction might differ in special cases depending on the actual geometry, e.g., straight boundary curves.

In \autoref{fig:testsuite_unibw}, some outliers are noted for certain geometries, e.g.: Algoim shows less accuracy on the test cases 6 (\autoref{fig:testsuite_unibw_TC6}) and 8 (\autoref{fig:testsuite_unibw_TC8}) -- kink and tangential connection between two boundaries -- and Algoim, QUGaR, and QuaHOG show less accuracy on test case 18 (\autoref{fig:testsuite_unibw_TC18}) -- NURBS curve as parametric description. Remark that QuaHOGPE is able to resolve the NURBS boundary correctly since it is using a rational quadrature rule as intermediate rule. The finer mesh leads to an improvement for all cases and codes except for Algoim.
The reason why Algoim performs worse in some cases might be that difficult cutting situations including almost tangential cuts (such as in test case 6) occur more likely for the refined mesh. While adapting the integration parameters could improve the performance in such cases, finding optimal settings across all integrators and test cases is beyond the scope of this paper. Nevertheless, the relative error remains mostly below $10^{-5}$. Gridap is again only included in case of the finer mesh because it does not run for single elements.
Furthermore, Gridap, mlhp, ngsxfem, and Nutils are not able to run test case 8 \autoref{fig:testsuite_unibw_TC8} in our Matlab interface due to the used definition of piece-wise defined implicit boundaries. The failed cases are indicated by the respective missing markers in \autoref{fig:testsuite_unibw}. 
FCMLab and mlhp again coincide for almost all cases. Only for test cases 6 (\autoref{fig:testsuite_unibw_TC6}), their results differ, with FCMLab being more accurate for $\numElem=1$ and mlhp for $\numElem=8$.
QuaHOGPE is the only code demonstrating perfect accuracy for all 26 test cases; in particular including test case 18 which employs a NURBS curve. QuaHOG and QUGaR show optimal accuracy for all test cases except for the mentioned test case 18.

\subsubsection{Robustness test with shifted intersected parabolas}\label{sec:robustness}

\begin{figure}[b!]
	\centering
	\includegraphics{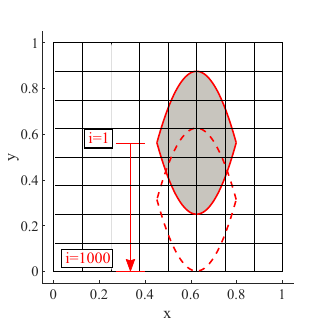}
	\caption{Problem setup of shifted intersected parabolas. The red, solid curve with the grey shading depicts the initial position at $i=1$. The red, dashed curve depicts the last position at $i=1000$.}
	\label{fig:example_moving_parabola_setup}
\end{figure}

\begin{figure*}
	\centering
	\def\plotwidthfactor{1}
	\includegraphics{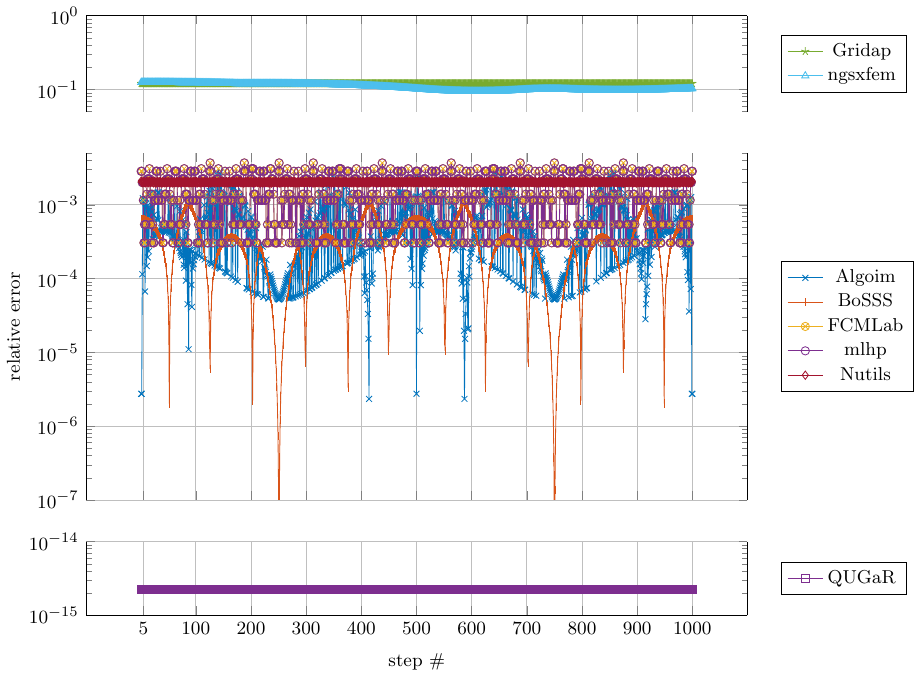}
	\caption{Error of area computation for shifted intersected parabolas. $n_{ele}=8$ per direction. FCMLab coincides with mlhp.}
	\label{fig:example_moving_parabola_results}
\end{figure*}

An important property of a cut element integration method is its robustness against small perturbations. This is tested herein by shifting two intersected parabolas through the domain as illustrated in \autoref{fig:example_moving_parabola_setup}. 
The background mesh contains eight elements per direction, i.e., $\numElem=8$. The geometry is shifted by 999 equally distanced steps, resulting in 1000 computed positions including the starting position. Thereby, no cutting situation is exactly repeated since the resulting step size is no fraction of the mesh size. 
The following values are chosen for $\numQuadPointsSetting$: $\numQuadPointsSetting=2$ for FCMLab and mlhp, $\numQuadPointsSetting=1$ for all codes with a linear boundary approximation, and $\numQuadPointsSetting=\lceil(\curveDegree+1)/2\rceil=2$ for all remaining codes, whereby $\curveDegree$ is the degree of the boundary curve. The chosen values are motivated by the studies performed in the particular code parts in \autoref{sec:review}.
The relative error of the area computation is investigated and illustrated in \autoref{fig:example_moving_parabola_results}.
It can be nicely observed that the graph almost but not exactly shows a repetition of the results after step $i=500$ (cf.~in particular the graph for Algoim). 

QuaHOG and QuaHOGPE are not considered in this study because they are mesh-free methods and, therefore, shifting the geometry through the background mesh would not affect their results.
Please note that the error is given in a logarithmic scale such that a larger graphical oscillation not necessarily resembles a higher absolute oscillation.

The ngsxfem code uses a triangular mesh since the geometry is defined by two level-set functions. This entails a graph that does not show a repetition of the results after step $i=500$ since the cutting patterns do not repeat.

In general, all codes prove quite robust in this test. The respective accuracies again depend on the element size as already discussed in \autoref{sec:review} and \autoref{sec:testsuite_unibw}. Still, QUGaR has to be highlighted because of its extraordinary accuracy (machine precision) and no oscillations at all.

\begin{table*}
	\centering
	\caption{Accuracy of the area computation for shifted intersected parabolas. $\bar{e}$ is the mean value of the relative error, $\sigma_e$ is the standard deviation of the relative error, $\bar{n}_{QP,tot}$ is the mean value of the number of quadrature points in all cut elements, and $\sigma_{n_{QP,cut}}$ is the standard deviation of the number of quadrature points in all cut elements. The mean values and the standard deviations are computed w.r.t.~the sum of all steps.}\label{tab:example_moving_parabola}
	\begin{tabular}{@{}p{2cm}p{2cm}p{2cm}p{2cm}p{2cm}@{}}
		\toprule
		Name & $\bar{e}$ & $\sigma_e$ & $\bar{n}_{QP,cut}$ & $\sigma_{n_{QP,tot}}$ \\
		\midrule
		Algoim & $5.72\text{e}\!-\!04$ & $5.44\text{e}\!-\!04$ & $87.62$ & $5.80$ \\ 
		BoSSS & $3.31\text{e}\!-\!04$ & $2.58\text{e}\!-\!04$ & $320.56$ & $2.96$ \\ 
		FCMLab & $1.50\text{e}\!-\!03$ & $9.21\text{e}\!-\!04$ & $623.66$ & $55.09$ \\ 
		Gridap & $1.20\text{e}\!-\!01$ & $8.43\text{e}\!-\!16$ & $44.10$ & $0.58$ \\ 
		mlhp & $1.50\text{e}\!-\!03$ & $9.21\text{e}\!-\!04$ & $255.94$ & $1.43$ \\ 
		ngsxfem & $1.08\text{e}\!-\!01$ & $1.09\text{e}\!-\!02$ & n/a & n/a \\ 
		Nutils & $2.02\text{e}\!-\!03$ & $2.56\text{e}\!-\!05$ & n/a & n/a \\ 
		QUGaR & $2.22\text{e}\!-\!15$ & $2.76\text{e}\!-\!30$ & $87.98$ & $0.36$ \\
		\bottomrule
	\end{tabular}
\end{table*}
\autoref{tab:example_moving_parabola} presents the mean relative error with its standard deviation and the mean total number of quadrature points with its standard deviation. It can be observed that Gridap and ngsxfem still have problems in resolving the geometry correctly for the given mesh. The results from FCMLab and mlhp again coincide w.r.t.~the computed errors. FCMLab and mlhp show a similar error compared to Nutils. However, the triangulation on the lowest level of the quadtree guarantees a more stable method, as indicated by the smaller standard deviation for Nutils. A clear reduction of the number of quadrature points can be observed in the case of mlhp compared to the FCMLab due to the applied moment fitting approach. The high standard deviation $\sigma_{n_{QP,cut}}$ for FCMLab shows the significant influence of the concrete cutting situation on the number of points for a pure quadtree approach. On the other hand, the stable number of points in case of Gridap and QUGaR has to be highlighted. In this example, QUGaR shows brilliant results w.r.t. its accuracy and its employed number of points.

\subsubsection{Monomial as integrand}
\label{sec:monomials}

In the preceding sections, the geometrical error was investigated by a simple computation of the area for different geometries. Now, monomials are used as integrands. This section discusses the afore introduced test case geometry 2 with a quadratic boundary curve (cf.~\autoref{fig:geo_area_single_q2}) and a monomial of degree six as integrand (cf. \autoref{sec:three_geometries} for further details regarding the setup). Some of the following results were already shown in the particular code descriptions in \autoref{sec:review} and are collected for a better comparison at this point. No results for ngsxfem are illustrated if the graphs are plotted against $\numQuadPoints$ because it was not possible to retrieve the quadrature points from this code. Further, results for monomials of degree one to five as integrand can be found in the specific code parts in \autoref{sec:review} and are omitted here for brevity.


\begin{figure*}
	\centering
	\begin{subfigure}{0.48\textwidth}
		\def\tkzscale{1}
		\def\plotwidthfactor{0.85} 
		\centering
		\includegraphics{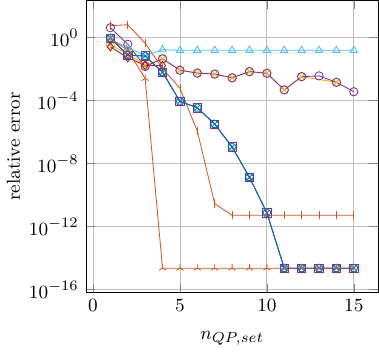}
	\end{subfigure}
	\hfill
	\begin{subfigure}{0.48\textwidth}
		\def\tkzscale{1}
		\def\plotwidthfactor{0.85} 
		\centering
		\includegraphics{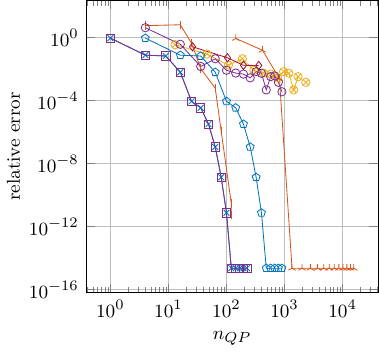}
	\end{subfigure}

	\includegraphics{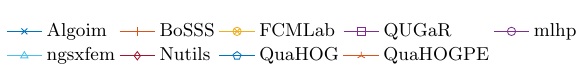}

	\caption{Study of $\numQuadPointsSetting$. Test case geometry 2 (cf.~\autoref{fig:geo_area_single_q2}). Monomial $p=6$. Same data: left with $\numQuadPointsSetting$ and right with $\numQuadPoints$ on $x$-axis. In left plot: Algoim exactly coincides with QuaHOG and QUGaR, and FCMLab mostly coincides with mlhp.} 
	\label{fig:example_unibw4_integrand_p6_nQP}
\end{figure*}
\subsubsection{Monomial as integrand}
\label{sec:monomials}

In the preceding sections, the geometrical error was investigated by a simple computation of the area for different geometries. Now, monomials are used as integrands. This section discusses the already introduced test case geometry 2 with a quadratic boundary curve (cf.~\autoref{fig:geo_area_single_q2}) and a monomial of degree six as integrand. This setup was already introduced in \autoref{sec:three_geometries}. Some of the results were already shown in the particular code descriptions in \autoref{sec:review} and are collected for a better comparison at this point. No results for ngsxfem are illustrated if the graphs are plotted against $\numQuadPoints$ because it was not possible to retrieve the quadrature points from this code. Further, results for monomials of degree one to five as integrand can be found in the specific code parts in \autoref{sec:review} and are omitted here for brevity.


At first, a study of $\numQuadPointsSetting$ for a single element is presented in \autoref{fig:example_unibw4_integrand_p6_nQP}. Therein, the results from Nutils are restricted to a maximum of $\numQuadPointsSetting=4$ due to code specific reasons as outlined in \autoref{sec:nutils}. From \autoref{fig:example_unibw4_integrand_p6_nQP}, a distinction between codes that introduce a larger geometrical error (FCMLab, mlhp, ngsxfem, and Nutils), e.g., by triangulation, and codes that treat the geometry accurately can be observed (Algoim, BoSSS, QUGaR, QuaHOG, and QuaHOGPE). In consequence, a lower value for $\numQuadPointsSetting$ is recommended for methods with a higher geometrical error since increasing $\numQuadPointsSetting$ would not further improve the overall error. 
The right plot shall highlight that the actual number of quadrature points cannot be solely anticipated from the parameter $\numQuadPointsSetting$. 
QuaHOGPE for example results in many points due its intermediate quadrature rule for which an appropriate number is automatically determined and not controlled by $\numQuadPointsSetting$ which only influences its antiderivative quadrature rule as explained in \autoref{sec:QuaHOG}. 
Hence, Algoim, BoSSS, and QUGaR result in the best ratio between total number of points and accuracy. QuaHOG and QuaHOGPE require more points but achieve also very high accuracies. A direct comparison with the mesh-free method of QuaHOG is only reasonable since a single element is studied. Furthermore, FCMLab and mlhp show an almost identical accuracy for a given $\numQuadPointsSetting$ since mlhp uses the quadtree method from FCMLab as a reference in its moment fitting equations. Small deviation between the two codes are observed in particular for lower values of $\numQuadPointsSetting$. Regardless, the successfully reduced actual number of quadrature points with mlhp in contrast to the FCMLab can be observed in the right plot.


\begin{figure*}
	\centering	
	\begin{subfigure}{0.48\textwidth}
		\def\tkzscale{1}
		\def\plotwidthfactor{0.85} 
		\centering
		\includegraphics{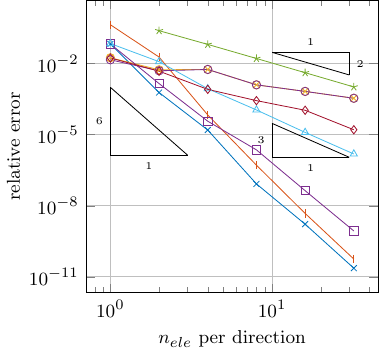}
				\label{fig:example_unibw4_integrand_href_nQP3_nele}
	\end{subfigure}
	\hfill
	\begin{subfigure}{0.48\textwidth}
		\def\tkzscale{1}
		\def\plotwidthfactor{0.85} 
		\centering
		\includegraphics{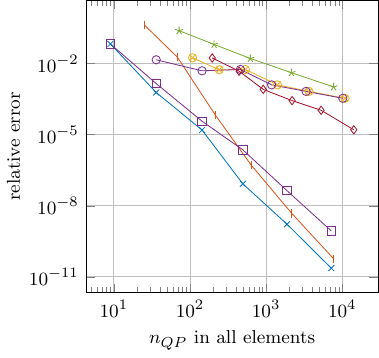}
				\label{fig:example_unibw4_integrand_href_nQP3_nQP}
	\end{subfigure}
	
	\includegraphics{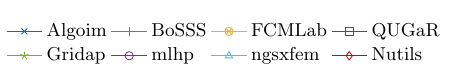}
	
%
	
	\caption{$h$-refinement study for test case geometry 2 (cf.~\autoref{fig:geo_area_single_q2}) and monomial of order $\integrandDegree=6$. $\numQuadPointsSetting=3$. Same data: left with $\numElem$ per direction and right with total $\numQuadPoints$ on $x$-axis. FCMLab and mlhp almost coincide in left plot.}	
		\label{fig:example_unibw4_integrand_href_nQP3}
\end{figure*}

\begin{figure*}
	\centering
	\begin{subfigure}{0.48\textwidth}
		\def\tkzscale{1}
		\def\plotwidthfactor{0.85} 
		\centering
		\includegraphics{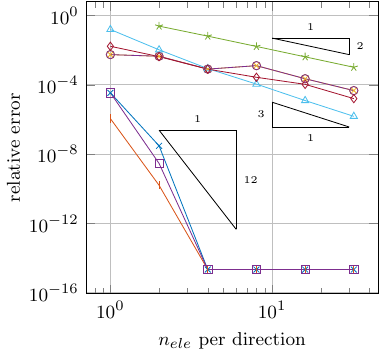}
				\label{fig:example_unibw4_integrand_href_nQP6_nele}
	\end{subfigure}
	\hfill
	\begin{subfigure}{0.48\textwidth}
		\def\tkzscale{1}
		\def\plotwidthfactor{0.85} 
		\centering
		\includegraphics{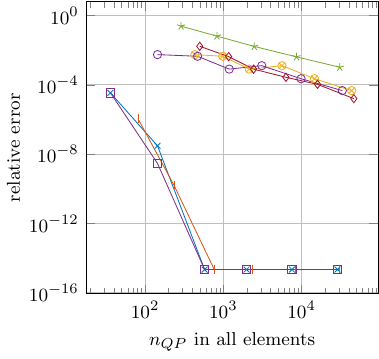}
		\label{fig:example_unibw4_integrand_href_nQP6_nQP}
	\end{subfigure}

	\includegraphics{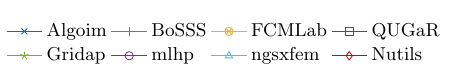}

%
	
	\caption{$h$-refinement study for test case geometry 2 (cf.~\autoref{fig:geo_area_single_q2}) and monomial of order $\integrandDegree=6$. $\numQuadPointsSetting=6$. Same data: left with $\numElem$ per direction and right with total $\numQuadPoints$ on $x$-axis. FCMLab and mlhp coincide in left plot.}	
	\label{fig:example_unibw4_integrand_href_nQP6}
\end{figure*}
In addition to the study of $\numQuadPointsSetting$, a $h$-refinement study is performed. Two different settings with $\numQuadPointsSetting=3$ and $\numQuadPointsSetting=6$ are shown in \autoref{fig:example_unibw4_integrand_href_nQP3} and \autoref{fig:example_unibw4_integrand_href_nQP6}, respectively. In both figures, the results are once plotted against $n_{ele}$ per direction and once against the total number of quadrature points $\numQuadPoints$.
The setting $\numQuadPointsSetting=6$ is considered because it demonstrates the possibility of higher convergence rates for certain methods. Please be aware of the different scales of the vertical axes when comparing results from \autoref{fig:example_unibw4_integrand_href_nQP3} with results from \autoref{fig:example_unibw4_integrand_href_nQP6}

\begin{figure*}
	\centering
	\begin{subfigure}{0.32\textwidth}
		\centering
		{\includegraphics[width=1.0\textwidth,trim={5 5 15 20},clip]{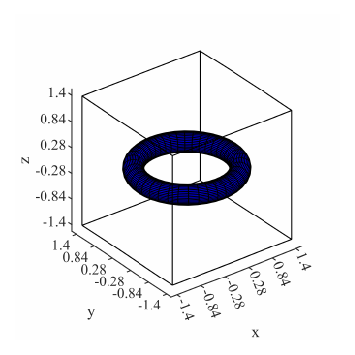}}
		\caption{Isometric view}
	\end{subfigure}
	\hfill
	\begin{subfigure}{0.32\textwidth}
		\centering
		{\includegraphics[width=1.0\textwidth,trim={5 10 15 20},clip]{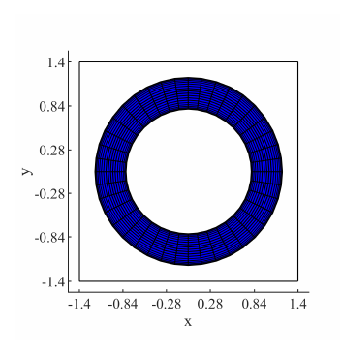}}
		\caption{Top view}
	\end{subfigure}
	\hfill
	\begin{subfigure}{0.32\textwidth}
		\centering
		{\includegraphics[width=1.0\textwidth,trim={5 10 15 20},clip]{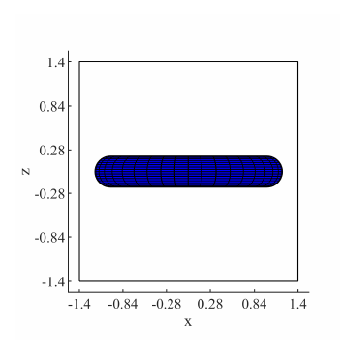}}
		\caption{Side view}
	\end{subfigure}
	\caption{Torus. The active domain is indicated by the blue surface. The lines on the surface are for illustration purposes only. The background domain is indicated by the black box and consists of $\Omega=\{\bm{x}\in [-1.4,1.4]\times [-1.4,1.4]\times [-1.4,1.4]\}$.}
	\label{fig:geo_torus}
\end{figure*}

The two choices of $\numQuadPointsSetting$ yield identical results for all codes that are limited by their geometrical approximation, i.e., Gridap, ngsxfem, and Nutils. Gridap and Nutils, which approximate the boundary linearly, show second-order convergence. The ngsxfem code performs its approximation in two steps: at first, the boundary is linearized and, then, this approximation is improved by a subsequent mapping of the background mesh which is quadratic for the considered geometry since we decided to choose the degree of this reparametrization according to the degree of the boundary curve. Consequently, ngsxfem displays third-order convergence. The expected convergence rate for these three codes is of order $\mathcal{O}^{\reparamDegree+1}$ (cf.~\autoref{eq:integrationErrorApprox} and \cite{Lehrenfeld2017}).
Whereas, Algoim, BoSSS, and QUGaR are able to obtain higher-order convergence rates of order $\mathcal{O}^{2\numQuadPointsSetting}$ since they accurately consider the actual boundary description (cf.~\autoref{eq:integrationErrorApprox} and \cite{Saye2022}).
\begin{remark}
	\autoref{eq:integrationErrorApprox} states as expected convergence rate $\mathcal{O}^{\quadDegree+1}$ for the case that the exact geometry is considered where $\quadDegree$ is the quadrature order. This results in $\mathcal{O}^{2\cdot \numQuadPointsSetting}$ by applying our in \autoref{sec:three_geometries} defined relationship $\quadDegree=2\cdot \numQuadPointsSetting-1$.
\end{remark}
FCMLab and mlhp reveal a second-order convergence with only small influence of $\numQuadPointsSetting$. The quadtree based methods (FCMLab, mlhp, and Nutils) show interestingly a better ratio between total number of quadrature points to relative error than Gridap. The triangulation on the lowest level in case of Nutils proves as a good tool to stabilize the convergence compared to the FCMLab and mlhp. Furthermore, it is again observed that mlhp successfully reduces $\numQuadPoints$ by its moment fitting compared to the FCMLab. Another observation in the comparison between the FCMLab and mlhp is that the difference in number of points also depends on the ratio of trimmed to untrimmed elements, which always decreases with further mesh refinement. For $\numQuadPointsSetting=6$, Algoim, BoSSS, and QUGaR reach machine precision with the third refinement. ngsxfem could also reach higher convergence rates with an increased 
order of its mesh transformation which is not studied herein; respective results can be found in \cite{Lehrenfeld2017}.


\subsection{Three-dimensional geometries}
\label{sec:numerical_examples_3d}

In \autoref{sec:torus}, a convergence study on the volume computation of a torus is carried out. As last example, the 
efficacy of the codes w.r.t. monomials is studied in 3D for an elliptic cylinder in \autoref{sec:monomials_3d}.
The following eight 
codes support three-dimensional computations: Algoim, BoSSS, FCMLab, mlhp, ngsxfem, Nutils and QuESo. QuESo is the only code which solely supports 3D simulations and, therefore, did not appear in the examples discussed so far.

\begin{figure*}[h]
	\centering
	\begin{subfigure}{0.48\textwidth}
		\def\tkzscale{1}
		\def\plotwidthfactor{0.85} 
		\centering
		\includegraphics{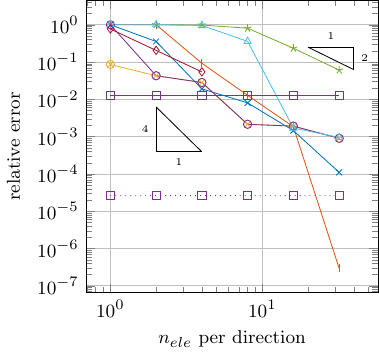}
	\end{subfigure}
	\hfill
	\begin{subfigure}{0.48\textwidth}
		\def\tkzscale{1}
		\def\plotwidthfactor{0.85} 
		\centering
		\includegraphics{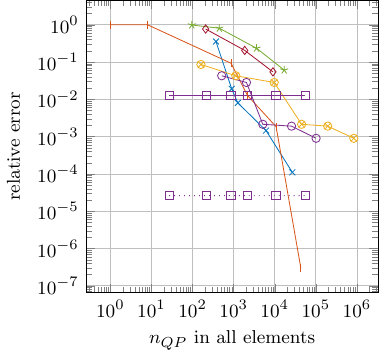}
	\end{subfigure}

	\includegraphics{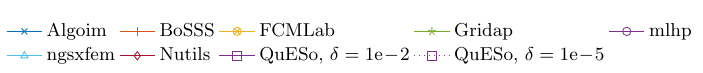}

%
	
	\caption{Volume computation for the torus given in \autoref{fig:geo_torus}. $h$-refinement study. Same data: left with $\numElem$ per direction and right with total $\numQuadPoints$ on $x$-axis. Nutils failed for the last three meshes with 8, 16, and 32 elements per direction.} 
	\label{fig:torus_results}
\end{figure*}

\subsubsection{Volume computation of a torus}\label{sec:torus}
In the following example, the volume computation of the torus depicted in \autoref{fig:geo_torus} is studied. The boundary face has degree two and is parametrically described by a NURBS surface. For the sake of brevity, only a $h$-refinement study is presented herein. The following values are chosen for $\numQuadPointsSetting$ in agreement with \autoref{sec:testsuite_unibw}: $\numQuadPointsSetting=2$ for FCMLab and mlhp, $\numQuadPointsSetting=1$ for all codes with a linear boundary approximation, and $\numQuadPointsSetting=\lceil(\curveDegree+1)/2\rceil=2$ for all remaining codes.

\autoref{fig:torus_results} presents the same data twice but with differing $x$-axis; once plotted against $\numElem$ per direction in the left figure, and once against the total number of quadrature points $\numQuadPoints$ in the right figure. The observations are similar to the 2D-examples. Second-order convergence is achieved by Gridap and Nutils. 
Algoim does not initially reach the expected fourth order convergence rate but tends towards it for the last refinement step. 
BoSSS performs similarly but even exceeds the expected rate for the last refinement step. The ngsxfem code partly falls below and partly exceeds its expected third-order rate. Furthermore, it can be observed that Gridap and ngsxfem require a certain mesh refinement to start resolving the geometry. FCMLab and mlhp show the same accuracy, but differ w.r.t.~the number of quadrature points.
Nutils fails for this example for the three finest meshes with 8, 16, and
32 elements per direction. The error message is that `leftover unmatched edges indicate the edges of baseref are not watertight'. 
The QuESo code, which is exclusively written for 3D problems, is not affected by the mesh refinement at all. The reason is that the geometry resolution is decided a-priori through the chord-tolerance $\delta$ in the STL generation. Then, the STL-geometry is very accurately integrated in QuESo which, however, does not influence the purely geometrical error in case of a volume computation. Therefore, we show results for two chord-tolerances to illustrate the influence of this parameter.

\subsubsection{Monomials as integrands for an elliptic cylinder}\label{sec:monomials_3d}

\begin{figure*}
	\centering
	\begin{subfigure}{0.32\textwidth}
		\centering
		{\includegraphics[width=\textwidth,trim={5 10 15 25},clip]{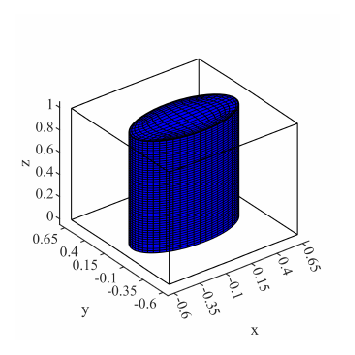}}
		\caption{Isometric view}
	\end{subfigure}
	\hfill
	\begin{subfigure}{0.32\textwidth}
		\centering
		{\includegraphics[width=\textwidth,trim={5 10 15 20},clip]{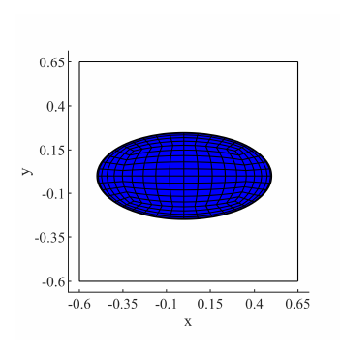}}
		\caption{Top view}
	\end{subfigure}
	\hfill
	\begin{subfigure}{0.32\textwidth}
		\centering
		{\includegraphics[width=\textwidth,trim={5 20 15 30},clip]{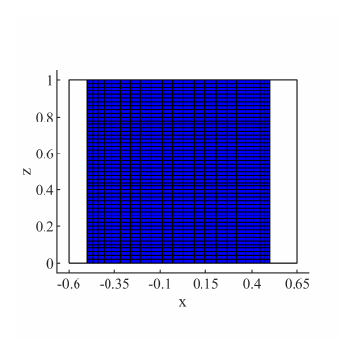}}
		\caption{Side view}
	\end{subfigure}
	\caption{Elliptic cylinder. The active domain is indicated by the blue surface. The lines on the surface are for illustration purposes only. The background domain is indicated by the black box and consists of $\Omega=\{\bm{x}\in [-0.6,0.65]\times [-0.6,0.65]\times [0,1]\}$.}
	\label{fig:geo_cylinder}
\end{figure*}

In this last example, monomials are used as integrands analogously to the 2D-example in \autoref{sec:monomials}. The considered geometry of an elliptic cylinder is illustrated in \autoref{fig:geo_cylinder}. 
The monomials $f$ have the same degree $\integrandDegree$ in all directions: $f=x^{\integrandDegree} y^{\integrandDegree} z^{\integrandDegree}$. 
Furthermore, the monomials are again shifted and scaled to the bounding box of the active domain as analogically done for the 2D case in \autoref{sec:three_geometries}.

\begin{figure*}
	\centering
	\begin{subfigure}{0.48\textwidth}
		\def\tkzscale{1}
		\def\plotwidthfactor{0.85} 
		\centering
		\includegraphics{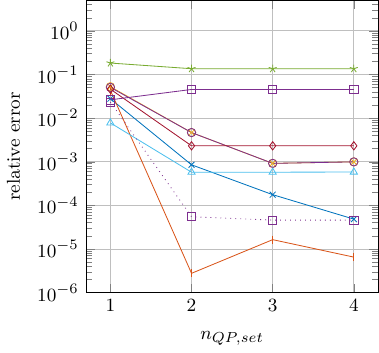}
		\caption{$\integrandDegree=3$}
		\label{fig:cylinder_integrand_p3}	
	\end{subfigure}
	\hfill	
	\begin{subfigure}{0.48\textwidth}
		\def\tkzscale{1}
		\def\plotwidthfactor{0.85} 
		\centering
		\includegraphics{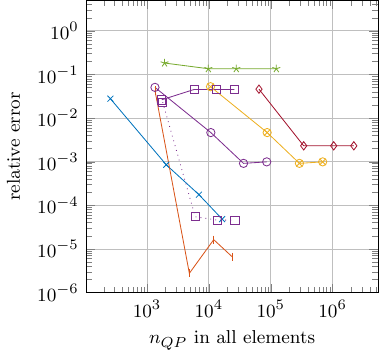}
		\caption{$\integrandDegree=3$}
		\label{fig:cylinder_integrand_p3_nQP}	
	\end{subfigure}
		
	\begin{subfigure}{0.48\textwidth}
		\def\tkzscale{1}
		\def\plotwidthfactor{0.85} 
		\centering
		\includegraphics{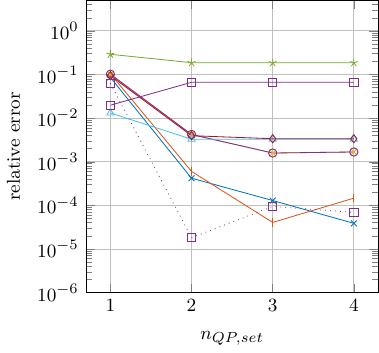}
		\caption{$\integrandDegree=5$}
		\label{fig:cylinder_integrand_p5}	
	\end{subfigure}
	\hfill
	\begin{subfigure}{0.48\textwidth}
		\def\tkzscale{1}
		\def\plotwidthfactor{0.85} 
		\centering
		\includegraphics{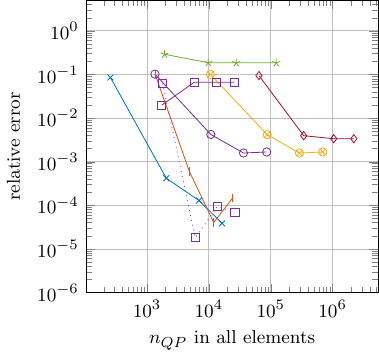}
		\caption{$\integrandDegree=5$}
		\label{fig:cylinder_integrand_p5_nQP}	
	\end{subfigure}
	
	
	
	\includegraphics{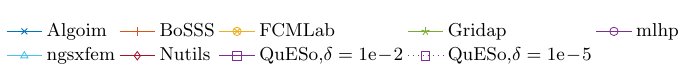}

%
	
	\caption{Study of property $\numQuadPointsSetting$ for the elliptic cylinder (cf. \autoref{fig:geo_cylinder}) with monomials of degree $\integrandDegree$ as integrands.  Same data: left with $\numQuadPointsSetting$ and right with total $\numQuadPoints$ on $x$-axis. $\numElem=8$ per direction.}
	\label{fig:cylinder_results}
\end{figure*}

At first, the influence of the number of quadrature points $\numQuadPointsSetting$ is studied again (cf.~\autoref{sec:review} and \autoref{sec:monomials}). The computation is performed for one given mesh with eight elements per direction. Monomials of degree zero to five were considered. However, the computations revealed little dependency on $\numQuadPointsSetting$ in contrast to the 2D-examples. The reason is that the codes for which a significant influence was observed in 2D are not available in 3D. The other codes are dominated by the geometrical approximation error for the given mesh meaning that they are not able to resolve the geometry accurately. Therefore, only the two monomial degrees $\integrandDegree=3$ and $\integrandDegree=5$ are illustrated in \autoref{fig:cylinder_results}.


The results are again plotted against $\numQuadPointsSetting$ (\autoref{fig:cylinder_integrand_p3} and \autoref{fig:cylinder_integrand_p5}) and $\numQuadPoints$ in all elements (\autoref{fig:cylinder_integrand_p3_nQP} and \autoref{fig:cylinder_integrand_p5_nQP}), respectively. All codes except of Algoim level off at a particular error limited by their geometrical approximation which depends on the mesh resolution. Accordingly, lower errors were observed for finer meshes in additional studies which are omitted here for brevity. Due to this limitation by the geometrical approximation, a higher $\numQuadPointsSetting$ does not necessarily improve the accuracy even though this might be expected from the value of $\integrandDegree$. 
Further observations to be made are analogical to \autoref{sec:torus}.

\begin{figure*}
	\centering
	\begin{subfigure}{0.48\textwidth}
		\def\tkzscale{1}
		\def\plotwidthfactor{0.85} 
		\centering
		\includegraphics{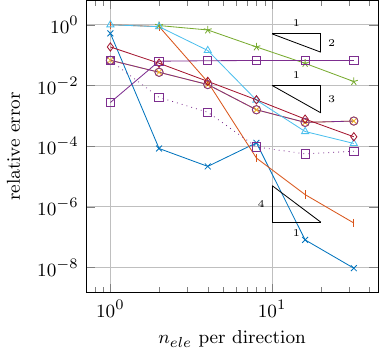}
		\label{fig:cylinder_integrand_p5_href}	
	\end{subfigure}
	\hfill	
	\begin{subfigure}{0.48\textwidth}
		\def\tkzscale{1}
		\def\plotwidthfactor{0.85} 
		\centering
		\includegraphics{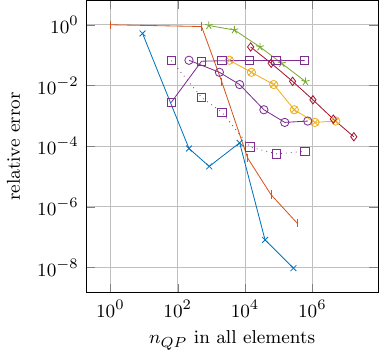}
		\label{fig:cylinder_integrand_p5_href_nQP}	
	\end{subfigure}
	
	\includegraphics{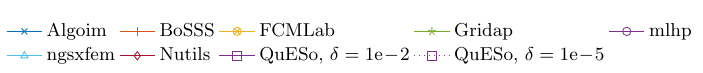}
	
%
	
	\caption{$h$-refinement study for the elliptic cylinder (cf. \autoref{fig:geo_cylinder}) with monomial of order $\integrandDegree=5$ as integrand. Same data: left with $\numElem$ per direction and right with total $\numQuadPoints$ on $x$-axis. FCMLab and mlhp coincide in left plot.}
	\label{fig:cylinder_results_href}
\end{figure*}

In addition, a $h$-refinement study is performed for $\integrandDegree=5$. The number of quadrature points is chosen as $\numQuadPointsSetting=n_{GP}= \left\lceil(\integrandDegree + 1)/2\right\rceil=3$ (common Gauss rule; $n_{GP}$ is the number of Gauss points) for all codes since no strong recommendation for this setting could be obtained from \autoref{fig:cylinder_results}. In \autoref{fig:cylinder_results_href}, the results are again plotted against $\numQuadPointsSetting$ in the left figure and against the total number of quadrature points $\numQuadPoints$ in the right figure. 
Gridap and BoSSS reach again their expected second respectively fourth order convergence rate.
Algoim shows pre-asymptotic behaviour for meshes with $\numElem$ being less than $10^1$. Nevertheless, it reaches the lowest relative error of all codes for further refinement.
The ngsxfem code again partly falls below and exceeds it expected third order rate.
Further observations to be made agree with the previous sections. 

\section{Conclusion and Outlook}
\label{sec:conclusion}

Open-source codes improve reproducibility and accelerate research. 
In this work, we identify, review, and compare a comprehensive set of freely available codes dealing with cut elements.
They cover a large variety of methods for the numerical integration over cut elements. Numerical benchmark tests computing domain integrals were carried out in 2D and 3D in order to test versatility, robustness, accuracy, and efficiency of the different tools. All codes have proven to be robust and ready to use for the tested scenarios. In particular, the required quadrature order, respectively, the number of quadrature points $\numQuadPointsSetting$, was investigated for three basic 2D test cases, and for two 3D examples.

A Matlab interface is provided as open-source repository that allows running all considered codes from a common test environment \cite{CutElementIntegration2025LatestVersion}. Hereby, it was possible to retrieve the final quadrature points from most of the codes. This enables their reuse or embedding in other solvers and programs. Certain cross-language bindings occasionally required operating system specific steps and additional dependencies.

The review revealed that the open-source programs can be distinguished in codes which tessellate the boundary and codes that preserve the (possibly) higher-order boundary or at least a higher-order approximation of it. The first group introduces a geometrical error leading to limited second-order convergence, whereas the second group could also reach higher-order convergence. In the end, it has to be remembered that it depends on the problem whether a very accurate integration is at all needed\footnote{Strang and Fix \cite[pg. 181]{strang_analysis_2008}: `What degree of accuracy in the integration formula is required for convergence? It is not required
that every polynomial which appears be integrated exactly.'}.

Despite the variety of available codes, a serious gap remains: No open-source codes are available dealing with parametric boundary descriptions in combination with volumes even though some work was published in this field \cite{antolin_isogeometric_2019,Antolin2022}. Such software would be of particular interest in a CAD-based workflow where geometries are often provided as Boundary Representations (B-Rep) \cite{breitenberger_cad-integrated_2016}.

Not all possible settings for each code were studied aiming for a certain conciseness. Exploring the following parameters could further illuminate the accuracy-versus-cost trade-offs and the robustness margins:
\begin{itemize}
	\item reparametrization degree
	\item number of subdivision levels
	\item moment fitting tolerances
\end{itemize}
Furthermore, the different treatments of boundary integrals in the context of immersed methods could be reviewed and compared in future studies.

Finally, we invite readers to utilize the test environment \cite{CutElementIntegration2025LatestVersion} to benchmark their own implementations against the codes compared in this work.

\backmatter

\bmhead*{Acknowledgements}
We gratefully thank Pablo Antolin for providing us with a Matlab interface for his reparametrization method in QUGaR. Furthermore, we appreciate the open and helpful communication with several authors of the reviewed codes.

\section*{Declarations}

\bmhead*{Funding}
Florian Kummer, 
Benjamin Marussig, 
Irina Shishkina, Guilherme H.~Teixeira, and Teoman Toprak 
kindly acknowledge the financial support by the joint DFG/FWF Collaborative Research Centre CREATOR (CRC – TRR361/F90) at TU Darmstadt, TU Graz and JKU Linz. Chen Miao kindly acknowledges the financial support by the Deutsche Forschungsgemeinschaft (DFG, German Research Foundation) – 422800359.
Additionally, the work of Irina Shishkina, Teoman Toprak and Florian Kummer is supported by the Graduate School CE within the Centre for Computational Engineering at TU Darmstadt.

\bmhead*{Code and data availability}
The code that was developed and employed for all computations within this work is publicly available on \cite{CutElementIntegration2025LatestVersion}. \autoref{sec:supplemental_repository} outlines which script was used to produce which results.

\begin{appendices}
\section{Geometries from testsuite with diverse cutting scenarios}
\label{sec:app_testsuite_unibw}

\autoref{fig:geo_testsuite_unibw} presents all geometries used in \autoref{sec:testsuite_unibw}.
\begin{figure}[h!]
	\centering

	\begin{subfigure}{0.24\textwidth}
		\centering
		{\includegraphics[width=\textwidth,trim={0 10 0 22},clip]{testCaseId9.pdf}}
		\caption{Test case 1}
	\end{subfigure}%
	\begin{subfigure}{0.24\textwidth}
		\centering
		\includegraphics[width=\textwidth,trim={0 10 0 22},clip]{testCaseId10.pdf}
		\caption{Test case 2}
	\end{subfigure}

	\begin{subfigure}{0.24\textwidth}
		\centering
		\includegraphics[width=\textwidth,trim={0 10 0 22},clip]{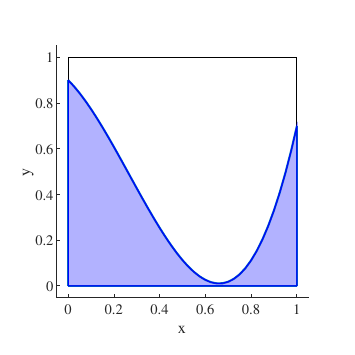}
		\caption{Test case 3}
	\end{subfigure}%
	\begin{subfigure}{0.24\textwidth}
		\centering
		\includegraphics[width=\textwidth,trim={0 10 0 22},clip]{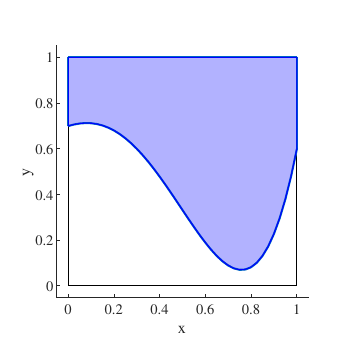}
		\caption{Test case 4}
	\end{subfigure}

	\begin{subfigure}{0.24\textwidth}
		\centering
		\includegraphics[width=\textwidth,trim={0 10 0 22},clip]{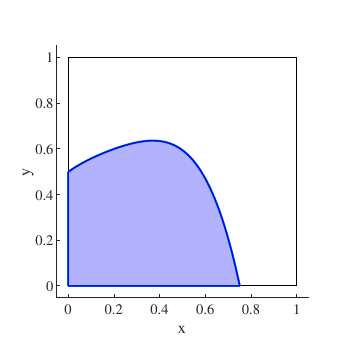}
		\caption{Test case 5}
	\end{subfigure}%
	\begin{subfigure}{0.24\textwidth}
		\centering
		\includegraphics[width=\textwidth,trim={0 10 0 22},clip]{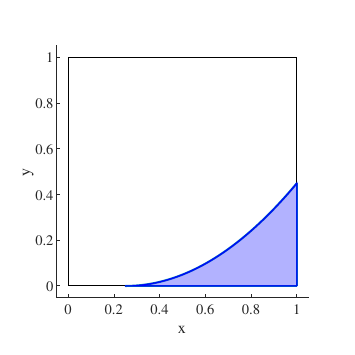}
		\caption{Test case 6}
		\label{fig:testsuite_unibw_TC6}
	\end{subfigure}

	\begin{subfigure}{0.24\textwidth}
		\centering
		\includegraphics[width=\textwidth,trim={0 10 0 22},clip]{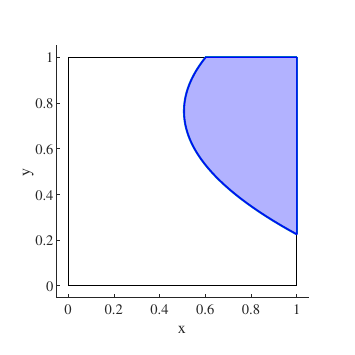}
		\caption{Test case 7}
	\end{subfigure}%
	\begin{subfigure}{0.24\textwidth}
		\centering
		\includegraphics[width=\textwidth,trim={0 10 0 22},clip]{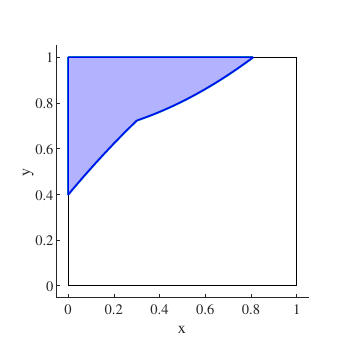}
		\caption{Test case 8}
		\label{fig:testsuite_unibw_TC8}
	\end{subfigure}
		\begin{subfigure}{0.24\textwidth}
		\centering
		\includegraphics[width=\textwidth,trim={0 10 0 22},clip]{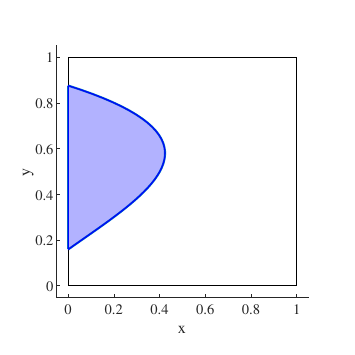}
		\caption{Test case 9}
	\end{subfigure}%
	\begin{subfigure}{0.24\textwidth}
		\centering
		\includegraphics[width=\textwidth,trim={0 10 0 22},clip]{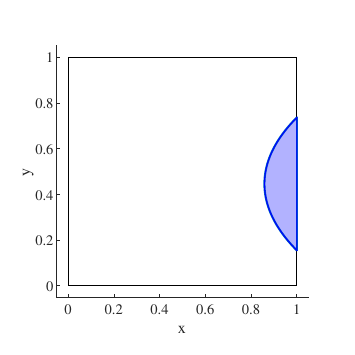}
		\caption{Test case 10}
	\end{subfigure}%

	\caption{Geometries from testsuite with diverse cutting scenarios}
\end{figure}
\begin{figure}[h!]
	\ContinuedFloat
	\centering
	
	\begin{subfigure}{0.24\textwidth}
		\centering
		\includegraphics[width=\textwidth,trim={0 10 0 22},clip]{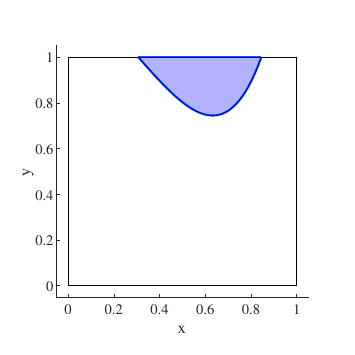}
		\caption{Test case 11}
	\end{subfigure}%
	\begin{subfigure}{0.24\textwidth}
		\centering
		\includegraphics[width=\textwidth,trim={0 10 0 22},clip]{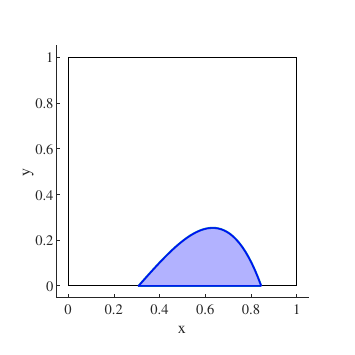}
		\caption{Test case 12}
	\end{subfigure}%
		
	\begin{subfigure}{0.24\textwidth}
		\centering
		\includegraphics[width=\textwidth,trim={0 10 0 22},clip]{testCaseId21.pdf}
		\caption{Test case 13}
	\end{subfigure}%
	\begin{subfigure}{0.24\textwidth}
		\centering
		\includegraphics[width=\textwidth,trim={0 10 0 22},clip]{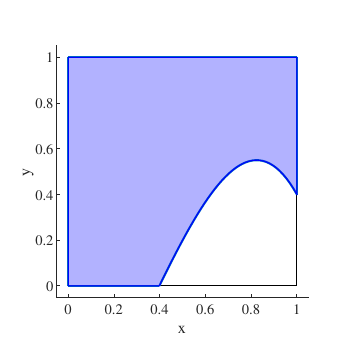}
		\caption{Test case 14}
	\end{subfigure}%

	\begin{subfigure}{0.24\textwidth}
		\centering
		\includegraphics[width=\textwidth,trim={0 10 0 22},clip]{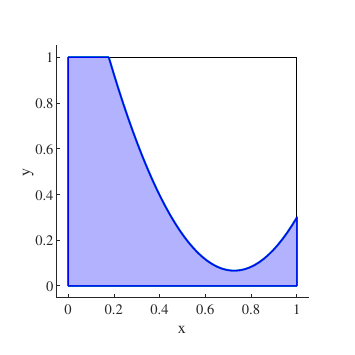}
		\caption{Test case 15}
	\end{subfigure}%
	\begin{subfigure}{0.24\textwidth}
		\centering
		\includegraphics[width=\textwidth,trim={0 10 0 22},clip]{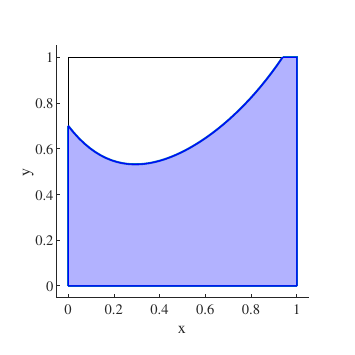}
		\caption{Test case 16}
	\end{subfigure}
	
	\begin{subfigure}{0.24\textwidth}
		\centering
		\includegraphics[width=\textwidth,trim={0 10 0 22},clip]{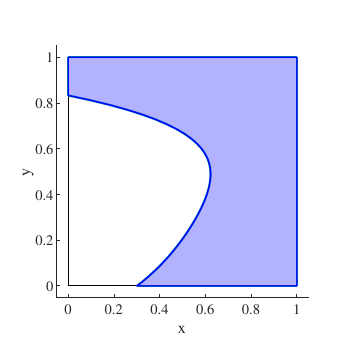}
		\caption{Test case 17}
	\end{subfigure}%
	\begin{subfigure}{0.24\textwidth}
		\centering
		\includegraphics[width=\textwidth,trim={0 10 0 22},clip]{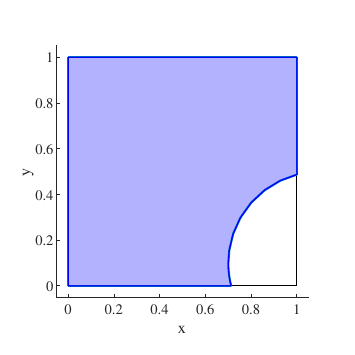}
		\caption{Test case 18}
		\label{fig:testsuite_unibw_TC18}
	\end{subfigure}

	\begin{subfigure}{0.24\textwidth}
		\centering
		\includegraphics[width=\textwidth,trim={0 10 0 22},clip]{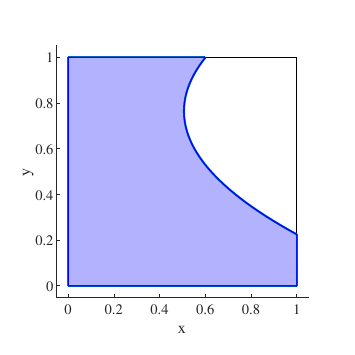}
		\caption{Test case 19}
	\end{subfigure}%
	\begin{subfigure}{0.24\textwidth}
		\centering
		\includegraphics[width=\textwidth,trim={0 10 0 22},clip]{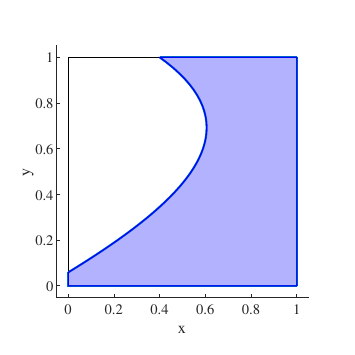}
		\caption{Test case 20}
	\end{subfigure}
	
	\caption{Geometries from testsuite with diverse cutting scenarios (continued)}
\end{figure}

\begin{figure}[h!]
	\ContinuedFloat
	\centering

	\begin{subfigure}{0.24\textwidth}
		\centering
		\includegraphics[width=\textwidth,trim={0 10 0 22},clip]{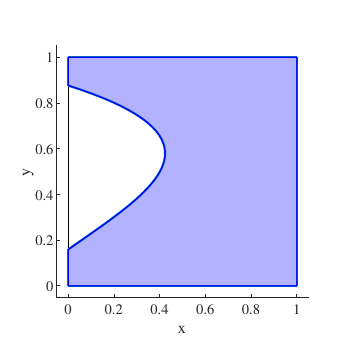}
		\caption{Test case 21}
	\end{subfigure}%
	\begin{subfigure}{0.24\textwidth}
		\centering
		\includegraphics[width=\textwidth,trim={0 10 0 22},clip]{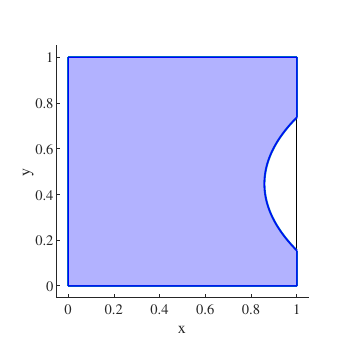}
		\caption{Test case 22}
	\end{subfigure}

	\begin{subfigure}{0.24\textwidth}
		\centering
		\includegraphics[width=\textwidth,trim={0 10 0 22},clip]{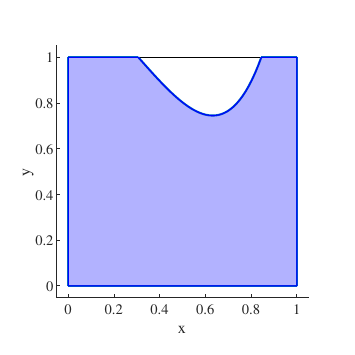}
		\caption{Test case 23}
	\end{subfigure}%
	\begin{subfigure}{0.24\textwidth}
		\centering
		\includegraphics[width=\textwidth,trim={0 10 0 22},clip]{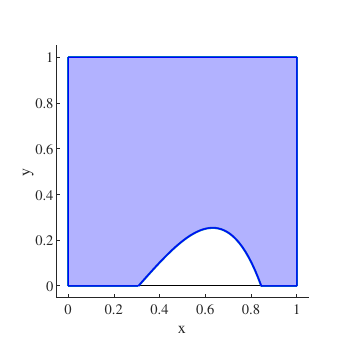}
		\caption{Test case 24}
	\end{subfigure}

	\begin{subfigure}{0.24\textwidth}
		\centering
		\includegraphics[width=\textwidth,trim={0 10 0 22},clip]{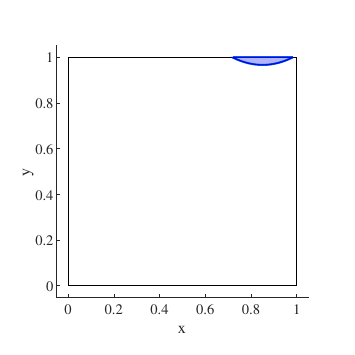}
		\caption{Test case 25}
	\end{subfigure}%
	\begin{subfigure}{0.24\textwidth}
		\centering
		\includegraphics[width=\textwidth,trim={0 10 0 22},clip]{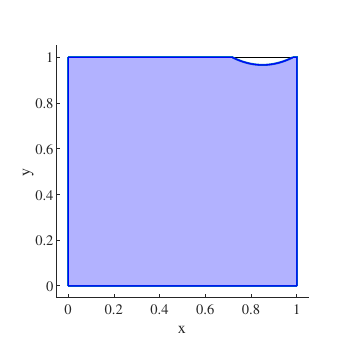}
		\caption{Test case 26}
	\end{subfigure}

	\caption{Geometries from testsuite with diverse cutting scenarios (continued)}
	\label{fig:geo_testsuite_unibw}
\end{figure}

\section{Supplemental open-source repository}
\label{sec:supplemental_repository}

All results presented in this paper were produced by a supplemental open-source repository \cite{CutElementIntegration2025LatestVersion}. \autoref{tab:repo_examples} links the generated results with the respective scripts. They are all stored in the folder `publications/Paper\_CutElemComp'. Be aware that certain settings might have been varied to produce particular results as stated throughout this paper. 

\begin{table*}[h]
	\caption{Link between filenames of scripts in the open-source repository and herein presented results. All files are Matlab scripts and, thereby, have the file ending `.m'.}
	\label{tab:repo_examples}
	\centering
	\begin{tabular}{@{}p{8cm}p{6cm}@{}}
		\toprule
		Filename & Result reference \\
		\midrule
		plot\_geometries & \Autorefs{fig:geo_area_single}, \ref{fig:geo_testsuite_unibw} \\ 
		CutElemComp\_example\_unibw3 & \Autorefs{fig:algoim_area}, \ref{fig:quahog_area}, \ref{fig:bosss_area}, \ref{fig:mlhp_area}, \ref{fig:fcmlab_area}, \ref{fig:nutils_area}, \ref{fig:qugar_area} \\ 
		CutElemComp\_example\_unibw4 & \Autorefs{fig:algoim_area}, \ref{fig:quahog_area}, \ref{fig:bosss_area}, \ref{fig:mlhp_area}, \ref{fig:fcmlab_area}, \ref{fig:nutils_area}, \ref{fig:qugar_area} \\ 
		CutElemComp\_example\_unibw16 & \Autorefs{fig:algoim_area}, \ref{fig:quahog_area}, \ref{fig:bosss_area}, \ref{fig:mlhp_area}, \ref{fig:fcmlab_area}, \ref{fig:nutils_area}, \ref{fig:qugar_area} \\ 
		CutElemComp\_example\_unibw3\_integrand & \Autorefs{fig:algoim_integrand}, \ref{fig:quahog_integrand}, \ref{fig:quahogpe_integrand}, \ref{fig:bosss_integrand}, \ref{fig:mlhp_integrand}, \ref{fig:fcmlab_integrand}, \ref{fig:nutils_integrand}, \ref{fig:ngsxfem_integrand}, \ref{fig:qugar_integrand} \\ 
		CutElemComp\_example\_unibw4\_integrand & \Autorefs{fig:algoim_integrand}, \ref{fig:quahog_integrand}, \ref{fig:quahogpe_integrand}, \ref{fig:bosss_integrand}, \ref{fig:mlhp_integrand}, \ref{fig:fcmlab_integrand}, \ref{fig:nutils_integrand}, \ref{fig:qugar_integrand}, \ref{fig:example_unibw4_integrand_p6_nQP} \\ 
		CutElemComp\_example\_unibw16\_integrand & \Autorefs{fig:algoim_integrand}, \ref{fig:quahog_integrand}, \ref{fig:quahogpe_integrand}, \ref{fig:bosss_integrand}, \ref{fig:mlhp_integrand}, \ref{fig:fcmlab_integrand}, \ref{fig:nutils_integrand}, \ref{fig:qugar_integrand} \\ 
		CutElemComp\_example\_unibw3\_href & \Autorefs{fig:mlhp_href}, \ref{fig:fcmlab_href}, \ref{fig:nutils_href}, \ref{fig:gridap_href} \\ 
		CutElemComp\_example\_unibw4\_href & \Autorefs{fig:mlhp_href}, \ref{fig:fcmlab_href}, \ref{fig:nutils_href}, \ref{fig:gridap_href}, \ref{fig:ngsxfem_href} \\ 
		CutElemComp\_example\_unibw16\_href & \Autorefs{fig:mlhp_href}, \ref{fig:fcmlab_href}, \ref{fig:nutils_href}, \ref{fig:gridap_href}, \ref{fig:ngsxfem_href} \\ 
		CutElemComp\_example\_testsuite\_unibw & \autoref{fig:testsuite_unibw} \\ 
		CutElemComp\_example\_parabola\_moving & \autoref{fig:example_moving_parabola_results}, \autoref{tab:example_moving_parabola} \\ 
		CutElemComp\_example\_unibw4\_integrand\_href & \Autorefs{fig:example_unibw4_integrand_href_nQP3}, \ref{fig:example_unibw4_integrand_href_nQP6} \\ 
		CutElemComp\_example\_unibw16\_QP & \autoref{fig:example_unibw16_QP} \\ 
		CutElemComp\_example\_torus\_1 & \autoref{fig:torus_results} \\ 
		CutElemComp\_example\_cylinder\_polynomial & \autoref{fig:cylinder_results} \\ 
		CutElemComp\_example\_cylinder\_polynomial\_href & \autoref{fig:cylinder_results_href} \\		
		\botrule
	\end{tabular}
\end{table*}

\end{appendices}

\bibliography{mybibfile}

\end{document}